\documentclass{aa}
\usepackage{graphicx}
\usepackage{txfonts}
\begin{document}
   \title{The INTEGRAL Galactic bulge monitoring program: the first 1.5 years}

   \author{E. Kuulkers\inst{1}
          \and
          S.E. Shaw\inst{2,}\inst{3}
          \and
	  A. Paizis\inst{4}
          \and 
          J. Chenevez\inst{5}
          \and
          S. Brandt\inst{5}
          \and
          T.J.-L. Courvoisier\inst{3,}\inst{6}
          \and
          A. Domingo\inst{7}
          \and
	  K. Ebisawa\inst{8}
          \and
	  P. Kretschmar\inst{1}
          \and
          C.B. Markwardt\inst{9,}\inst{10}
          \and
	  N. Mowlavi\inst{3}
          \and
	  T. Oosterbroek\inst{11}
	  \and
	  A. Orr\inst{12}
	  \and
          D. R\'isquez\inst{7}
	  \and
	  C. Sanchez-Fernandez\inst{1}
	  \and
	  R. Wijnands\inst{13}
	  }

   \authorrunning{E.~Kuulkers et al.}
   \titlerunning{INTEGRAL Galactic bulge monitoring program}

   \offprints{E. Kuulkers}

   \institute{ISOC, ESA/ESAC, Urb.\ Villafranca del Castillo,
           P.O.\ Box 50727, 28080 Madrid, Spain
              \email{Erik.Kuulkers@esa.int}
          \and
              School of Physics and Astronomy, University of Southampton, SO17 1BJ,
              Southampton, UK
          \and
              INTEGRAL Science Data Centre (ISDC), 16 Chemin d'Ecogia, CH-1290 Versoix, Switzerland
          \and
	      INAF-IASF, Sezione di Milano, Via Bassini 15, 20133 Milano, Italy
          \and
              Danish National Space Center, Juliane Maries Vej 30, DK-2100 Copenhagen, Denmark
          \and
              Observatoire de Gen\`eve, 51 Chemin des Mailletes, CH-1290 Sauverny, Switzerland
          \and
              Laboratorio de Astrof\'isica Espacial y F\'isica Fundamental, INTA, Apartado 50727, 
              28080 Madrid, Spain
          \and
	      Center for PLAnning and INformation systems, Institute of Space and Astronautical Science,
              Yoshinodai 3-1-1 Sagamihara, Kanagawa, 229-8510, Japan
          \and
	      Department of Astronomy, University of Maryland, College Park, MD 20742, USA
          \and
	      X-ray Astrophysics Laboratory, Mail Code 662, NASA Goddard Space Flight Center, Greenbelt,
              MD 20771, USA
          \and
	      Science Payload and Advanced Concepts Office, ESA-ESTEC, Postbus 299, NL-2200 AG, Noordwijk, The Netherlands
          \and
              Research and Scientific Support Department, ESA-ESTEC, Postbus 299, NL-2200 AG, Noordwijk, The
              Netherlands
          \and
	      Astronomical Institute ``Anton Pannekoek'', University of Amsterdam, Kruislaan 403, 1098 SJ Amsterdam, The Netherlands
             }

   \date{Received; accepted}

  \abstract
   {}
   {The Galactic bulge region is a rich host of variable high-energy point sources.
Since 2005, February 17 we are monitoring the source activity in the Galactic bulge region regularly and frequently,
i.e., about every three days, with the instruments onboard {\em INTEGRAL}. 
Thanks to the large field of view, the imaging capabilities and the sensitivity at hard X-rays,
we are able to present for the first time a detailed homogeneous (hard) X-ray view of a sample of 
76 sources in the Galactic bulge region.}
   {We describe the successful monitoring program and show the first results from the start of the monitoring 
up to 2006, April 21, i.e., for a period of about one and a half year, during three visibility seasons.
We focus on the short (hour), medium
(month) and long-term (year) variability in the hard X-ray bands, i.e., 20--60\,keV and 60--150\,keV.
When available, we discuss the simultaneous observations 
in the soft X-ray, 3--10\,keV and 10--25\,keV, bands.}
   {Almost all the sources in the Galactic bulge region we detect in the 20--60\,keV and 60--150\,keV bands are
variable. During the last two and a half weeks 
of the third visibility season most of the 
known persistent (hard) X-ray sources in the Galactic Center region were not detected. 
Of our sample of sources, per visibility season we detect 32/33 sources in the 20--60\,keV band 
and 8/9 sources in the 60--150\,keV band above a signal to noise of 7. 
On average, we find per visibility season one active bright 
($\gtrsim$100\,mCrab, 20--60\,keV) black-hole candidate X-ray transient and three
active weaker ($\lesssim$25\,mCrab, 20--60\,keV) neutron star X-ray transients.

Most of the time a clear anti-correlation can be seen between the soft and hard X-ray emission
in some of the X-ray bursters.
Hard X-ray flares or outbursts in X-ray bursters, which have a duration of the order of weeks are accompanied
by soft X-ray drops.
On the other hand, hard X-ray drops can be accompanied 
by soft X-ray flares/outbursts.

During the course of our program we found a number of new sources, IGR\,J17354$-$3255, IGR\,17453$-$2853, IGR\,J17454$-$2703, 
IGR\,J17456$-$2901b, IGR\,J17536$-$2339, and IGR\,J17541$-$2252.
We report here on some of the high-energy properties of these sources.}
   {The high-energy light curves of all the 
sources in the field of view, and the high-energy images of the region, are made available through the WWW,
as soon as possible after the observations have been performed, at {\tt http://isdc.unige.ch/Science/BULGE/}.}


   \maketitle
%

\section{Introduction}

The bulge of our Galaxy hosts a variety of
hard X-ray and $\gamma$-ray point sources
(e.g., Knight et al.\ 1985, Skinner et al.\ 1987, Churazov et al.\ 1994; 
see, e.g., Bird et al.\ 2006, B\'elanger et al.\ 2006, 
Revnivtsev et al.\ 2004a, 
for observations made by {\em INTEGRAL}, the {\bf Inte}rnational 
{\bf G}amma-{\bf R}ay {\bf A}strophysics {\bf L}aboratory;
Winkler et al.\ 2003).
Among them are persistent and transient neutron-star and
black-hole (candidate) binaries, as well as magnetic cataclysmic variables and AGN. 
Due to the variability these sources possess on time scales of milliseconds to days
(quasi-periodic oscillations, pulsations, [absorption] dips, eclipses, type I and
type II X-ray bursts, orbital variations, flares) and weeks to years (orbital
variations, outburst cycles, on/off states), the region never looks exactly the same.

Hard X-ray ($\gtrsim$20\,keV) emission emerging from the Galactic bulge sources
mainly comes from highly energetic processes occurring in the course of
accretion from a donor star on to the compact object, i.e., a white dwarf,
neutron star or black hole. Since most of the emission
comes from the region close to these compact objects, studying the 
hard X-ray and $\gamma$-ray emission will give us more insight
into the accretion processes under extreme conditions, as well as
possibly identifying the nature of the compact accretor.

From 17 February 2005 onwards, whenever the Galactic bulge region 
was visible by {\em INTEGRAL}, we have been monitoring this region
approximately every 3 days.
This paper serves mainly as a description of this program in more detail 
and we show the first results, spanning times scales
between half an hour and one and a half year.
Preliminary announcements of some of these results were made by 
Bodghee et al.\ (2005), Brandt et al.\ (2005), Chenevez et al.\ (2006a,b),
Kretschmar et al.\ (2005), Kuulkers et al.\ (2005a,b, 2006a), 
Mowlavi et al.\ (2005), Shaw et al.\ (2005a,b,c, 2006) and Turler et al.\ (2006).
A preliminary report of our program was presented by Kuulkers et al.\ (2006b).

\section{INTEGRAL and the Galactic bulge monitoring program}

Since X-rays typically above 10\,keV are difficult to focus
using currently known reflecting material, one has to revert to other means of
imaging techniques when investigating crowded regions. Coded masks are effective imagers
in this energy range, as, e.g., shown by the first hard X-ray (20--30\,keV)
images of the region around the Galactic Center (Skinner et al.\ 1987).

{\em INTEGRAL} is an ESA scientific mission dedicated to fine spectroscopy 
($E/\Delta E$$\simeq$500; SPI; Vedrenne et al.\ 2003) and fine imaging (angular 
resolution: 12' FWHM, source location accuracy: $\simeq$1--3'; IBIS; Ubertini et al.\ 2003)
of celestial $\gamma$-ray sources in the energy range 15\,keV to 10\,MeV
with simultaneous monitoring in the X-ray (3--35\,keV, angular resolution: 3'; JEM-X; 
Lund et al.\ 2003) and optical (V-band, 550\,nm; OMC; Mas-Hesse et al.\ 2003) energy ranges.
All the instruments onboard {\em INTEGRAL}, except the OMC, have coded masks.

{\em INTEGRAL} has already spent a considerable amount of observing time
on the Galactic bulge region, providing a deep insight (see, e.g., B\'elanger et al.\ 2006).
Our program, however, was initiated to monitor, for the first time with {\em INTEGRAL}, this
region frequently on a regular basis at hard X-rays and $\gamma$-rays. 
The main aim is to investigate the source variability and transient activity on 
time scales of hours to days to weeks to months
simultaneously at relatively soft ($\lesssim$10\,keV) and hard ($\gtrsim$10\,keV)
energies. One complete hexagonal dither pattern (7 pointings of $\simeq$1800\,s each, i.e.,
1 on-axis pointing, 6 off-source pointings in a hexagonal pattern around
the nominal target location, each 2$^{\circ}$ apart)
is performed during each {\em INTEGRAL} revolution (orbit) around the earth, i.e., 
roughly every 3 days.
This yields a total coverage around the Galactic Center of 29$^{\circ}$ with 
IBIS/ISGRI and about 6$^{\circ}$ with JEM-X for a total exposure of 12.6\,ks.
With the fully and partially coded field of view we cover about half of the 
Galactic bulge X-ray binary population with IBIS/ISGRI (see also, e.g., in 't Zand 2001, 
in 't Zand et al.\ 2004).

The hexagonal dither pattern is done whenever the region is visible by {\em INTEGRAL}
(about two times per year for a total period of about 4 months).
We refer to each visibility period as `season'; in this paper we report on observations
made during three seasons.
As a service to the scientific community, the JEM-X light curves (3--10\,keV
and 10--25\,keV) and the IBIS/ISGRI light
curves (20--60\,keV and 60--150\,keV) are made
publicly available as soon as possible after the
observations are performed, both in graphic and in ASCII form. 
In addition, IBIS/ISGRI and JEM-X mosaic images (i.e., composite images made of overlapping images) 
of each hexagonal dither observation are provided, with information on
the detected sources. Finally, all
IBIS/ISGRI 20--60\,keV mosaic images per revolution
are stacked into a movie, showing the
ever-changing hard X-ray/$\gamma$-ray sky.
The results, as well as more information about
the program, can be retrieved from the {\em INTEGRAL} Galactic bulge Monitoring WWW
page hosted at the {\em INTEGRAL} Science Data Center (ISDC) in Switzerland: {\tt http://isdc.unige.ch/Science/BULGE/}.

Similar Galactic bulge monitoring programs at various soft and hard X-ray energies
have been performed (see, e.g., in 't Zand 2001). For example,
{\em GRANAT}/SIGMA (30\,keV--1\,MeV) performed regular observations of the Galactic Center region
between 1990 and 1994 in comparable energy ranges, i.e., above $>$35\,keV, and 
with comparable visibility periods throughout the year (see, e.g., Churazov et al.\ 1993). 
The sensitivity, angular resolution and source location accuracy 
of {\em GRANAT}/SIGMA (respectively, 5$\sigma$ detection level of $\simeq$10\,mCrab at 40--100\,keV for an exposure time
of $\simeq$9\,Ms at the Galactic Center, $\simeq$15' and $\simeq$2--3'; see
Revnivtsev et al.\ 2004b) are, however, not as good as {\em INTEGRAL}/IBIS (5$\sigma$ detection level of 
$\simeq$1\,mCrab at 40--100\,keV for an exposure time of $\simeq$1.5\,Ms at the Galactic Center, 
see Bird et al.\ 2006).

Some monitoring programs are currently ongoing, such as the {\em RXTE} Galactic bulge Scans (Swank \&\ Markwardt 2001;
Markwardt 2006).
However, the {\em RXTE}/PCA (2--60\,keV) and HEXTE (15--250\,keV) do only have 
a 2$^{\circ}$ collimator and 
no imaging resolution, providing only information on a given source for a short time
when the instrument scans over it; moreover, in the Galactic Center region
itself there is significant source confusion. There are currently other instruments 
in operation at similar energy ranges as those covered by IBIS/ISGRI, 
such as {\em Swift}/BAT (15--150\,keV with a field-of view of 2 steradians; Barthelmy 2000). 
However, they have a worse imaging capability, again leading to source confusion 
in the Galactic Center region ({\em Swift}/BAT PSF angular resolution is 22'), and
a lower sensitivity (3$\sigma$ detection level of $\simeq$27\,mCrab at 15--50\,keV
for an exposure of 736\,s; see Krimm et al.\ 2006).

In this paper we show the results of three full seasons of our AO-3 monitoring,
about 1.5 years, i.e., from 2005, February 18 -- April 19 ({\em INTEGRAL} revolutions 287--307;
MJD\,53419--53479), 2005, August 16 -- October 26 (revolutions 347--370; MJD\,53598--54034), 
and 2006, February 9 -- April 21 (revolutions 406--429; MJD\,53775--53846).
We note that the program continues in {\em INTEGRAL} AO-4
(see Kuulkers et al.\ 2006c), and 
we intend to extend this program into further Announcement of
Opportunities with {\em INTEGRAL}.
In the next Section we explain how the data analysis is done
(Sect.~\ref{data_analysis}). We then summarize  
the observations per season by focusing on the averaged (or mosaiced) 
images and corresponding source fluxes and detection significances (Sect.~\ref{images}). 
The long-term hard X-ray ($>$20\,keV) light curves of the monitoring program
are shown in Sect.~\ref{light_curves}. In Sect.~\ref{off} we focus on an 
interesting period when there was almost no activity in the region around the Galactic center.
New {\rm INTEGRAL} sources found by our program are discussed in Sect.~\ref{IGR}.
Results from the monitoring with the OMC are given in Sect.~\ref{omc}.
Finally, we summarize and make our concluding remarks in Sect.~\ref{discussions}.

\section{Data analysis}
\label{data_analysis}

In the current paper we only consider data from IBIS/ISGRI 
(Lebrun et al.\ 2003), JEM-X and OMC. We do not consider the data from
IBIS/PICsIT (Labanti et al.\ 2003) or SPI instruments:
either the angular resolution is limited (SPI: 2.5$^{\circ}$) and therefore the various sources
in the Galactic bulge region close to each other complicate the
analysis (see, e.g., Bouchet et al.\ 2005), or the sources are too weak to be detected (IBIS/PICsIT,
which operates at $\gtrsim$175\,keV).

The {\em INTEGRAL} IBIS and JEM-X data reduction is performed using
the Off-line Scientific Analysis ({\tt OSA};
see Courvoisier et al.\ 2003), v5.1. In our analysis we use an 
input source catalog, containing a total of 76 sources (see Table~\ref{sourcelist}). 
Most of these sources had been previously detected by IBIS/ISGRI (see, e.g., Revnivtsev et al.\ 2004a; 
Bird et al.\ 2006). To these sources we added (transient) sources that 
have been detected in the mean time, as well as those found in the
pre-{\em INTEGRAL} era but currently not detected (yet) by {\em INTEGRAL}.
To make sure that our results do not depend on the selected input catalogue, we compared 
our results with the ones obtained using the full
ISDC Reference Catalog (version 23.0) which contains all known 
high-energy sources (see Ebisawa et al.\ 2003); the imaging and light curve results 
were consistent. 
The classifications of the sources are mainly based on the information 
given by Bird et al.\ (2006), with updates where noted (Table~\ref{sourcelist}). 
The black-hole (candidate) binaries were selected from 
McClintock \&\ Remillard (2006) and Remillard et al.\ (2006a),
except for the new transient source XTE\,J1817$-$330 (see Remillard et al.\ 2006b).
The X-ray burster list is based on the list given by \mbox{in 't Zand}
et al.\ (2004), except for XTE\,J1739$-$285 (Brandt et al.\ 2005)
and IGR\,J17254$-$3257 (Brandt et al.\ 2006).
SLX\,1744$-$299 and SLX\,1744$-$300 are separated by only $\simeq$3' (Skinner et al.\ 1990)
and therefore inseparable with current hard X-ray instruments like IBIS/ISGRI.
3' is at the resolution limit of JEM-X and the soft X-ray emission from the two sources is seen to be different from a
point source by JEM-X. However, it is not able to separate the emission confidently,
when both sources are active. We therefore regard them as one source, SLX\,1744$-$299/300. 
IGR\,J17460$-$3047 was a source reported by Bird et al.\ (2004), that turned out to be 
an artefact (Bird et al.\ 2006). For the time being this source has been included in our list. We can 
confirm that in our monitoring program it was never detected by either IBIS/ISGRI or JEM-X
(see Sect.~\ref{images}).

\tabcolsep=1mm
\begin{table*}
\caption{List of sources in the {\em INTEGRAL}
Galactic bulge monitoring program, ordered by Galactic longitude
(from left to right in the Hammer-Aitoff projection, i.e.,
from 180$^{\circ}$--0$^{\circ}$, 360$^{\circ}$--180$^{\circ}$).
The source type and the reference from where this information is taken are given. 
We further provide the Galactic as well as the J2000.0 equatorial coordinates; these
are taken from the SIMBAD Astronomical Database (except for the sources detected
by {\em ASCA}, i.e., the ``AX\,J'' sources, which are taken from Sakano et al.\ 2002).
For some sources we give alternative names frequently used in the literature. 
Source type classification (after Bird et al.\ 2006): ? = unknown type, A = atoll source (neutron
star), AGN = active galactic nucleus, B = burster (neutron star), Be =
B-type emission-line star, BH = black hole (dynamically confirmed),
BHC = black-hole candidate, Cluster = cluster of galaxies, CV =
cataclysmic variable, D = dipping source, G = globular cluster source,
HMXB = high-mass X-ray binary, LMXB = low-mass X-ray binary, Mol cloud
= molecular cloud, NS = neutron star, QSO = quasar, SGR = soft
gamma-ray repeater, Sy = Seyfert galaxy, Symb = Symbiotic binary, T =
transient source, XP = X-ray pulsar (neutron star), Z = Z-type source
(neutron star). References:
[1] Bird et al.\ (2006), [2] Lowes et al.\ (2002), [3] Forman et al.\ (1976), 
[4] Levine et al.\ (2005), [5] Revnivtsev et al.\ (2004a),
[6] Lutovinov et al.\ (2004), [7] Chakrabarty \&\ Roche (1997),
[8] in 't Zand (2005), [9] McClintock \&\ Remillard (2006), 
[10] Reynolds et al.\ (1999), [11] Grebenev \&\ Sunyaev (2004b),
[12] Grebenev et al.\ (2005c), [13] Kretschmar et al.\ (2004),
[14] Remillard et al.\ (2006), [15] Brandt et al.\ (2005), 
[16] Cocchi et al.\ (1999), [17] Muno et al.\ (2003b),
[18] in 't Zand et al.\ (1997), [19] in 't Zand et al.\ (2003),
[20] Jonker \&\ van der Klis (2001), 
[21] Brandt et al.\ (2006), [22] Bhattacharyya et al.\ (2006), 
[23] Zurita Heras et al.\ (2006), [24] Thompson et al.\ (2006),
[25] Grebenev et al.\ (2005a), [26] Bird et al.\ (2004).}
\begin{tabular}{l|lllrrlll}
\hline
\multicolumn{1}{l|}{Source} &
\multicolumn{1}{l}{Type} &
\multicolumn{1}{l}{Ref.} &
\multicolumn{1}{l}{Fig.} &
\multicolumn{1}{c}{l$_{\rm II}$} &
\multicolumn{1}{c}{b$_{\rm II}$} &
\multicolumn{1}{c}{R.A.} &
\multicolumn{1}{c}{Dec} &
\multicolumn{1}{l}{Comment} \\
\hline
GX\,17+2                 & LMXB, B, Z         & 1    & \ref{xrb} &  16.432 & $-$0.710 & 18 16 01.4 & $-$14 02 11 & \\
SAX\,J1818.6$-$1703      & ?,T                & 1    & &  14.078 &  $-$0.710 & 18 18 39    & $-$17 03.1    & \\
GX\,13+1                 & LMXB, B, A         & 1    & \ref{xrb} &  13.517 &    +0.106 & 18 14 31.55 & $-$17 09 26.7 & \\
PKS\,1830$-$211          & AGN, QSO           & 1    & &  12.166 &  $-$5.712 & 18 33 39.89 & $-$21 03 39.8 & \\
SGR\,1806$-$20           & SGR                & 1    & &   9.996 &  $-$0.242 & 18 08 39.32 & $-$20 24 39.5 & \\
SAX\,J1805.5$-$2031      & ?,T                & 2    & &   9.554 &    +0.340 & 18 05 34    & $-$20 30.8    & \\
IGR\,J18027$-$2016       & HMXB, T, XP        & 1    & &   9.417 &    +1.044 & 18 02 39.9  & $-$20 17 13   & SAX\,J1802.7$-$2017 \\
GS\,1826$-$24            & LMXB, B            & 1    & \ref{xrb},\ref{misc2} &   9.272 &  $-$6.088 & 18 29 28.2  & $-$23 47 49   & Ginga\,1826$-$24 \\
GX\,9+1                  & LMXB, A            & 1    & \ref{misc} &   9.077 &    +1.154 & 18 01 32.3  & $-$20 31 44   & \\
GX\,9+9                  & LMXB, A            & 1    & &   8.513 &    +9.038 & 17 31 44.2  & $-$16 57 42   & 3A\,1728$-$169 \\
1RXS\,J175113.3$-$201214 & ?                  & 1    & &   8.145 &    +3.408 & 17 51 13.4  & $-$20 12 14   & \\
H1745$-$203              & LMXB, G, B         & 1,3  & &   7.729 &    +3.798 & 17 48 53.40 & $-$20 21 43.0 & in NGC\,6440 \\
IGR\,J17597$-$2201       & LMXB, B, D         & 1    & &   7.581 &    +0.775 & 17 59 46    & $-$22 00.9    & XTE\,J1759$-$220 \\
XTE\,J1818$-$245         & LMXB?, T           & 4    & \ref{misc} &   7.441 &  $-$4.196 & 18 18 25.2  & $-$24 32 31   & \\
1RXS\,J174607.8$-$213333 & ?                  & 5    & &   6.367 &    +3.734 & 17 46 07.80 & $-$21 33 33.0 & \\
GX\,5$-$1                & LMXB, Z            & 1    & \ref{misc},\ref{GX_5-1_isgri_jmx},\ref{GX_5-1_jmx_isgri_cd} &   5.077 &  $-$1.019 & 18 01 08.2  & $-$25 04 45   & \\
V1223\,Sgr               & CV                 & 1    & &   4.958 & $-$14.355 & 18 55 02.24 & $-$31 09 48.5 & 1H 1853$-$312 \\
GRS\,1758$-$258          & LMXB, BHC          & 1    & \ref{bhc} &   4.508 &  $-$1.361 & 18 01 12.3  & $-$25 44 36   & \\
IGR\,J17544$-$2619       & HMXB?, T           & 1    & \ref{misc},\ref{OMC_lightcurve} &   3.235 &  $-$0.338 & 17 54 25.7  & $-$26 19 58   & \\
H1820$-$303              & LMXB, G, B, A      & 1    & \ref{xrb} &   2.788 &  $-$7.913 & 18 23 40.45 & $-$30 21 40.1 & 4U\,1820$-$303; in NGC\,6624 \\
IGR\,J17331$-$2406       & ?,T                & 6    & &   2.607 &    +4.927 & 17 33 08    & $-$24 06.8    & \\
GX\,3+1                  & LMXB, B, A         & 1    & \ref{xrb} &   2.294 &    +0.794 & 17 47 56.0  & $-$26 33 49   & \\
GX\,1+4                  & Symb, XP           & 1,7  & \ref{misc},\ref{misc2} &   1.937 &    +4.795 & 17 32 02.16 & $-$24 44 44.0 & \\
XTE\,J1807$-$294         & LMXB, T, XP        & 1    & &   1.935 &  $-$4.273 & 18 06 59.8  & $-$29 24 30   & \\
AX\,J1749.2$-$2725       & HMXB, XP           & 5    & &   1.699 &    +0.108 & 17 49 11.6  & $-$27 25 36   & \\
AX\,J1749.1$-$2733       & HMXB?              & 8    & &   1.585 &    +0.051 & 17 49 09.0  & $-$27 33 14   & \\
XB\,1832$-$330           & LMXB, G, B, T      & 1    & \ref{xrb} &   1.531 & $-$11.372 & 18 35 44.1  & $-$32 59 29   & in NGC\,6652 \\ 
SLX\,1735$-$269          & LMXB, B            & 1    & \ref{xrb} &   0.785 &    +2.398 & 17 38 16.00 & $-$27 00 16.0 & \\
XTE\,J1748$-$288         & LMXB, T, BHC       & 9    & &   0.676 &  $-$0.222 & 17 48 05.06 & $-$28 28 25.8 & \\
IGR\,J17475$-$2822       & Mol cloud?         & 1    & &   0.601 &  $-$0.040 & 17 47 12    & $-$28 26.6    & \\      
EXMS\,B1709$-$232        & ?                  & 10    & &   0.594 &    +9.269 & 17 12 31    & $-$23 21.2    & \\      
IGR\,J17507$-$2856       & ?,T                & 11   & &   0.576 &  $-$0.959 & 17 50 44    & $-$28 56.3    & \\
Oph\,Cluster             & Cluster            & 1    & &   0.564 &    +9.272 & 17 12 26.0  & $-$23 22 33   & \\
IGR\,J17419$-$2802       & ?,T                & 12   & &   0.345 &    +1.164 & 17 41 56.0  & $-$28 01 54   & \\
1E\,1743.1$-$2843        & LMXB               & 1    & \ref{misc} &   0.251 &  $-$0.026 & 17 46 19.20 & $-$28 44 07.0 & \\
SAX\,J1747.0$-$2853      & LMXB, B, T         & 1    & \ref{xrb} &   0.207 &  $-$0.239 & 17 47 02.60 & $-$28 52 58.9 & \\
IGR\,J17407$-$2808       & ?,T                & 13   & &   0.115 &    +1.341 & 17 40.7     & $-$28 08      & \\
SLX\,1737$-$282          & LMXB, B            & 1    & \ref{xrb} & 359.995 &    +1.201 & 17 40 57.00 & $-$28 18 36.0 & \\
IGR\,J17456$-$2901       & ?                  & 1    & & 359.930 &  $-$0.048 & 17 45 38.5  & $-$29 01 15   & AX\,J1745.6$-$2900, Sgr A* \\
V2400\,Oph               & CV                 & 1    & & 359.867 &    +8.739 & 17 12 36.45 & $-$24 14 44.6 & RX\,J1712.6$-$2414 \\
XTE\,J1817$-$330         & LMXB, T, BHC       & 14   & \ref{bhc} & 359.817 & $-$7.996 & 18 17 43.54 & $-$33 01 07.8 & \\
XTE\,J1739$-$285         & LMXB, T, B         & 1,15 & \ref{xrb} & 359.714 &    +1.298 & 17 39 53.95 & $-$28 29 46.8 & \\
GRS\,1741.9$-$2853       & LMXB, T, B         & 5,16,17 & \ref{xrb} & 359.612 &    +0.734 & 17 41 50    & $-$28 52 42   & AX\,J1745.0-2855 \\
SAX\,J1744.7$-$2916      & ?                  & 18   & & 359.600 &  $-$0.009 & 17 44 42    & $-$29 16.9    & \\
KS\,1741$-$293           & LMXB, T, B         & 1    & \ref{xrb} & 359.584 &  $-$0.087 & 17 44 58    & $-$29 20.2    & \\
1A\,1742$-$294           & LMXB, B            & 1    & \ref{xrb},\ref{1A_1742-294_isgri_jmx} & 359.559 &  $-$0.389 & 17 46 05.5  & $-$29 30 55   & \\
SLX\,1744$-$299/300      & LMXB, B            & 1    & \ref{xrb} & 359.296 &  $-$0.889 & 17 47 25.9  & $-$29 59 58   & \\
1E\,1740.7$-$2942        & LMXB, BHC          & 1    & \ref{bhc} & 359.116 &  $-$0.106 & 17 43 54.83 & $-$29 44 42.6 & \\
GRS\,1734$-$292          & AGN, Sy1           & 1    & & 358.844 &    +1.395 & 17 37 24.3  & $-$29 10 48   & GRS\,1734$-$294 in [1] \\
GRS\,1747$-$312          & LMXB, G, T, B      & 1,19 & & 358.555 &  $-$2.168 & 17 50 45.5  & $-$31 17 32   & in Terzan~6\\
IGR\,J17460$-$3047       & ?                  & 1    & & 358.494 &  $-$1.094 & 17 46 19    & $-$30 47.5    & Artefact in [26] \\
IGR\,J17391$-$3021       & HMXB, NS, Be?, T   & 1    & & 358.068 &    +0.445 & 17 39 11.58 & $-$30 20 37.6 & XTE\,J1739$-$302 \\
IGR\,J17285$-$2922       & BHC?, T            & 1    & & 357.630 &    +2.923 & 17 28.5     & $-$29 22      & \\         
H1743$-$322              & LMXB, T, BHC       & 1    & \ref{bhc} & 357.255 &  $-$1.833 & 17 46 15.61  & $-$32 13 59.9    & IGR\,J17464$-$3213 \\     
\hline
\end{tabular}
\label{sourcelist}
\end{table*}
\begin{table*}
\begin{tabular}{l|lllrrlll}
\multicolumn{9}{l}{{\bf Table 1.} (continued)} \\
\multicolumn{9}{l}{~} \\
\hline
\multicolumn{1}{l|}{Source} &
\multicolumn{1}{l}{Type} &
\multicolumn{1}{l}{Ref} &
\multicolumn{1}{l}{Fig.} &
\multicolumn{1}{c}{l$_{\rm II}$} &
\multicolumn{1}{c}{b$_{\rm II}$} &
\multicolumn{1}{c}{R.A.} &
\multicolumn{1}{c}{Dec} &
\multicolumn{1}{l}{Comment} \\
\hline
IGR\,J17488$-$3253       & ?                  & 1    & & 356.962 &  $-$2.662 & 17 48 54.71 & $-$32 54 44.0 & \\
3A\,1822$-$371           & LMXB, XP, D        & 1,20 & \ref{misc} & 356.850 & $-$11.291 & 18 25 46.8  & $-$37 06 19   & \\  
SLX\,1746$-$331          & LMXB?, BHC         & 5,9 & & 356.807 &  $-$2.973 & 17 49 48.50 & $-$33 12 18.3 & \\
XTE\,J1710$-$281         & LMXB, T, B         & 1    & & 356.357 &    +6.922 & 17 10 12.3  & $-$28 07 54   & \\
4U\,1722$-$30            & LMXB, G, B, A      & 1    & \ref{xrb} & 356.320 &    +2.298 & 17 27 33.2  & $-$30 48 07   & XB\,1724$-$30; in Terzan~2 \\
IGR\,J17200$-$3116       & ?,T                & 1    & & 355.022 &    +3.346 & 17 20 06.10 & $-$31 17 02.0 & 1RXS\,J172006.1$-$311702 \\
MXB\,1730$-$335          & LMXB, G, B, T      & 1    & \ref{xrb} & 354.841 &  $-$0.158 & 17 33 24.10 & $-$33 23 16.0 & The Rapid Burster; in Liller 1 \\
XTE\,J1720$-$318         & LMXB, T, BHC       & 1    & & 354.597 &    +3.087 & 17 19 58    & $-$31 46.8    & \\    
GX\,354$-$0              & LMXB, B, A         & 1    & \ref{xrb},\ref{GX354-0_isgri_jmx} & 354.302 &  $-$0.150 & 17 31 57.4  & $-$33 50 05   & 4U 1728$-$337 \\
IGR\,J17254$-$3257       & LMXB?, B           & 1,21 & & 354.280 &    +1.472 & 17 25 25.50 & $-$32 57 17.5 & 1RXS\,J172525.5$-$325717 \\
1A\,1744$-$361           & LMXB, T, B, A?     & 1,22 & & 354.140 &  $-$4.204 & 17 48 19.22 & $-$36 07 16.6 & XTE\,J1748$-$361\\
1H\,1746$-$370           & LMXB, G, B, A      & 1    & & 353.531 &  $-$5.005 & 17 50 12.7  & $-$37 03 08   & in NGC\,6441 \\
XTE\,J1743$-$363         & ?,T                & 1    & \ref{misc} & 353.392 &  $-$3.402 & 17 43 00.0  & $-$36 20 41   & \\
4U\,1705$-$32            & LMXB, B            & 1    & & 352.794 &    +4.681 & 17 08 54.40 & $-$32 18 57.5 & \\
IGR\,J17252$-$3616       & HMXB, XP, T        & 23,24 & \ref{misc} & 351.510 &  $-$0.356 & 17 25 14    & $-$36 16.4   & EXO\,1722$-$363 \\
IGR\,J17098$-$3628       & ?,T                & 25   & \ref{misc},\ref{misc2} & 349.555 &    +2.066 & 17 09 48    & $-$36 28.2    & \\ 
IGR\,J17091$-$3624       & BHC?               & 1    & & 349.519 &    +2.215 & 17 09 06    & $-$36 24.7    & \\
GX\,349+2                & LMXB, Z            & 1    & \ref{misc} & 349.104 &    +2.748 & 17 05 44.5  & $-$36 25 23   & Sco X-2 \\  
SAX\,J1712.6$-$3739      & LMXB, T, B         & 1    & & 348.935 &    +0.928 & 17 12 34.00 & $-$37 38 36.0 & \\
4U\,1700$-$377           & HMXB               & 1    & \ref{misc},\ref{misc2} & 347.754 &    +2.173 & 17 03 56.77 & $-$37 50 38.9 & \\
GRO\,J1655$-$40          & LMXB, T, BH, D     & 9    & \ref{bhc} & 344.982 &    +2.456 & 16 54 00.14 & $-$39 50 44.9 & \\
OAO\,1657$-$415          & HMXB, XP           & 1    & \ref{misc} & 344.354 &    +0.311 & 17 00 47.9  & $-$41 40 23   & \\
\hline
\end{tabular}
\end{table*} 

The data from IBIS/ISGRI are processed
until the production of images in the 20--60 and 60--150\,keV energy ranges
per single pointing.\footnote{A single pointing with {\em INTEGRAL} is often refered to as a 
Science Window, or ScW for short.} 
We force the flux extraction of each of the 
catalogue sources, regardless of the detection significance of the source.
This method is essential in order to clean the images from the ghosts of all the active sources in 
the field, but does not make any threshold selection and all the positive 
fluxes are recorded (see Goldwurm et al.\ 2003
for a detailed description of the IBIS analysis software).
In order to detect fainter sources, we then mosaic the images from the 
seven single pointings (i.e., one hexagonal dither observation) 
and search for all catalog sources, as well as possible new ones.
In the case a new source is found, it can be added to the input catalogue, and 
a re-analysis is necessary to extract its source fluxes.
To achieve even higher sensitivity, we also produced one mosaic image per season as well as  
a global (3 seasons) mosaic image from the whole AO-3 monitoring program.

For JEM-X the analysis is run through the imaging step to the light-curve step.
Light curves with a time bin of the same length as each single pointing are
produced for every catalog source inside a radius of 5$^{\circ}$.
The analysis software used to extract light curves has a
known problem in crowded fields like the Galactic bulge, as the contribution from
bright sources (such as GX\,5$-$1) is not modeled in sufficient detail which affects
the results for weak sources (typically less than about 100\,mCrab) close
to very bright ones. Although no detailed quantitative analysis has been
done yet, comparisons with the results from JEM-X imaging as well as with
results from other X-ray instruments in the same energy range, leads us
to the conclude that there is an uncertainty in the fluxes by up to a
factor of $\simeq$2.
Again, the images from the seven single pointings
are mosaiced in order to create the final image. No further 
automatic source detection is done for the moment at the mosaic level; however, the images
are visually examined for possible new sources.

Type~I X-ray bursts (see, e.g., Strohmayer \&\ Bildsten 2006) are mainly
seen in the soft X-ray band, since the observed black-body temperatures are
$\simeq$1--3\,keV. We use JEM-X to search for such events.
Per single pointing we compute the average detector count rate
and the corresponding standard deviation. Whenever the difference
between the count rate in a 1-second bin and the average count rate
exceeds four times the standard deviation value, we flag it as
a potential start of a burst. Recontructed images and source light curves 
are generated within the good-time interval covering the burst event, and are
visually checked in order to identify the originating source of the event.
The light curves are also visually checked to see whether they adhere to the basic characteristics of a 
type~I X-ray burst, i.e., emitting mainly at the lowest energies
(e.g., to exclude solar flares), with a light-curve shape consistent with a fast rise -- with respect to the decay -- and 
exponential-like decay. Whenever the statistics do allow we also check whether the events 
show evidence for a spectral softening, due to cooling of the neutron surface,
during the decay. 
In the present paper we only briefly mention the occurrence of type~I X-ray bursts,
for those sources which were in the field of view of JEM-X, when discussing the long-term light curves. 
A more detailed (time-resolved) burst analysis is in progress (Sanchez-Fernandez et al., in preparation).
We here note that an account of type~I X-ray bursts, including those seen from sources in the 
Galactic bulge region, in earlier {\em INTEGRAL} data based on IBIS/ISGRI was reported by Chelovekov et al.\ (2006).

All fluxes used in this paper are given with respect to the Crab count rate
in the respective energy ranges (see Appendix~\ref{crab}), i.e., in units of mCrab.
Our sensitivity for one hexagonal dither observation
(i.e., all seven single pointings per revolution combined) is
typically between 5 and 15\,mCrab (5$\sigma$) for both JEM-X and IBIS/ISGRI. 
The actual sensitivity depends on factors such as source position 
(fully or partially coded field of view), background (instrument systematics, 
solar activity), number of pointings actually 
performed and usable, and energy (instrument response).
We refer to Appendix~\ref{sens} for more details.
Errors in this paper are quoted at the 1$\sigma$ level.
When a source is not detected, we provide 3$\sigma$ upper limits,
whenever appropriate. The latter are calculated by determining the variance value 
in the mosaic maps at a given source position, taking the square root of this value and multiply it by 3. 
Note that the variances respresent statistical values, and do not include any systematics.

From the end of August up to mid September 2005 (revolutions
349--357; MJD\,53605--53630) there was a high solar activity and 
various solar flares hampered the observations.
During some of the observations the instruments were even off
(revolutions 349, 355 and 356; MJD\,53605, 53624, 53626, respectively). 
Note that in a few revolutions no monitoring was done, either due
to Crab calibration (see Appendix~\ref{crab}) or target-of-opportunity observations
(revolutions 300, 352, 422--423; MJD\,53458--53460, MJD\,53613--53616, 
MJD\,53823--53828, respectively). 
Some single pointings were lost due to other reasons, such as ground station outages.
The total net observing time was 727\,ks.

\begin{table}
\caption{Sources monitored by the OMC in the Galactic bulge region. The ordering is the same as Table~1. Listed are the source name,
OMC identifier and the period of monitoring (period I covers the revolutions before 421, 
period II covers revolution 421 and onwards). 
Revolution 421 corresponds to 2006, March 25 or MJD\,53819.
The results for IGR\,J17544$-$2619 are shown in
Fig.~\ref{OMC_lightcurve}.} 
\begin{tabular}{l|lc}
\hline
\multicolumn{1}{l|}{Source} &
\multicolumn{1}{l}{OMC identifier} &
\multicolumn{1}{l}{Period} \\
\hline
GRS\,1758$-$258     & IOMC\,6846000121 & II   \\
IGR\,J17544$-$2619  & IOMC\,6849000050 & II   \\
GX\,3+1             & IOMC\,6836000093 & II   \\
SLX\,1735$-$269     & IOMC\,6835000362 & I,II \\
XTE\,J1748$-$288    & IOMC\,6840000094 & II   \\
IGR\,J17475$-$2822  & IOMC\,6840000103 & II   \\
1E\,1743.1$-$2843   & IOMC\,6840000034 & I,II \\
SAX\,J1747.0$-$2853 & IOMC\,6840000039 & II   \\
IGR\,J17407$-$2808  & IOMC\,6839000184 & I,II \\
SLX\,1737$-$282     & IOMC\,6839000170 & II   \\
IGR\,J17456$-$2901  & IOMC\,6840000098 & I,II \\
SAX\,J1744.7$-$2916 & IOMC\,6840000084 & I,II \\
KS\,1741$-$293      & IOMC\,6840000085 & I,II \\
1A\,1742$-$294      & IOMC\,6840000080 & I,II \\
SLX\,1744$-$299     & IOMC\,6840000042 & II   \\
1E\,1740.7$-$2942   & IOMC\,6840000077 & I,II \\
GRS\,1734$-$292     & IOMC\,6839000175 & I,II \\
IGR\,J17460$-$3047  & IOMC\,7377000086 & II   \\
IGR\,J17391$-$3021  & IOMC\,7376000097 & I,II \\
IGR\,J17285$-$2922  & IOMC\,6838000560 & I,II \\
H1743$-$322         & IOMC\,7381000052 & II   \\
IGR\,J17488$-$3253  & IOMC\,7381000160 & II   \\
SLX\,1746$-$331     & IOMC\,7381000150 & II   \\
4U\,1722$-$30       & IOMC\,7375000273 & I,II \\
\hline
\end{tabular}
\label{OMC_sourcelist}
\end{table}

For each INTEGRAL pointing, the OMC monitors the sources in its field of
view by means of shots of variable integration time. Typical values in the
range 10 to 200\,s (currently 10, 50 and 200\,s) are used
to optimize sensitivity, and to minimize read-out noise and cosmic-ray effects.
For the faintest objects, several 200\,s exposures in the same pointing
can be combined during data analysis on the ground.
Telemetry constraints do not allow to download the
entire OMC image. For this reason, windows are selected around the
proposed X-ray/$\gamma$-ray targets as well as other targets of interest in
the same field of view. Only sub-windows of the CCD with a size of 11x11 pixels
(3.2'x3.2') containing those objects are transmitted to ground.
For extended sources or sources with poor precision in their
coordinates, a mosaic of such sub-windows are set around
the source position to cover the entire error circle. 
The sources in the Galactic bulge region being monitored by the OMC are indicated in Table~\ref{OMC_sourcelist}. 
Since revolution 421 (March 25, 2006; MJD\,53819) onwards, the OMC instrument is
operating with a new Input Catalogue (version 0005). Using this catalogue,
OMC monitors all sources detected by IBIS/ISGRI (Bird et al.\ 2006) in the
Galactic bulge which fall in its field of view. 

The OMC data of the whole period has been processed with an updated OMC Off-line Scientific 
Analysis software.\footnote{The updated software will be included in {\sc OSA 6.0}.} 
The full analysis has been working in an automatic way since the third season of the Galactic bulge monitoring, 
i.e., revolutions 406--429 (February to April 2006; MJD\,53775--53846). 
Light curves are produced in a short period of time, usually less than 24\,hr after 
the observation is performed. 
Since the Galactic bulge region is a very crowded field for the OMC, in the flux
extraction process we force the photometric aperture to be centred at the 
source coordinates, which are taken from the OMC Input Catalogue. 
This allows us to monitor not only the
sources already detected by OMC, but also those previously undetected that 
could show, for example, bright flares detectable occasionally by OMC. Obviously, having 
accurate source coordinates is very important to obtain reliable results.

OMC's typical limiting magnitude in the Galactic bulge observations is between 
V=15 and 16\,mag (3$\sigma$). The actual value depends on sky background and
source contamination, which can be very important in this region. 
We obtain one photometric point per OMC shot for each source.
To increase the signal to noise of weak sources, we combine the individual
photometric points in the hexagonal dither observation.

\section{Results}

The results we present in this Section are organized in two main subsections,
one focussing on the seasonal and overall source behaviour of all sources
in our sample and another on the source variability.

\subsection{IBIS/ISGRI and JEM-X mosaic images}
\label{images}

\begin{figure*}
\centering
  \includegraphics[height=.46\textheight]{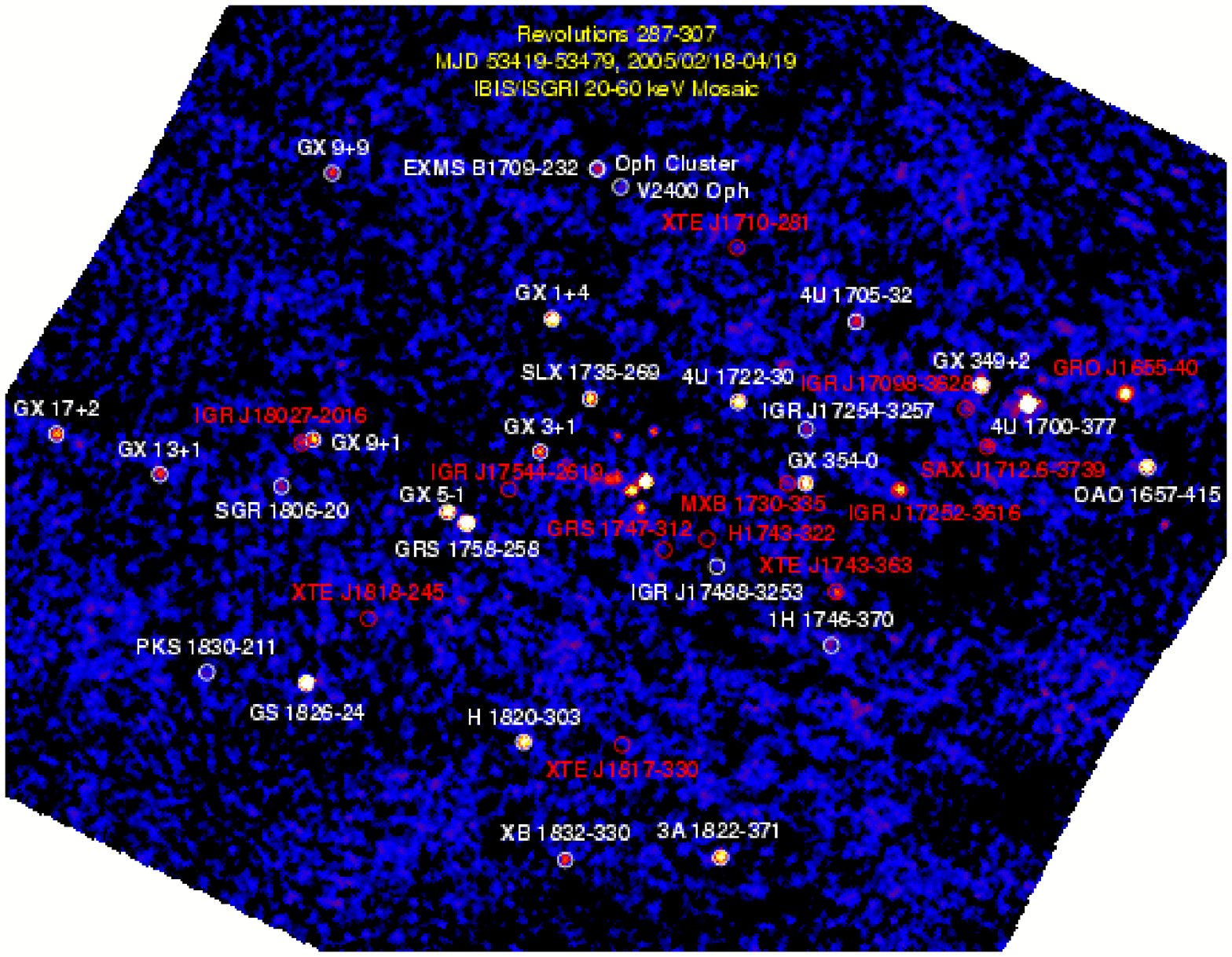}
  \includegraphics[height=.46\textheight]{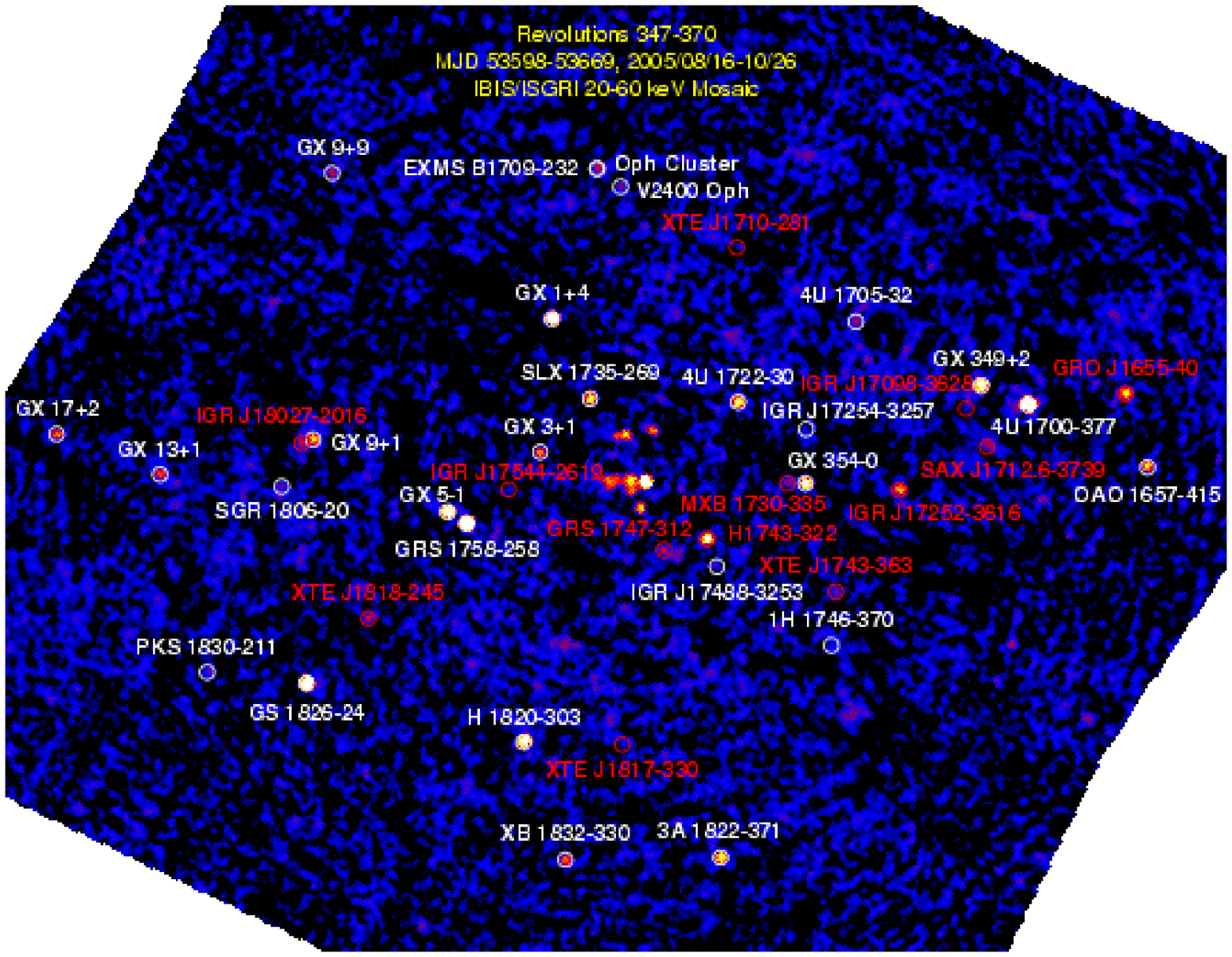}
  \caption{{\em INTEGRAL} IBIS/ISGRI (20--60\,keV) mosaic significance images 
from the first ({\it top}; total exposure of 234\,ks) and second season 
({\it bottom}; total exposure of 235\,ks), using the Hammer-Aitoff projection. 
Sources and their names are annotated, except for those in the 
Galactic Center region. The latter are displayed in Fig.~\ref{fig2}.
Transients (see Table~1) are indicated in red, 
the other sources in white. The annotated sources are those which reached
a significance higher than 7 in either a single revolution,
a season or all seasons together (see Table~\ref{significance}).
}
\label{fig1a}
\end{figure*}

\begin{figure*}
\centering
  \includegraphics[height=.46\textheight]{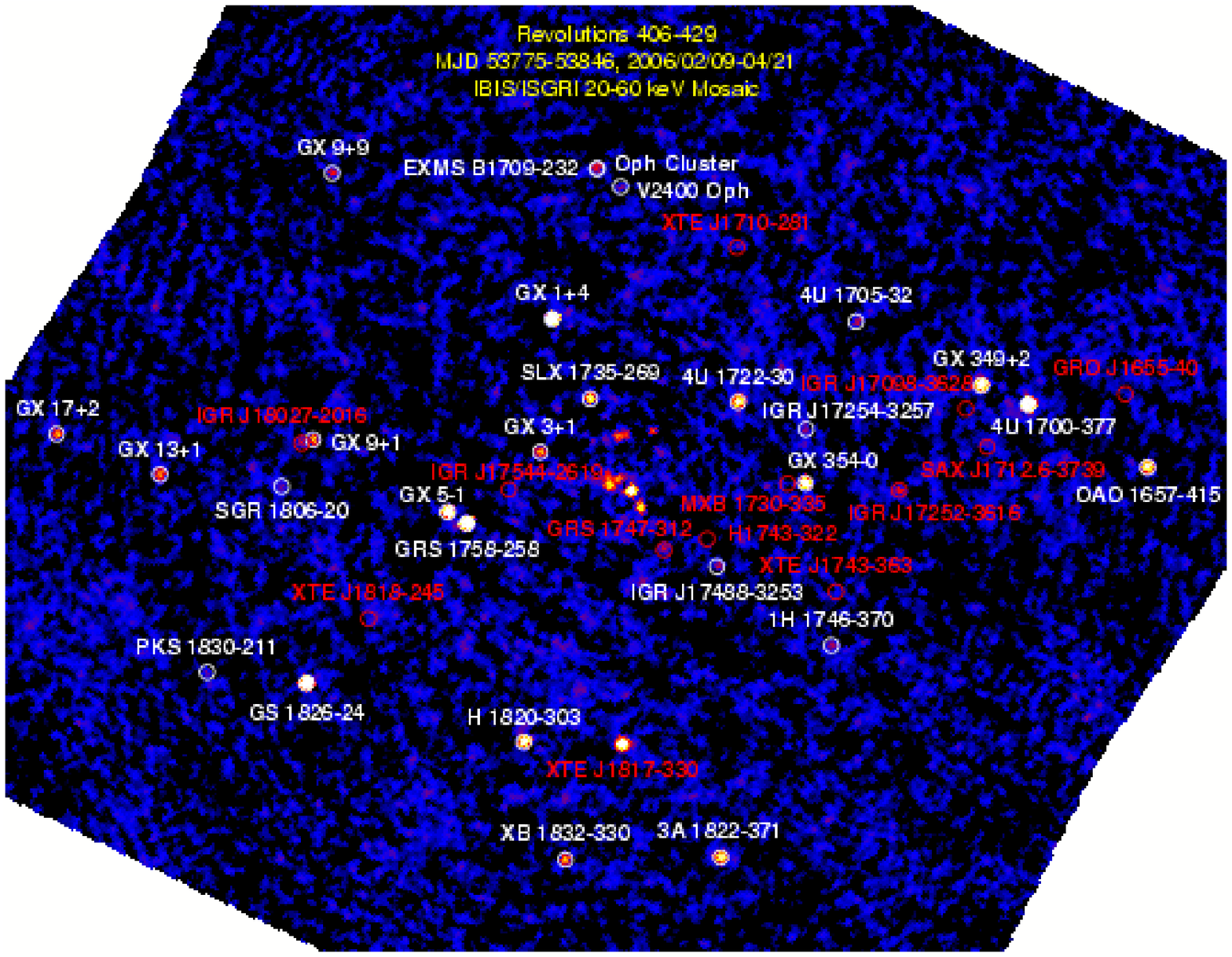}
  \includegraphics[height=.46\textheight]{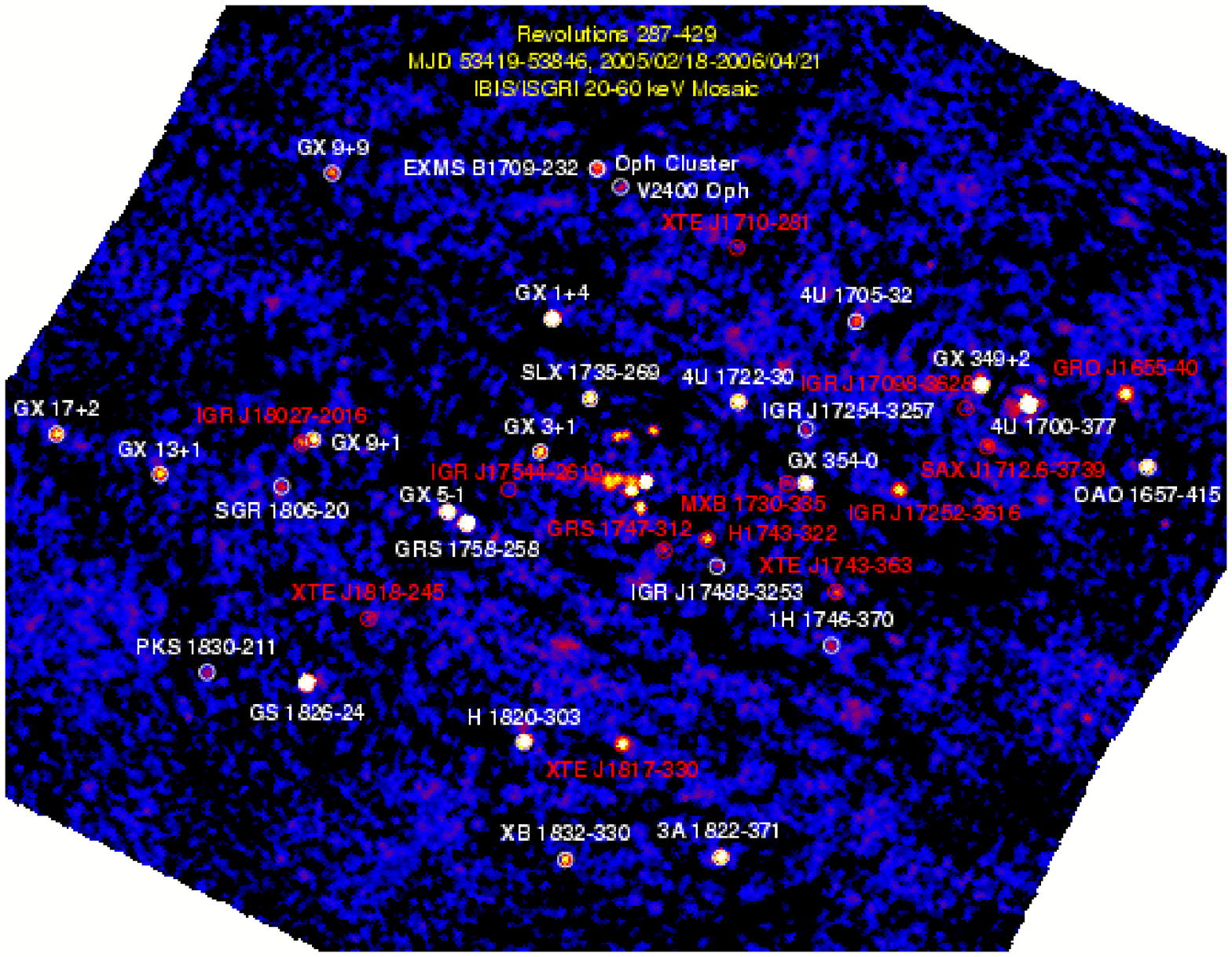}
  \caption{Same as Fig.~\ref{fig1a} but for the third season ({\it top}; total exposure of 
258\,ks) and all seasons together ({\it bottom}; total exposure of
727\,ks).
}
\label{fig1b}
\end{figure*}

\begin{figure*}
  \includegraphics[height=.3\textheight]{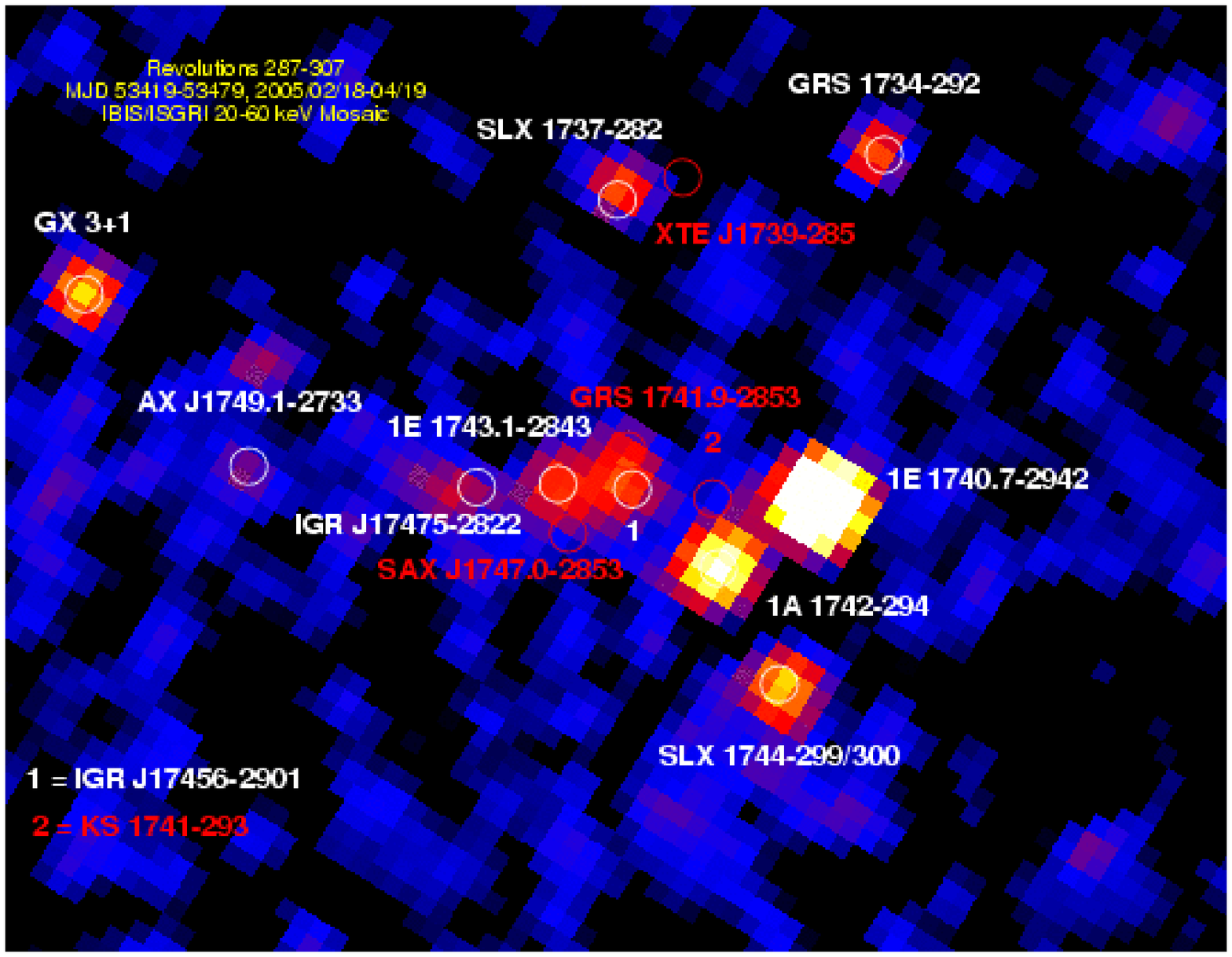}
  \includegraphics[height=.3\textheight]{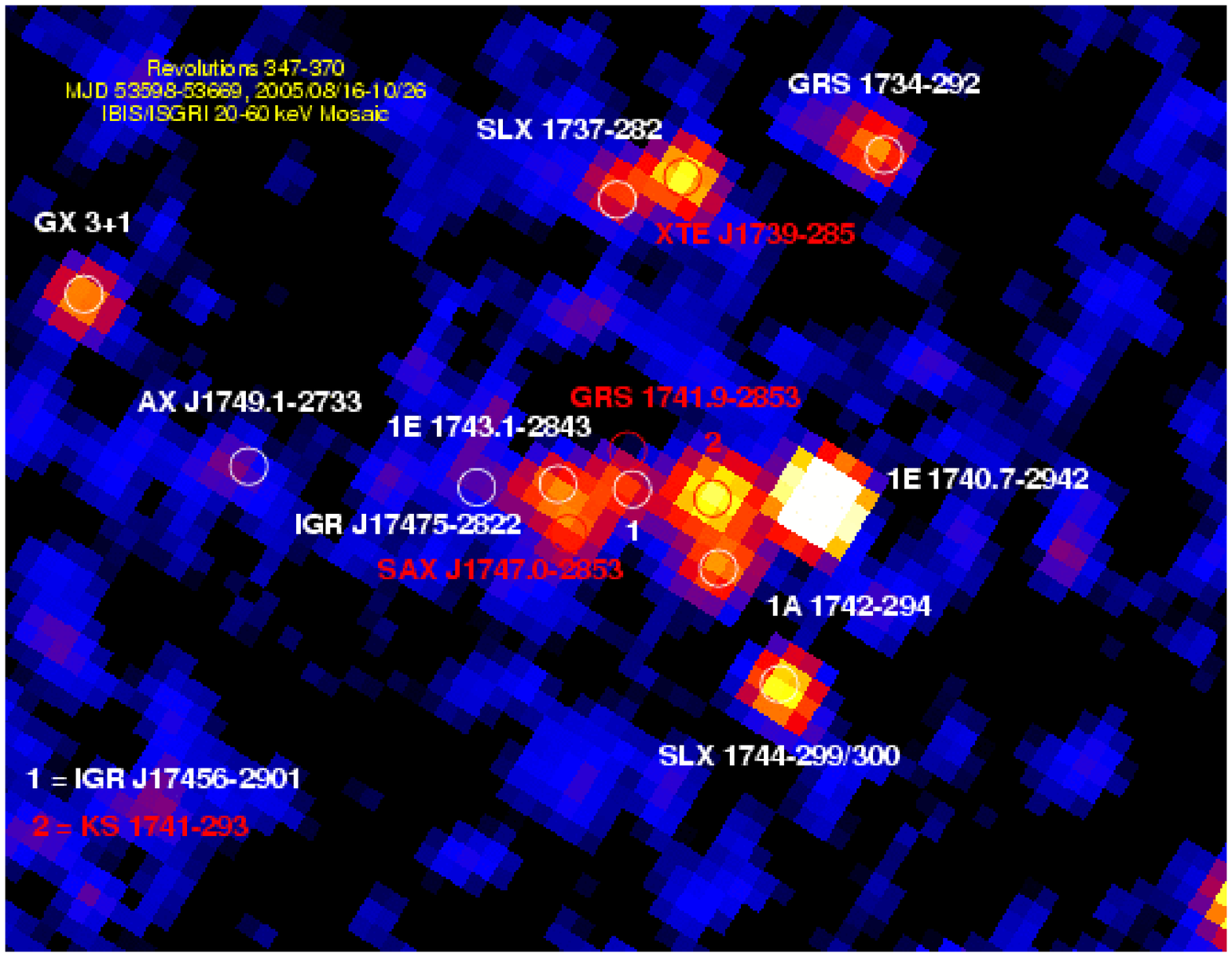}
  \includegraphics[height=.3\textheight]{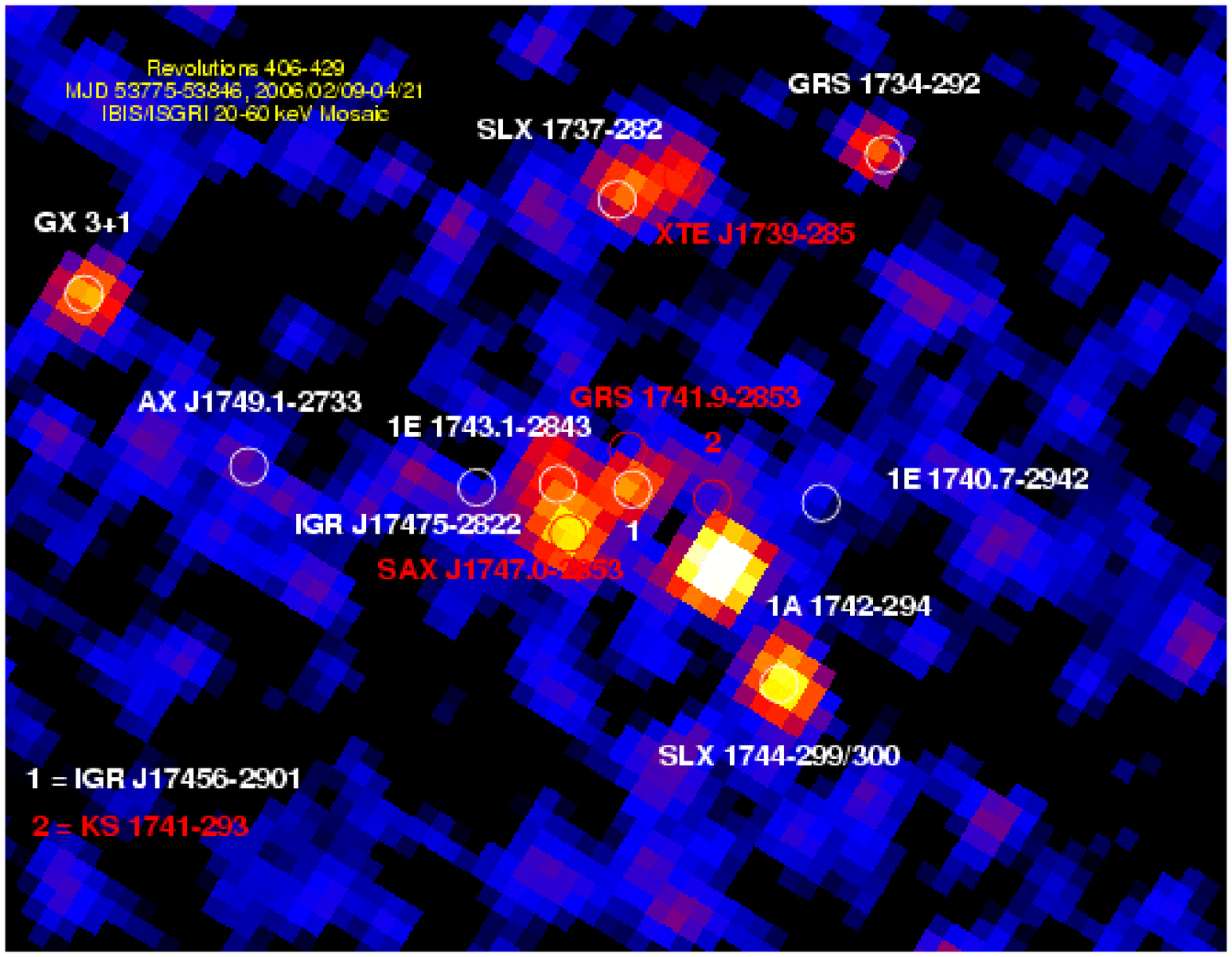}
  \includegraphics[height=.3\textheight]{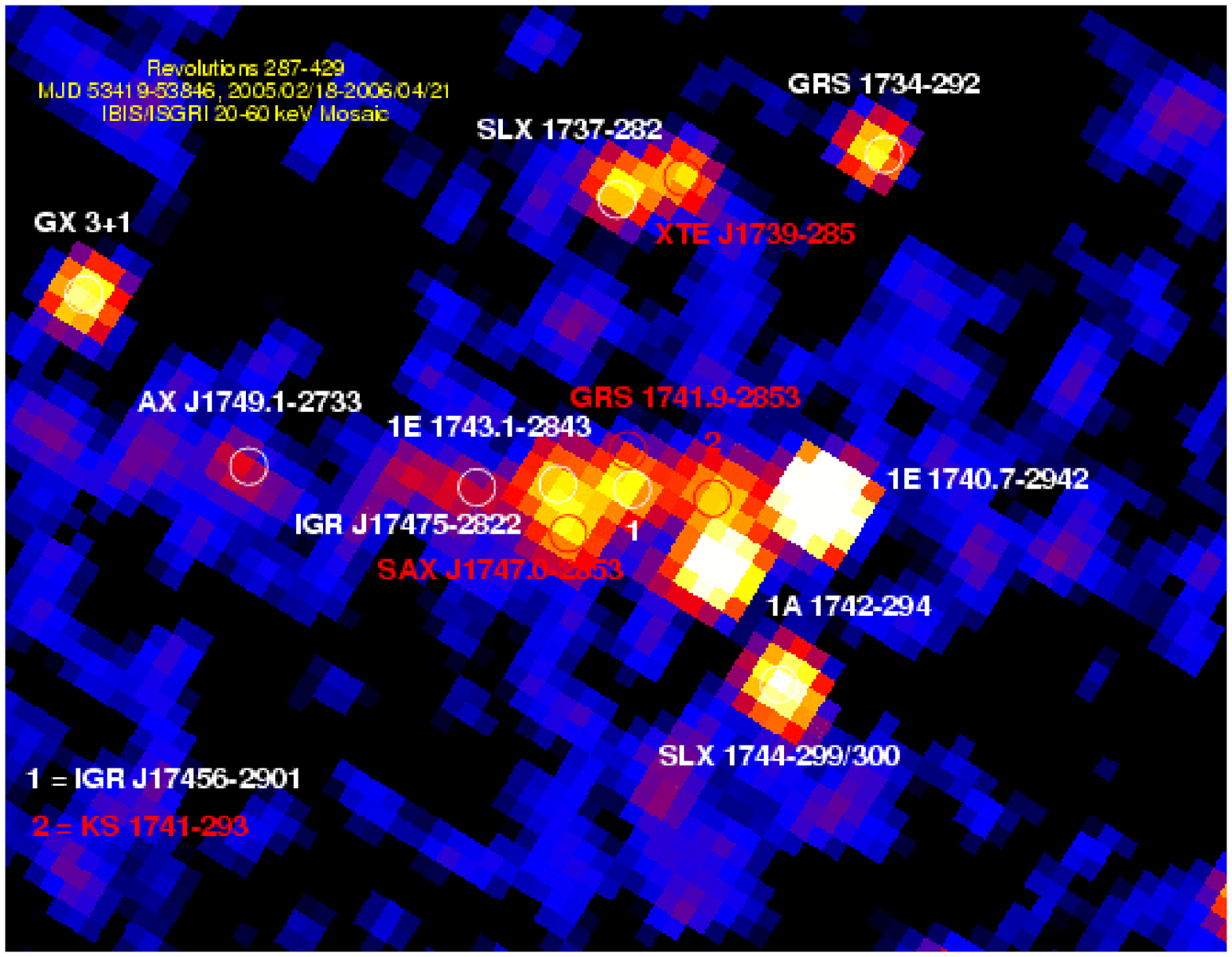}
  \caption{Same as Figs.~\ref{fig1a} and \ref{fig1b}, but now zoomed in on the Galactic Center region. 
}
\label{fig2}
\end{figure*}

\begin{figure*}
  \includegraphics[height=.37\textheight]{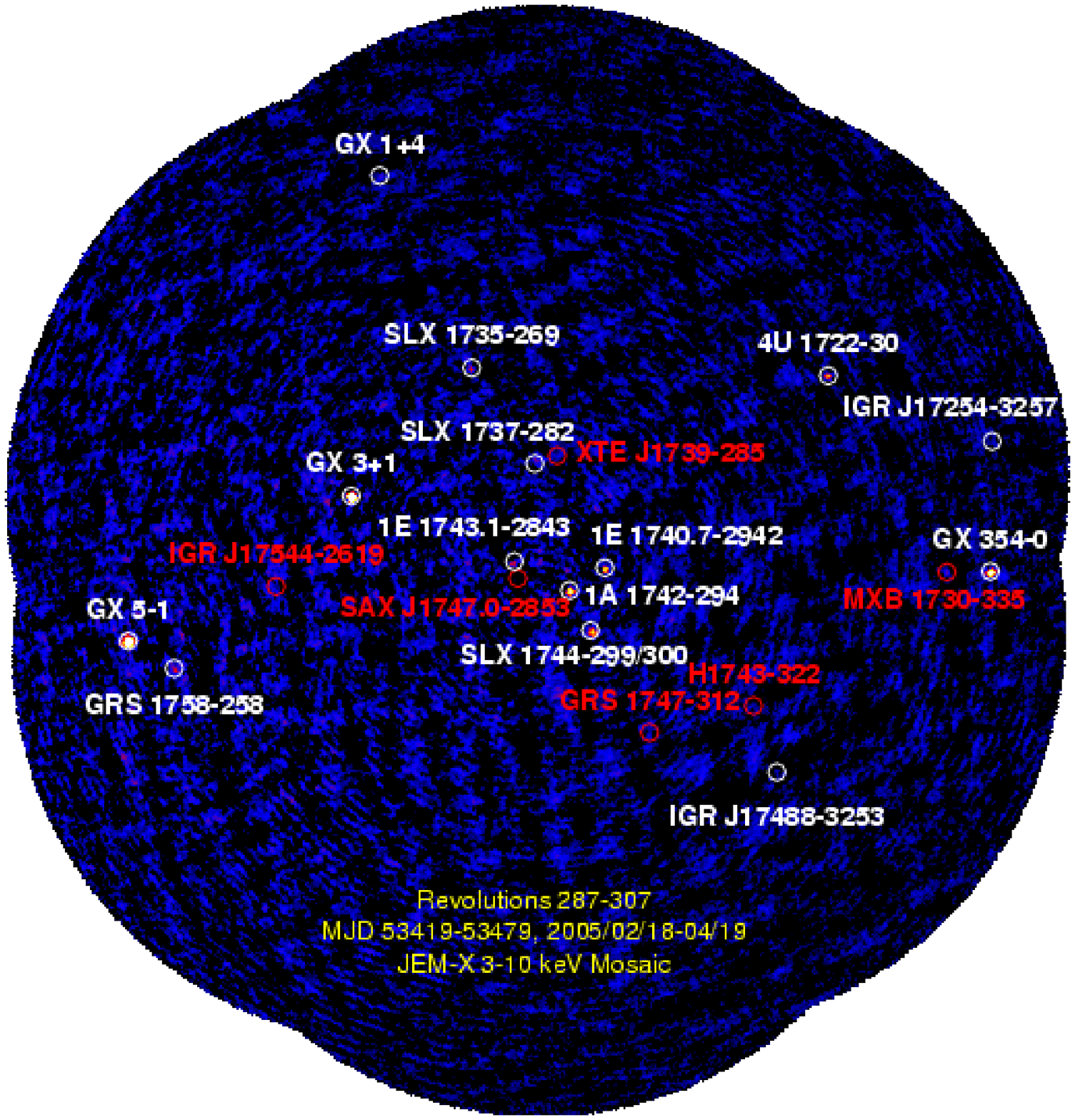}
  \includegraphics[height=.37\textheight]{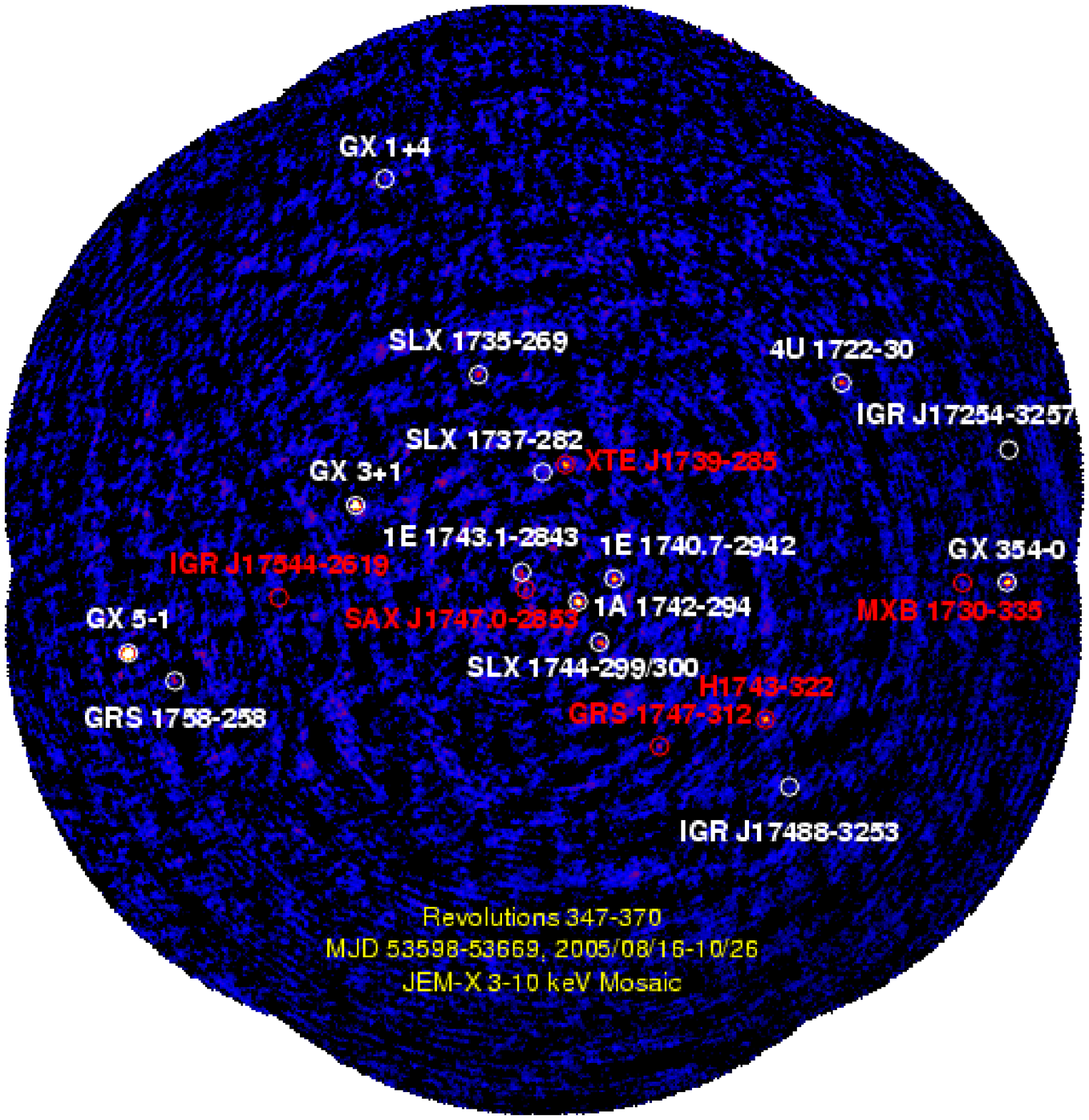}
  \includegraphics[height=.37\textheight]{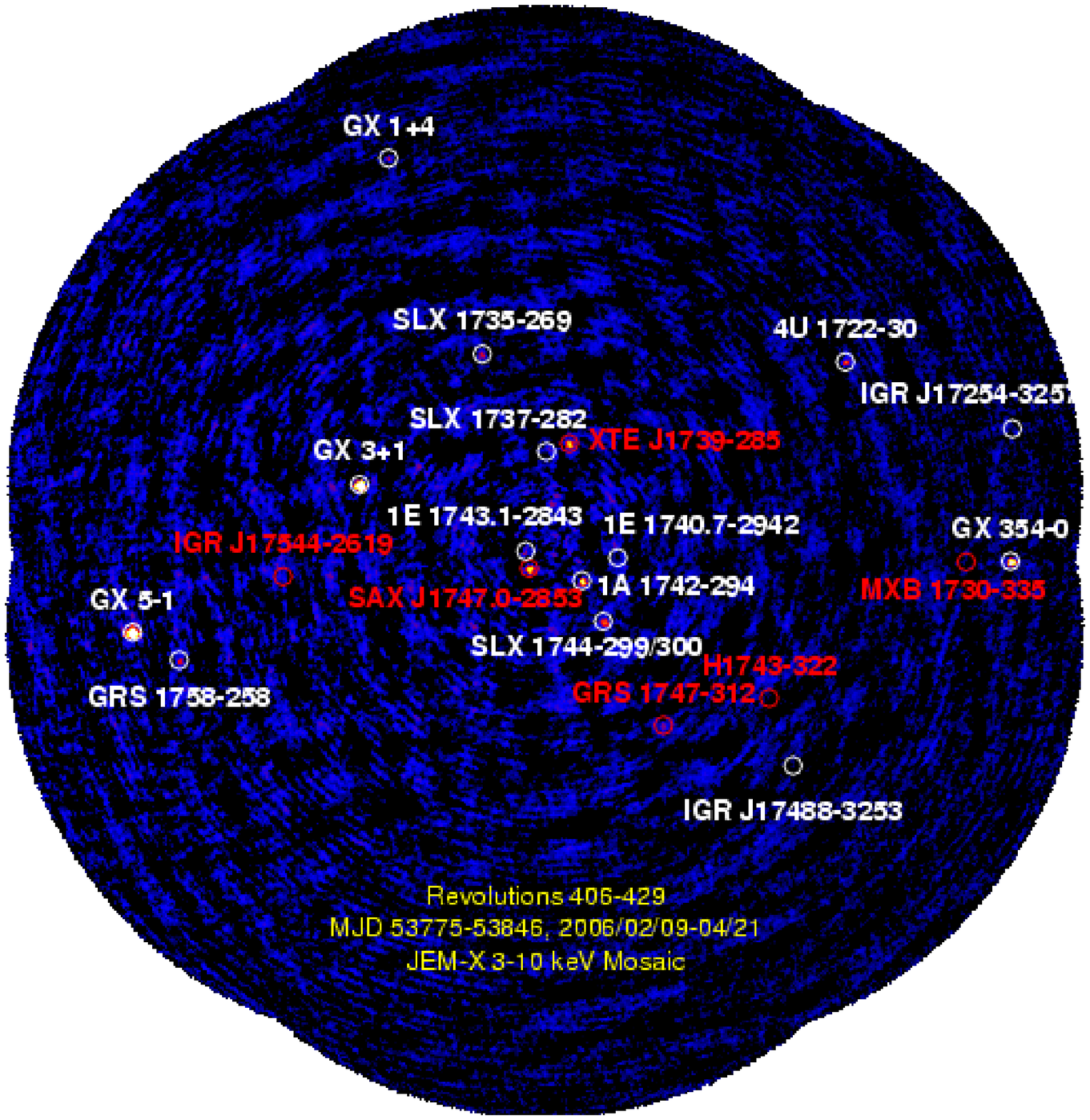}
  \includegraphics[height=.37\textheight]{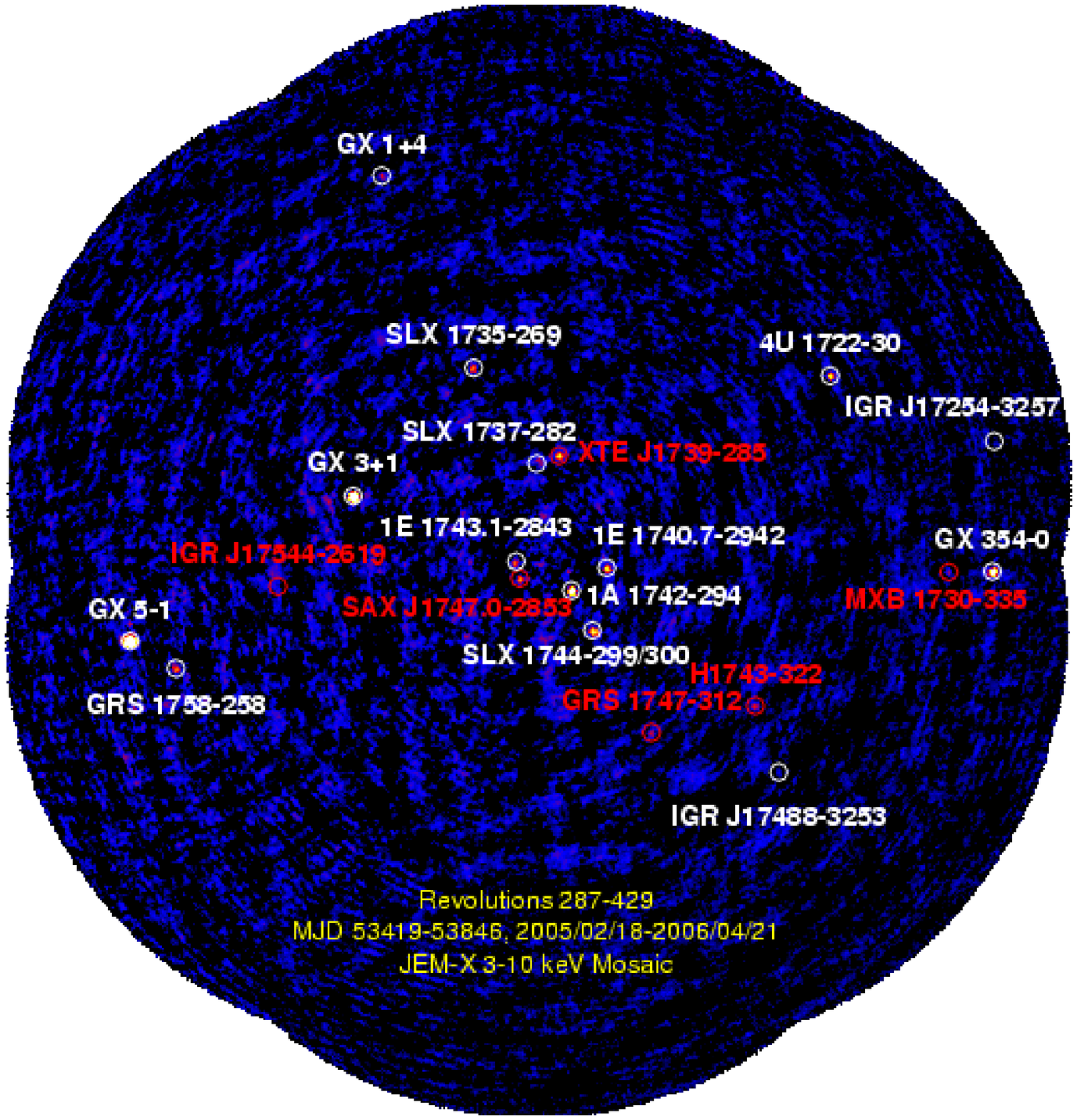}
  \caption{Same as Figs.~\ref{fig1a} and \ref{fig1b}, but for JEM-X (3--10\,keV). 
The same sources as in Figs.~\ref{fig1a} and \ref{fig2} are annotated
except for AX\,J1749.1$-$2733, IGR\,J17475$-$2822, 
IGR\,J17456$-$2901, GRS\,1741.9$-$2853, KS\,1741$-$293 and GRS\,1734$-$292
(all in the Galactic Center region) for reason of clarity.
}
\label{fig3}
\end{figure*}

We show four IBIS/ISGRI (20--60\,keV) mosaic significance images, 
one for each season and a total one, i.e., of all the pointings together, in
Figs.~\ref{fig1a} and \ref{fig1b}. 
Since the Galactic Center region (in the middle of the figure)
contains a considerable amount of hard X-ray sources close together,
we zoom in on this region in Fig.~\ref{fig2}.
Similarly, JEM-X (3--10\,keV) significance mosaic images for the 
three seasons and all the pointings together are shown in 
Fig.~\ref{fig3}.
Since annotating all the sources in our sample on the IBIS/ISGRI and JEM-X images would make
them unreadable, we chose to label only the sources with a
significance of 7 or higher in the 20--60\,keV data 
during either a single hexagonal dither observation,
during one season, or during all seasons combined (see Table~\ref{significance}). 
This results in the inclusion of fast transients that only appear within one or a couple of hexagonal 
dither observations, transients which show outbursts on week to month time
scales, and all persistent sources. To
avoid crowdedness of source names in the JEM-X images near the Galactic center 
we did not annotate all sources in that region.

The source detection significances and corresponding average fluxes are given in 
Tables~\ref{significance} (20--60\,keV) and \ref{significance_2} (60--150\,keV).
When a source is not detected we provide 3$\sigma$ upper limits.
We additionally give the highest detection 
significance reached of all the single hexagonal dither observations; these are used to determine
whether long-term light curves are shown or not (see Sect.~\ref{light_curves}).
In Table~\ref{significance_2} we show information only for those sources which were detected above 
a significance level of 7 in the 60--150\,keV band, either in a single hexagonal dither observation, 
in the average mosaic per season, or in the mosaic of all observations. 
In Table~\ref{jemx_mosaics} we give some information for each of the
JEM-X mosaic images, such as raw and effective exposure time at the center of the images and which sources
are visible above a detection limit of 3$\sigma$.
Unfortunately, the current JEM-X analysis software 
does not allow us to extract accurate flux information from
the mosaic images as is the case for IBIS/ISGRI.

\tabcolsep=1mm
\begin{table*}[ht]
\caption[]{This table lists the maximum detection significance reached during the
hexagonal dither observations (Sig$_{\rm rev, max}$), the 
detection significance per season (Sig$_{\rm sX}$; where X=1,2,3 for the three seasons) 
and the detection significance for all observations together (Sig$_{\rm all}$), together 
with the corresponding fluxes ($F_{\rm s,X}$) and errors in the fluxes for the 
IBIS/ISGRI 20--60\,keV band. The ordering of sources is the same as used for Table~1.
When a source was not detected it is indicated with `---'; we then provide a 3$\sigma$ (statistical) upper limit
on the flux. Whenever Sig$_{\rm rev, max}$
exceeds 7, the source long-term light curves are discussed in Sect.~\ref{light_curves}.
}
\begin{tabular}{l|ccccccccc}
\hline
\multicolumn{1}{l|}{} &
\multicolumn{1}{c}{Sig$_{\rm rev, max}$} &
\multicolumn{1}{c}{Sig$_{\rm s1}$} &
\multicolumn{1}{c}{$F_{\rm s1}$} &
\multicolumn{1}{c}{Sig$_{\rm s2}$} &
\multicolumn{1}{c}{$F_{\rm s2}$} &
\multicolumn{1}{c}{Sig$_{\rm s3}$} &
\multicolumn{1}{c}{$F_{\rm s3}$} &
\multicolumn{1}{c}{Sig$_{\rm all}$} &
\multicolumn{1}{c}{$F_{\rm all}$} \\
\multicolumn{1}{l|}{Source} &
\multicolumn{1}{c}{} &
\multicolumn{1}{c}{} &
\multicolumn{1}{c}{(mCrab)} &
\multicolumn{1}{c}{} &
\multicolumn{1}{c}{(mCrab)} &
\multicolumn{1}{c}{} &
\multicolumn{1}{c}{(mCrab)} &
\multicolumn{1}{c}{} &
\multicolumn{1}{c}{(mCrab)} \\
\hline
GX\,17+2                 &  12.2 &  21.2 &   52.4 $\pm$ 2.5 &  17.7 &   41.5 $\pm$ 2.3 &  20.1 &   45.7 $\pm$ 2.3 &  33.0 &   46.3 $\pm$ 1.4 \\
SAX\,J1818.6$-$1703      &   5.5 &  ---  & $<$2.6           &  ---  & $<$2.6           &  ---  & $<$2.6           &  ---  & $<$1.5           \\
GX\,13+1                 &   7.2 &  13.2 &   10.4 $\pm$ 0.8 &  10.9 &    8.5 $\pm$ 0.8 &  16.7 &   13.4 $\pm$ 0.8 &  24.0 &   11.0 $\pm$ 0.5 \\
PKS\,1830$-$211          &   4.7 &   5.6 &    4.3 $\pm$ 0.8 &   5.6 &    4.3 $\pm$ 0.7 &   5.2 &    4.3 $\pm$ 0.8 &   8.6 &    3.7 $\pm$ 0.4 \\
SGR\,1806$-$20           &   4.7 &   6.6 &    3.0 $\pm$ 0.4 &   5.7 &    2.4 $\pm$ 0.4 &   5.1 &    2.4 $\pm$ 0.5 &   9.6 &    2.4 $\pm$ 0.2 \\
SAX\,J1805.5$-$2031      &  ---  &   --- & $<$1.3           &  ---  & $<$1.3           &  ---  & $<$1.3           &  ---  & $<$0.8           \\
IGR\,J18027$-$2016       &   5.5 &   9.4 &    4.3 $\pm$ 0.5 &   8.1 &    3.7 $\pm$ 0.5 &   7.1 &    3.0 $\pm$ 0.4 &  15.9 &    3.7 $\pm$ 0.2 \\
GS\,1826$-$24            &  45.4 & 136.5 &   72.0 $\pm$ 0.5 & 162.3 &   86.6 $\pm$ 0.5 & 153.8 &   81.1 $\pm$ 0.5 & 259.2 &   79.9 $\pm$ 0.3 \\
GX\,9+1                  &   9.6 &  26.7 &   11.0 $\pm$ 0.4 &  18.9 &    7.9 $\pm$ 0.4 &  22.5 &    9.1 $\pm$ 0.4 &  40.7 &    9.8 $\pm$ 0.2 \\
GX\,9+9                  &   6.4 &  14.2 &    9.8 $\pm$ 0.7 &   8.7 &    6.1 $\pm$ 0.7 &  11.2 &    7.9 $\pm$ 0.7 &  19.8 &    7.9 $\pm$ 0.4 \\
1RXS\,J175113.3$-$201214 &  ---  &   5.8 &    2.4 $\pm$ 0.4 &  ---  & $<$1.3           &   5.4 &    2.4 $\pm$ 0.4 &   6.6 &    1.8 $\pm$ 0.2 \\
H1745$-$203              &  ---  &   --- & $<$1.2           &  ---  & $<$1.2           &  ---  & $<$1.2           &  ---  & $<$0.7           \\
IGR\,J17597$-$2201       &   6.1 &  ---  & $<$1.1           &  ---  & $<$1.1           &   4.9 &    1.8 $\pm$ 0.4 &  ---  & $<$0.6           \\
XTE\,J1818$-$245         &  15.6 &  ---  & $<$1.2           &  10.6 &    4.3 $\pm$ 0.4 &  ---  & $<$1.2           &   8.8 &    1.8 $\pm$ 0.2 \\
1RXS\,J174607.8$-$213333 &  ---  &   --- & $<$1.1           &  ---  & $<$1.2           &  ---  & $<$1.1           &  ---  & $<$0.7           \\
GX\,5$-$1                &  48.7 &  73.9 &   23.2 $\pm$ 0.3 & 104.9 &   33.5 $\pm$ 0.3 & 108.6 &   34.8 $\pm$ 0.3 & 166.8 &   30.5 $\pm$ 0.2 \\
V1223\,Sgr               &   4.5 &   5.3 &   10.4 $\pm$ 2.0 &  ---  & $<$5.6           &   4.5 &    7.9 $\pm$ 1.8 &   5.8 &    6.7 $\pm$ 1.1 \\
GRS\,1758$-$258          &  68.5 & 212.6 &   65.2 $\pm$ 0.3 & 230.2 &   72.6 $\pm$ 0.3 & 182.4 &   56.1 $\pm$ 0.3 & 361.7 &   64.6 $\pm$ 0.2 \\
IGR\,J17544$-$2619       &  15.3 &  ---  & $<$0.9           &  ---  & $<$0.9           &  ---  & $<$0.9           &  ---  & $<$0.5           \\
H1820$-$303              &  50.9 &  48.5 &   20.1 $\pm$ 0.4 &  70.7 &   29.3 $\pm$ 0.4 &  54.0 &   22.0 $\pm$ 0.4 & 100.0 &   23.8 $\pm$ 0.2 \\
IGR\,J17331$-$2406       &  ---  &   --- & $<$1.0           &  ---  & $<$1.0           &  ---  & $<$1.0           &  ---  & $<$0.6           \\
GX\,3+1                  &   7.4 &  22.3 &    6.7 $\pm$ 0.3 &  18.9 &    5.5 $\pm$ 0.3 &  20.4 &    6.1 $\pm$ 0.3 &  35.2 &    6.1 $\pm$ 0.2 \\
GX\,1+4                  &  56.2 &  84.8 &   26.8 $\pm$ 0.3 & 110.0 &   35.4 $\pm$ 0.3 & 170.9 &   55.5 $\pm$ 0.3 & 210.4 &   39.0 $\pm$ 0.2 \\
XTE\,J1807$-$294         &  ---  &   --- & $<$0.9           &  ---  & $<$1.0           &  ---  & $<$1.0           &  ---  & $<$0.6           \\
AX\,J1749.2$-$2725       &  ---  &   --- & $<$0.9           &   4.9 &    1.2 $\pm$ 0.3 &  ---  & $<$0.9           &  ---  & $<$0.5           \\
AX\,J1749.1$-$2733       &   6.8 &   5.6 &    1.8 $\pm$ 0.3 &  ---  & $<$0.9           &   3.8 &    1.2 $\pm$ 0.3 &   7.4 &    1.2 $\pm$ 0.2 \\
XB\,1832$-$330           &   7.6 &  14.3 &    7.9 $\pm$ 0.5 &  15.6 &    9.1 $\pm$ 0.5 &  19.9 &   11.0 $\pm$ 0.5 &  28.8 &    9.8 $\pm$ 0.3 \\
SLX\,1735$-$269          &  11.9 &  42.8 &   12.2 $\pm$ 0.3 &  39.0 &   11.6 $\pm$ 0.3 &  36.7 &   11.0 $\pm$ 0.3 &  68.8 &   11.6 $\pm$ 0.2 \\
XTE\,J1748$-$288         &  ---  &   --- & $<$0.9           &  ---  & $<$0.9           &  ---  & $<$0.9           &  ---  & $<$0.5           \\
IGR\,J17475$-$2822       &   5.2 &   9.0 &    2.4 $\pm$ 0.3 &  ---  & $<$0.9           &   4.5 &    1.2 $\pm$ 0.3 &   8.8 &    1.2 $\pm$ 0.1 \\
EXMS\,B1709$-$232        &   6.4 &   9.3 &    3.7 $\pm$ 0.4 &   7.9 &    3.0 $\pm$ 0.4 &  ---  & $<$1.2           &  15.1 &    3.7 $\pm$ 0.2 \\
IGR\,J17507$-$2856       &  ---  &   --- & $<$0.9           &  ---  & $<$0.8           &  ---  & $<$0.9           &  ---  & $<$0.5           \\
Oph\,Cluster             &   4.4 &  ---  & $<$1.2           &  ---  & $<$1.3           &  10.2 &    4.3 $\pm$ 0.4 &  ---  & $<$0.7           \\
IGR\,J17419$-$2802       &   5.0 &  ---  & $<$0.9           &  ---  & $<$0.9           &   5.0 &    1.2 $\pm$ 0.2 &  ---  & $<$0.5           \\
1E\,1743.1$-$2843        &   8.0 &  ---  & $<$0.9           &  16.5 &    4.3 $\pm$ 0.2 &  ---  & $<$0.8           &  24.4 &    4.3 $\pm$ 0.2 \\
SAX\,J1747.0$-$2853      &  15.2 &  ---  & $<$0.9           &  ---  & $<$0.8           &  21.8 &    6.1 $\pm$ 0.3 &  ---  & $<$0.5           \\
IGR\,J17407$-$2808       &   4.6 &  ---  & $<$0.9           &  ---  & $<$0.9           &  ---  & $<$0.9           &  ---  & $<$0.5           \\
SLX\,1737$-$282          &   7.0 &  14.1 &    4.3 $\pm$ 0.3 &  ---  & $<$0.9           &  15.2 &    4.3 $\pm$ 0.3 &  23.5 &    3.7 $\pm$ 0.2 \\
IGR\,J17456$-$2901       &   5.9 &  ---  & $<$0.9           &  ---  & $<$0.8           &  14.5 &    4.3 $\pm$ 0.3 &  ---  & $<$0.5           \\
V2400\,Oph               &   5.3 &   6.2 &    2.4 $\pm$ 0.4 &   6.1 &    2.4 $\pm$ 0.4 &   7.1 &    3.0 $\pm$ 0.4 &  11.2 &    2.4 $\pm$ 0.2 \\
XTE\,J1817$-$330         &  53.2 &  ---  & $<$1.0           &  ---  & $<$1.2           &  79.6 &   29.9 $\pm$ 0.4 &  47.7 &   10.4 $\pm$ 0.2 \\
XTE\,J1739$-$285         &  13.7 &  ---  & $<$0.9           &  22.3 &    6.1 $\pm$ 0.3 &  ---  & $<$0.9           &  ---  & $<$0.5           \\
GRS\,1741.9$-$2853       &   9.6 &  ---  & $<$0.9           &  ---  & $<$0.8           &  ---  & $<$0.9           &  13.9 &    2.4 $\pm$ 0.2 \\
SAX\,J1744.7$-$2916      &   5.1 &  ---  & $<$0.9           &  ---  & $<$0.9           &   4.9 &    1.2 $\pm$ 0.2 &  ---  & $<$0.5           \\
KS\,1741$-$293           &  19.3 &   5.3 &    1.2 $\pm$ 0.2 &  28.1 &    7.9 $\pm$ 0.3 &  ---  & $<$0.9           &  19.7 &    3.0 $\pm$ 0.2 \\
1A\,1742$-$294           &  25.5 &  41.5 &   11.6 $\pm$ 0.3 &  12.6 &    3.7 $\pm$ 0.3 &  63.1 &   18.3 $\pm$ 0.3 &  68.0 &   11.0 $\pm$ 0.2 \\
SLX\,1744$-$299/300      &   9.8 &  20.1 &    6.1 $\pm$ 0.3 &  26.6 &    7.9 $\pm$ 0.3 &  26.8 &    7.9 $\pm$ 0.3 &  42.0 &    7.3 $\pm$ 0.2 \\
1E\,1740.7$-$2942        &  47.1 & 162.4 &   45.7 $\pm$ 0.3 & 139.5 &   39.6 $\pm$ 0.3 &  ---  & $<$0.8           & 173.1 &   28.7 $\pm$ 0.2 \\
GRS\,1734$-$292          &   6.6 &  15.3 &    4.3 $\pm$ 0.3 &  16.9 &    4.9 $\pm$ 0.3 &  15.5 &    4.3 $\pm$ 0.3 &  28.0 &    4.9 $\pm$ 0.2 \\
GRS\,1747$-$312          &   6.0 &  ---  & $<$0.9           &   9.0 &    2.4 $\pm$ 0.3 &   9.2 &    2.4 $\pm$ 0.3 &  11.5 &    1.8 $\pm$ 0.2 \\
IGR\,J17460$-$3047       &  ---  &  ---  & $<$0.9           &  ---  & $<$0.9           &  ---  & $<$0.9           &  ---  & $<$0.5           \\
IGR\,J17391$-$3021       &   4.3 &  ---  & $<$0.9           &  ---  & $<$0.9           &  ---  & $<$0.9           &  ---  & $<$0.5           \\
IGR\,J17285$-$2922       &   4.0 &  ---  & $<$0.9           &  ---  & $<$0.9           &  ---  & $<$0.9           &  ---  & $<$0.5           \\
H1743$-$322              &  49.7 &  ---  & $<$0.9           &  50.4 &   15.2 $\pm$ 0.3 &  ---  & $<$0.9           &  27.9 &    4.9 $\pm$ 0.2 \\
IGR\,J17488$-$3253       &   4.4 &  ---  & $<$0.9           &  ---  & $<$0.9           &   8.5 &    2.4 $\pm$ 0.3 &   9.4 &    1.8 $\pm$ 0.2 \\
3A\,1822$-$371           &  14.1 &  39.3 &   22.6 $\pm$ 0.5 &  31.8 &   18.3 $\pm$ 0.6 &  35.1 &   20.1 $\pm$ 0.5 &  61.8 &   20.1 $\pm$ 0.3 \\
SLX\,1746$-$331          &  ---  &   --- & $<$0.9           &  ---  & $<$0.9           &  ---  & $<$0.9           &  ---  & $<$0.5           \\
XTE\,J1710$-$281         &   5.6 &   6.5 &    2.4 $\pm$ 0.4 &  ---  & $<$1.2           &   5.6 &    2.4 $\pm$ 0.4 &   8.5 &    1.8 $\pm$ 0.2 \\
4U\,1722$-$30            &  18.0 &  60.3 &   18.3 $\pm$ 0.3 &  36.8 &   11.6 $\pm$ 0.3 &  41.5 &   12.8 $\pm$ 0.3 &  80.1 &   14.0 $\pm$ 0.2 \\
IGR\,J17200$-$3116       &  ---  &   6.3 &    1.8 $\pm$ 0.3 &  ---  & $<$1.0           &  ---  & $<$1.0           &   6.6 &    1.2 $\pm$ 0.2 \\
MXB\,1730$-$335          &  12.2 &   6.5 &    1.8 $\pm$ 0.3 &   6.4 &    1.8 $\pm$ 0.3 &  ---  & $<$1.0           &   7.9 &    1.2 $\pm$ 0.2 \\
\hline
\end{tabular}
\label{significance}
\end{table*}
\begin{table*}[ht]
\begin{tabular}{l|ccccccccc}
\multicolumn{10}{l}{{\bf Table 3} (continued).} \\
\multicolumn{10}{l}{} \\
\hline
\multicolumn{1}{l|}{} &
\multicolumn{1}{c}{Sig$_{\rm rev, max}$} &
\multicolumn{1}{c}{Sig$_{\rm s1}$} &
\multicolumn{1}{c}{$F_{\rm s1}$} &
\multicolumn{1}{c}{Sig$_{\rm s2}$} &
\multicolumn{1}{c}{$F_{\rm s2}$} &
\multicolumn{1}{c}{Sig$_{\rm s3}$} &
\multicolumn{1}{c}{$F_{\rm s3}$} &
\multicolumn{1}{c}{Sig$_{\rm all}$} &
\multicolumn{1}{c}{$F_{\rm all}$} \\
\multicolumn{1}{l|}{Source} &
\multicolumn{1}{c}{} &
\multicolumn{1}{c}{} &
\multicolumn{1}{c}{(mCrab)} &
\multicolumn{1}{c}{} &
\multicolumn{1}{c}{(mCrab)} &
\multicolumn{1}{c}{} &
\multicolumn{1}{c}{(mCrab)} &
\multicolumn{1}{c}{} &
\multicolumn{1}{c}{(mCrab)} \\
\hline
XTE\,J1720$-$318         &  ---  &   --- & $<$1.0           &  ---  & $<$1.0           &  ---  & $<$1.0           &  ---  & $<$0.6           \\
GX\,354$-$0              &  48.1 &  69.2 &   22.0 $\pm$ 0.3 &  71.1 &   23.2 $\pm$ 0.3 &  83.0 &   26.8 $\pm$ 0.3 & 128.8 &   24.4 $\pm$ 0.2 \\
IGR\,J17254$-$3257       &  ---  &   6.7 &    1.8 $\pm$ 0.3 &  ---  & $<$1.0           &   6.2 &    1.8 $\pm$ 0.3 &   8.0 &    1.2 $\pm$ 0.2 \\
1A\,1744$-$361           &  ---  &   --- & $<$1.1           &  ---  & $<$1.1           &  ---  & $<$1.1           &  ---  & $<$0.6           \\
1H\,1746$-$370           &   5.5 &   6.2 &    2.4 $\pm$ 0.4 &  ---  & $<$1.2           &   6.5 &    2.4 $\pm$ 0.4 &   9.0 &    1.8 $\pm$ 0.2 \\
XTE\,J1743$-$363         &   9.1 &  14.4 &    5.5 $\pm$ 0.4 &   6.0 &    2.4 $\pm$ 0.4 &  ---  & $<$1.2           &  13.1 &    3.0 $\pm$ 0.2 \\
4U\,1705$-$32            &   5.1 &  10.3 &    4.3 $\pm$ 0.4 &   6.8 &    3.0 $\pm$ 0.4 &   8.0 &    3.0 $\pm$ 0.4 &  13.7 &    3.0 $\pm$ 0.2 \\
IGR\,J17252$-$3616       &  36.2 &  28.7 &   11.6 $\pm$ 0.4 &  23.9 &    9.8 $\pm$ 0.4 &  10.3 &    4.3 $\pm$ 0.4 &  36.8 &    8.5 $\pm$ 0.2 \\
IGR\,J17098$-$3628       &  23.8 &   8.7 &    4.3 $\pm$ 0.5 &  ---  & $<$1.5           &  ---  & $<$1.5           &   7.4 &    2.4 $\pm$ 0.3 \\
IGR\,J17091$-$3624       &  ---  &  ---  & $<$1.5           &  ---  & $<$1.5           &  ---  & $<$1.5           &  ---  & $<$0.9           \\
GX\,349+2                &  24.5 &  70.7 &   36.6 $\pm$ 0.5 &  51.7 &   28.0 $\pm$ 0.5 &  57.9 &   30.5 $\pm$ 0.5 & 103.3 &   31.7 $\pm$ 0.3 \\
SAX\,J1712.6$-$3739      &   5.6 &  10.1 &    5.5 $\pm$ 0.5 &   6.4 &    3.7 $\pm$ 0.5 &  ---  & $<$1.6           &  11.6 &    3.7 $\pm$ 0.3 \\
4U\,1700$-$377           & 194.5 & 344.5 &  214.0 $\pm$ 0.6 & 268.9 &  174.4 $\pm$ 0.7 & 216.9 &  137.2 $\pm$ 0.6 & 481.5 &  176.2 $\pm$ 0.4 \\
GRO\,J1655$-$40          &  71.7 &  55.9 &   59.8 $\pm$ 1.0 &  24.3 &   27.4 $\pm$ 1.2 &  ---  & $<$3.3           &  47.9 &   29.9 $\pm$ 0.6 \\
OAO\,1657$-$415          &  26.9 &  62.2 &   87.8 $\pm$ 1.4 &  27.2 &   39.6 $\pm$ 1.5 &  42.8 &   58.5 $\pm$ 1.3 &  75.9 &   62.8 $\pm$ 0.9 \\
\hline
\end{tabular}
\end{table*}

\tabcolsep=1mm
\begin{table*}[ht]
\caption[]{This table lists the maximum detection significance reached during the
hexagonal dither observations (Sig$_{\rm rev, max}$), the detection significance per season (Sig$_{\rm sX}$; where X=1,2,3 
for the three seasons) and the detection significance for all observations together (Sig$_{\rm all}$), together 
with the corresponding fluxes ($F_{\rm sX}$) and errors in the fluxes for the 
IBIS/ISGRI 60--150\,keV band. The ordering of sources is the same as used for Table~1.
We only list those sources for which either Sig$_{\rm rev, max}$, Sig$_{\rm sX}$ or Sig$_{\rm all}$ reached a value higher
than 7. When a source was not detected it is indicated with `---'; we then provide a 3$\sigma$ (statistical) upper limit
on the flux. Whenever Sig$_{\rm rev, max}$ exceeds 7, the source long-term light curves are discussed in Sect.~\ref{light_curves}.
}
\begin{tabular}{l|ccccccccc}
\hline
\multicolumn{1}{l|}{} &
\multicolumn{1}{c}{Sig$_{\rm rev, max}$} &
\multicolumn{1}{c}{Sig$_{\rm s1}$} &
\multicolumn{1}{c}{$F_{\rm s1}$} &
\multicolumn{1}{c}{Sig$_{\rm s2}$} &
\multicolumn{1}{c}{$F_{\rm s2}$} &
\multicolumn{1}{c}{Sig$_{\rm s3}$} &
\multicolumn{1}{c}{$F_{\rm s3}$} &
\multicolumn{1}{c}{Sig$_{\rm all}$} &
\multicolumn{1}{c}{$F_{\rm all}$} \\
\multicolumn{1}{l|}{Source} &
\multicolumn{1}{c}{} &
\multicolumn{1}{c}{} &
\multicolumn{1}{c}{(mCrab)} &
\multicolumn{1}{c}{} &
\multicolumn{1}{c}{(mCrab)} &
\multicolumn{1}{c}{} &
\multicolumn{1}{c}{(mCrab)} &
\multicolumn{1}{c}{} &
\multicolumn{1}{c}{(mCrab)} \\
\hline
GS\,1826$-$24       & 11.0 & 29.5 & 12.2 $\pm$ 0.4 &  33.7 & 13.4 $\pm$ 0.4 & 30.2 & 12.2 $\pm$ 0.4 &  53.6 & 12.8 $\pm$ 0.2 \\
GRS\,1758$-$258     & 28.9 & 95.3 & 23.8 $\pm$ 0.2 & 103.4 & 26.2 $\pm$ 0.2 & 79.5 & 19.5 $\pm$ 0.2 & 161.3 & 23.2 $\pm$ 0.1 \\
GX\,1+4             &  9.3 & 14.1 &  3.7 $\pm$ 0.2 &  18.1 &  4.9 $\pm$ 0.2 & 23.3 &  6.1 $\pm$ 0.2 &  31.8 &  4.9 $\pm$ 0.1 \\
XB\,1832$-$330      &  4.3 &  6.2 &  3.0 $\pm$ 0.4 &   4.1 &  1.8 $\pm$ 0.4 &  8.0 &  3.7 $\pm$ 0.4 &  10.2 &  2.4 $\pm$ 0.2 \\
SLX\,1735$-$269     &  5.2 & 10.0 &  2.4 $\pm$ 0.2 &   9.8 &  2.4 $\pm$ 0.2 & 10.0 &  2.4 $\pm$ 0.2 &  16.7 &  2.4 $\pm$ 0.1 \\
XTE\,J1817$-$330    & 11.1 &  --- & $<$3.4         &   --- & $<$3.4         & 22.1 &  6.7 $\pm$ 0.3 &  12.6 &  2.4 $\pm$ 0.2 \\
1A\,1742$-$294      &  5.0 &  6.1 &  1.2 $\pm$ 0.2 &   --- & $<$2.8         &  9.4 &  2.4 $\pm$ 0.2 &   7.6 &  1.2 $\pm$ 0.2 \\
SLX\,1744$-$299/300 &  4.4 &  --- & $<$2.7         &   6.7 &  1.8 $\pm$ 0.2 &  5.4 &  1.2 $\pm$ 0.2 &   8.2 &  1.2 $\pm$ 0.1 \\
1E\,1740.7$-$2942   & 19.8 & 58.8 & 14.0 $\pm$ 0.2 &  50.1 & 11.6 $\pm$ 0.2 &  --- & $<$2.7         &  61.9 &  8.5 $\pm$ 0.1 \\
H1743$-$322         & 13.2 &  --- & $<$2.7         &  15.4 &  3.7 $\pm$ 0.2 &  --- & $<$2.8         &   9.6 &  1.2 $\pm$ 0.1 \\
4U\,1722$-$30       &  5.4 & 11.7 &  3.0 $\pm$ 0.2 &   7.1 &  1.8 $\pm$ 0.2 &  6.9 &  1.8 $\pm$ 0.2 &  15.2 &  2.4 $\pm$ 0.1 \\
IGR\,J17098$-$3628  & 10.0 &  --- & $<$4.2         &  ---  & $<$4.5         &  --- & $<$4.3         &  ---  & $<$2.5         \\
4U\,1700$-$377      & 33.7 & 48.0 & 22.0 $\pm$ 0.4 &  40.0 & 19.5 $\pm$ 0.5 & 28.4 & 13.4 $\pm$ 0.5 &  67.7 & 18.3 $\pm$ 0.2 \\
GRO\,J1655$-$40     & 21.4 & 23.0 & 17.1 $\pm$ 0.7 &   9.4 &  7.3 $\pm$ 0.8 &  --- & $<$8.7         &  18.6 &  7.9 $\pm$ 0.4 \\
\hline
\end{tabular}
\label{significance_2}
\end{table*}

\begin{table}
\caption[]{Compilation of information for each of the JEM-X mosaic images
per season and for the whole program (Fig.~\ref{fig3}). We report the season 
which is refered to, the {\em INTEGRAL} revolutions corresponding to the season, the total 
number of single pointings (ScWs), the total exposure time at the center of each
mosaic ($t_{\rm exp}$), the effective exposure time ($t_{\rm eff}$) when taking 
into account vignetting and other effects inherent to the JEM-X instrument
(see Lund et al.\ 2003), and the list of sources which are detected above
the 3$\sigma$ level in the 3--10\,keV and 10--25\,keV bands during a season or 
in the average over the whole program. When the detection significance is lower 
than 3 or a source is not detected we indicate it with (--). 
The ordering of the sources is the same as in Table 1.}
\begin{tabular}{l|cccc}
\hline
Season & 1 & 2 & 3 & all \\
Revs & 287--307 & 347--370 & 406--429 & 287--429 \\
\#\ ScWs & 132 & 118 & 141 & 391 \\
$t_{\rm exp}$ (ks) & $\simeq$200 & $\simeq$180 & $\simeq$195 & $\simeq$578 \\
$t_{\rm eff}$ (ks) & $\simeq$135 & $\simeq$122 & $\simeq$130 & $\simeq$379 \\
\hline
\multicolumn{1}{l|}{Source} & \multicolumn{4}{c}{Detection significances [3--10]/[10--25]\,keV} \\
\hline
GX\,5$-$1 		& 150/62 	& 174/82 	& 158/79 	& 276/97 \\
GRS\,1758$-$258  	& 8.4/14 	& 8.4/14 	& 8.5/12        & 14/17 \\
GX\,3+1  		& 150/71	& 135/57  	& 154/76 	& 259/94 \\
GX\,1+4  		& 3.8/6.9 	& 7.5/10 	& 8.6/11.1 	& 11/15 \\
SLX\,1735$-$269  	& 8.3/9.9 	& 11/8.4 	& 7.3/15	& 15/11 \\
1E\,1743.1$-$2843       & 9.0/7.7	& 10/8.6 	& 5.6/4.1 	& 14/9.2 \\
SAX\,J1747.0$-$2853     & --/-- 	& 9.0/-- 	& 23/14 	& 16/7.4 \\
SLX\,1737$-$282 	& 4.7/4.0	& --/-- 	& 3.9/5.1 	& 7.3/6.6 \\
XTE\,J1739$-$285  	& --/-- 	& 20/13 	& 21/12 	& 24/10 \\
1A\,1742$-$294  	& 28/24 	& 39/28 	& 25/31 	& 57/38 \\
SLX\,1744$-$229/300     & 18/12    	& 15/11 	& 12/10 	& 26/15 \\
1E\,1740.7$-$2942 	& 20/24 	& 19/32 	& --/-- 	& 22/29 \\
GRS\,1747$-$312  	& --/-- 	& 7.6/4.4	& 5.7/4.0 	& 8.0/3.9 \\
GRS\,1734$-$292 	& --/-- 	& --/-- 	& --/-- 	& 4.6/5.3 \\
H1743$-$322 	        & --/-- 	& 18/8.9 	& --/-- 	& 9.3/3.7 \\
4U\,1722$-$30  	        & 14/13 	& 14/11 	& 11/7.0 	& 23/14 \\
MXB\,1730$-$335 	& 4.4/-- 	& 5.1/3.4 	& --/-- 	& 5.7/-- \\
GX\,354$-$0  	        & 44/29 	& 32/20   	& 41/28   	& 72/35 \\
\hline
\end{tabular}
\label{jemx_mosaics}
\end{table}

Most of the bright well-known persistent and transient X-ray binaries are easily detected in 
either or all of the four JEM-X and IBIS/ISGRI 20--60\,keV mosaic images. 
Note that 1E\,1740.7$-$2942, normally one of the brightest sources in the 
Galactic center region, turned off during the whole third season
(see Sects.~\ref{off} and \ref{bhc}). Some of the fast transients were not detected in the
mosaics per season or for the whole period (such as IGR\,J17544$-$2619).
Some of the weaker sources are detected during every season's average 
(e.g., IGR\,J18027$-$2016, GX\,9+9, GRS\,1734$-$292);
some are only detected during one or two season averages (e.g., 1E\,1743.1$-$2843, IGR\,17456$-$2901). 
The latter is also true for some of the transient sources, such as IGR\,J17475$-$2822,
GRS\,1747$-$312 [in Terzan 6], and SAX\,J1712.6$-$3739.
We note that IGR\,17456$-$2901 (or a source coincident with it) was 
detected during the third season when the 1E\,1740.7$-$2942 and neighbouring sources were off
(see Sect.~\ref{off}). Sources detected only in the average over 
the whole monitoring period include various kinds of sources, such as
PKS\,1830$-$211 (AGN), V2400\,Oph (cataclysmic variable), XTE\,J1710$-$281 (low-mass X-ray binary).

From Table~\ref{significance_2} we see that only about a dozen sources are visible 
at 60--150\,keV. Different kinds of sources are detected at these energies,
such as 1E\,1740.7$-$2942 (black-hole candidate), GRO\,J1655$-$40
(transient black-hole binary), GS\,1826$-$24 (X-ray burster), and 4U\,1700$-$377
(high-mass X-ray binary).

EXMS\,B1709$-$232 was detected in the first two seasons, while 
the Oph\,Cluster was only detected in the third season.
These two sources are only 1.7' apart and can, therefore, not be resolved by IBIS/ISGRI.
For this reason we attribute the observed flux from only one source
which we label Oph\,Cluster (see also Bird et al.\ 2006).

Of the weak sources mentioned above, only the persistent sources 1E\,1743.1$-$2843
(low-mass X-ray binary) and GRS\,1734$-$292 (AGN), and the transient low-mass X-ray binary GRS\,1747$-$312, are detected
by JEM-X (see Table~\ref{jemx_mosaics}).
GRS\,1747$-$312 was also seen in JEM-X mosaics of our individual 
hexagonal dither observations near the start of the third season (Chenevez et al.\ 2006a), consistent
with the fact that it shows outbursts roughly every 4.5 months (in 't Zand et al.\ 2003).
It was barely detected in the IBIS/ISGRI mosaic of the individual 
hexagonal dither observation at the same time (at 6$\sigma$, $\simeq$11\,mCrab, 20--60\,keV).

\subsection{Source variability}
\label{light_curves}

\begin{figure}
\centering
  \includegraphics[height=.3\textheight,angle=-90]{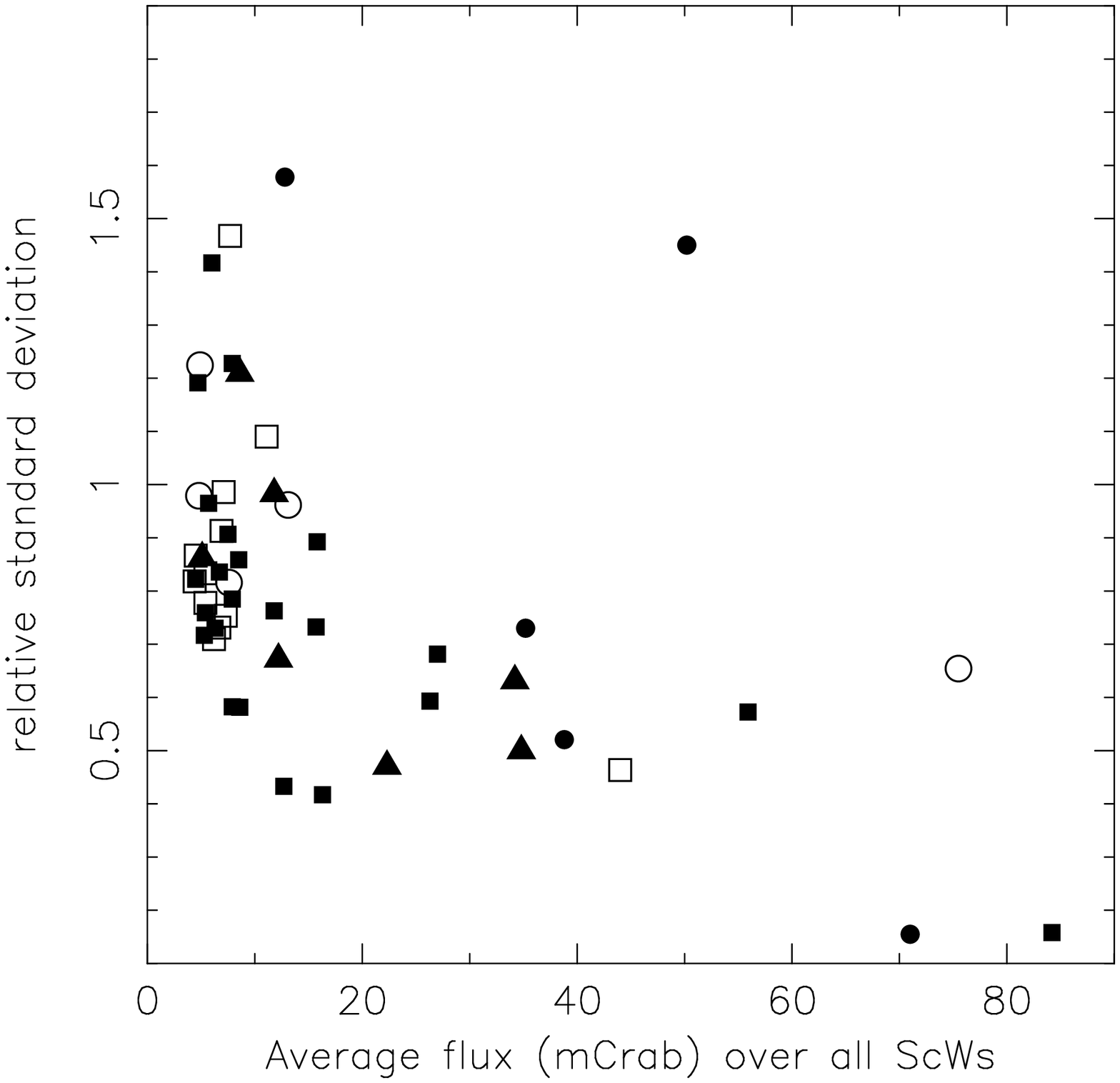}
  \caption{Relative standard deviation ($\sigma$/$\overline{F_{\rm A}}$) versus mean flux ($\overline{F_{\rm A}}$) for the 
20--60\,keV band. The different symbols refer to the different kinds of sources as outlined in Table~\ref{sourcelist};
filled circles: black-hole (candidate) binaries, filled squares: X-ray bursters, filled triangles: other low-mass X-ray binaries,
open circles: X-ray pulsars, open squares: miscellaneous sources. 4U\,1700$-$377 is outside the boundaries of the plot.}
\label{variance}
\end{figure}

To further explore the overall hard X-ray variability of the sources we 
computed the mean flux over the whole program, as well as the error in the mean and
standard deviation, both for the 20--60\,keV and the 60--150\,keV bands.
They are given in Table~\ref{average}.
The mean flux is calculated by averaging the flux values (weighted by their errors) from all the single pointings. 
This approach allows us to evaluate the source variability that is expressed in the standard 
deviation given in Table~\ref{average}.
Note that in this case the mean flux is not the same as the average value obtained from
the mosaic image of the three seasons together, discussed in Sect.~\ref{images} and shown in Table~\ref{significance}.
They are slightly overestimated with respect to the values obtained from the mosaic image, as discussed in Appendix~\ref{ave_mosa}.

In general, Table~\ref{average} shows that the standard deviation values ($\sigma$) strongly correlate with the mean fluxes
($\overline{F}$). It is, therefore, more appropriate to evaluate the relative standard deviations ($\sigma$/$\overline{F}$).
In Fig.~\ref{variance} we plot the relative standard deviation versus the mean flux in the 20--60\,keV band ($\overline{F_{\rm A}}$) 
for those sources which reached a significance higher than 7 in either one or more single hexagonal pointing, or in one
season, or in all three seasons together (see Table~\ref{significance}).
We grouped the sources among source type: 
black-hole (candidate) binaries, X-ray bursters, X-ray pulsars, other low-mass X-ray binaries (i.e., those which are
not members of the former groups), and other, miscellaneous, sources.
The different source types are shown with different symbols. All types of sources span more or less the same 
variability range. The transient sources are the most variable sources (i.e., $\sigma$/$\overline{F_{\rm A}}$$\gtrsim$1).
Most bright ($\overline{F_{\rm A}}$$\gtrsim$20\,mCrab) persistent sources have $\sigma$/$\overline{F_{\rm A}}$
values of around 0.5. Two persistent bright sources are outliers and vary little on long time scales; they 
have $\sigma$/$\overline{F_{\rm A}}$$\simeq$0.15. These are GRS\,1758$-$258, a black-hole candidate, and 
GS\,1826$-$24, an X-ray burster.

\tabcolsep=1mm
\begin{table}[ht]
\caption[]{Mean ($\overline{F}$) observed fluxes in the 20--60\,keV (denoted by A) 
and 60--150\,keV (denoted by B) bands, respectively, over the whole AO-3 monitoring program.
The error given for $\overline{F}$ is the error in the mean. 
"$\sigma$" denotes the standard deviation of the averaged values and gives a measure of the 
variability of a source.
The ordering of sources is the same as used for Table~1.}
\begin{tabular}{l|cccc}
\hline
\multicolumn{1}{l|}{} &
\multicolumn{1}{c}{$\overline{F_{\rm A}}$} &
\multicolumn{1}{c}{$\sigma$} &
\multicolumn{1}{c}{$\overline{F_{\rm B}}$} &
\multicolumn{1}{c}{$\sigma$} \\
\multicolumn{1}{l|}{Source} &
\multicolumn{2}{c}{(mCrab)} &
\multicolumn{2}{c}{(mCrab)} \\
\hline
GX\,17+2                 &  55.9 $\pm$ 1.9 & 32.0 & 44.3 $\pm$ 6.9 & 42.8 \\
SAX\,J1818.6$-$1703      &  11.1 $\pm$ 0.9 & 13.5 & 29.0 $\pm$ 2.5 & 31.7 \\
GX\,13+1                 &  15.8 $\pm$ 0.7 & 14.1 & 30.7 $\pm$ 2.3 & 34.8 \\
PKS\,1830$-$211          &  11.1 $\pm$ 0.7 & 12.1 & 30.0 $\pm$ 2.2 & 35.4 \\
SGR\,1806$-$20           &   7.1 $\pm$ 0.4 &  7.0 & 17.8 $\pm$ 1.3 & 14.6 \\
SAX\,J1805.5$-$2031      &   5.2 $\pm$ 0.5 &  4.8 & 17.0 $\pm$ 1.2 & 13.6 \\
IGR\,J18027$-$2016       &   7.6 $\pm$ 0.4 &  6.2 & 17.6 $\pm$ 1.3 & 19.4 \\
GS\,1826$-$24            &  84.2 $\pm$ 0.4 & 13.3 & 53.5 $\pm$ 1.1 & 27.0 \\
GX\,9+1                  &  12.2 $\pm$ 0.3 &  8.2 & 16.0 $\pm$ 1.2 & 24.7 \\
GX\,9+9                  &  11.8 $\pm$ 0.6 & 11.6 & 24.5 $\pm$ 2.0 & 27.4 \\
1RXS\,J175113.3$-$201214 &   6.2 $\pm$ 0.4 &  4.8 & 16.2 $\pm$ 1.3 & 15.8 \\
H1745$-$203              &   4.9 $\pm$ 0.5 &  4.6 & 15.7 $\pm$ 1.3 & 17.8 \\
IGR\,J17597$-$2201       &   5.5 $\pm$ 0.4 &  4.1 & 14.1 $\pm$ 1.1 & 14.0 \\
XTE\,J1818$-$245         &   8.6 $\pm$ 0.4 & 10.4 & 18.4 $\pm$ 1.2 & 17.2 \\
1RXS\,J174607.8$-$213333 &   4.5 $\pm$ 0.4 &  7.0 & 15.3 $\pm$ 1.2 & 14.2 \\
GX\,5$-$1                &  34.2 $\pm$ 0.2 & 21.6 & 13.1 $\pm$ 1.0 & 10.7 \\
V1223\,Sgr               &  15.1 $\pm$ 1.7 & 11.3 & 45.6 $\pm$ 4.9 & 26.0 \\
GRS\,1758$-$258          &  71.0 $\pm$ 0.2 & 11.0 & 97.9 $\pm$ 0.7 & 21.6 \\
IGR\,J17544$-$2619       &   4.9 $\pm$ 0.3 &  6.0 & 11.8 $\pm$ 1.0 & 10.2 \\
H1820$-$303              &  26.3 $\pm$ 0.3 & 15.6 & 17.1 $\pm$ 1.2 & 18.4 \\
IGR\,J17331$-$2406       &   4.4 $\pm$ 0.3 &  4.0 & 13.4 $\pm$ 1.0 & 16.6 \\
GX\,3+1                  &   7.9 $\pm$ 0.2 &  4.6 & 11.1 $\pm$ 1.0 &  9.8 \\
GX\,1+4                  &  44.0 $\pm$ 0.2 & 20.4 & 26.0 $\pm$ 0.8 & 15.7 \\
XTE\,J1807$-$294         &   4.4 $\pm$ 0.3 &  8.2 & 12.9 $\pm$ 1.0 & 11.7 \\
AX\,J1749.2$-$2725       &   3.6 $\pm$ 0.3 &  3.1 &  8.6 $\pm$ 1.0 & 10.4 \\
AX\,J1749.1$-$2733       &   4.8 $\pm$ 0.3 &  4.7 & 10.0 $\pm$ 1.1 & 24.0 \\
XB\,1832$-$330           &  11.8 $\pm$ 0.5 &  9.0 & 25.9 $\pm$ 1.5 & 22.4 \\
SLX\,1735$-$269          &  12.7 $\pm$ 0.2 &  5.5 & 15.7 $\pm$ 0.8 & 17.2 \\
XTE\,J1748$-$288         &   4.1 $\pm$ 0.3 &  5.4 &  9.7 $\pm$ 1.0 & 12.9 \\
IGR\,J17475$-$2822       &   4.5 $\pm$ 0.3 &  3.9 & 11.3 $\pm$ 1.0 & 11.4 \\
EXMS\,B1709$-$232        &   7.3 $\pm$ 0.4 &  5.5 & 17.7 $\pm$ 1.4 & 15.8 \\
IGR\,J17507$-$2856       &   4.2 $\pm$ 0.3 &  3.5 & 11.1 $\pm$ 1.0 &  9.0 \\
Oph\,Cluster             &   5.4 $\pm$ 0.5 &  4.2 & 10.3 $\pm$ 2.0 & 10.3 \\
IGR\,J17419$-$2802       &   4.2 $\pm$ 0.3 &  3.9 & 12.0 $\pm$ 1.0 & 10.3 \\
1E\,1743.1$-$2843        &   5.1 $\pm$ 0.3 &  4.4 &  9.9 $\pm$ 1.0 & 11.2 \\
SAX\,J1747.0$-$2853      &   7.5 $\pm$ 0.3 &  6.8 & 10.3 $\pm$ 0.9 & 11.8 \\
IGR\,J17407$-$2808       &   4.2 $\pm$ 0.3 &  3.8 & 10.0 $\pm$ 1.0 & 10.6 \\
SLX\,1737$-$282          &   5.4 $\pm$ 0.3 &  4.1 & 10.7 $\pm$ 1.0 & 10.8 \\
IGR\,J17456$-$2901       &   5.4 $\pm$ 0.4 &  4.5 &  7.0 $\pm$ 1.4 &  8.6 \\
V2400\,Oph               &   6.7 $\pm$ 0.4 &  4.9 & 15.1 $\pm$ 1.2 & 12.8 \\
XTE\,J1817$-$330         &  35.2 $\pm$ 0.5 & 25.7 & 34.9 $\pm$ 1.6 & 21.5 \\
XTE\,J1739$-$285         &   7.9 $\pm$ 0.3 &  6.2 & 12.2 $\pm$ 1.0 & 13.0 \\
GRS\,1741.9$-$2853       &   4.7 $\pm$ 0.3 &  5.6 & 10.4 $\pm$ 1.0 & 10.0 \\
SAX\,J1744.7$-$2916      &   4.3 $\pm$ 0.4 &  5.0 & 10.7 $\pm$ 1.2 & 12.5 \\
KS\,1741$-$293           &   7.9 $\pm$ 0.3 &  9.7 & 12.3 $\pm$ 1.0 & 19.4 \\
1A\,1742$-$294           &  15.7 $\pm$ 0.2 & 11.5 & 14.8 $\pm$ 0.8 & 12.9 \\
SLX\,1744$-$299/300      &   8.6 $\pm$ 0.2 &  5.0 & 13.7 $\pm$ 0.9 & 12.4 \\
1E\,1740.7$-$2942        &  38.8 $\pm$ 0.2 & 20.2 & 46.9 $\pm$ 0.7 & 26.3 \\
GRS\,1734$-$292          &   6.2 $\pm$ 0.2 &  4.4 & 12.7 $\pm$ 0.9 & 17.0 \\
IGR\,J17460$-$3047       &   3.6 $\pm$ 0.3 &  3.5 & 10.9 $\pm$ 1.0 &  9.2 \\
GRS\,1747$-$312          &   5.3 $\pm$ 0.5 &  3.8 & 12.9 $\pm$ 1.5 &  9.8 \\
IGR\,J17391$-$3021       &   4.3 $\pm$ 0.3 &  5.3 & 11.2 $\pm$ 1.0 & 11.4 \\
IGR\,J17285$-$2922       &   4.1 $\pm$ 0.3 &  3.2 & 12.7 $\pm$ 1.0 & 10.2 \\
H1732$-$322              &  12.8 $\pm$ 0.3 & 20.2 & 17.1 $\pm$ 0.9 & 19.2 \\
IGR\,J17488$-$3253       &   4.4 $\pm$ 0.3 &   3.6 & 12.9 $\pm$ 0.9 &  10.8 \\
3A\,1822$-$371           &  22.3 $\pm$ 0.4 &  10.5 & 20.1 $\pm$ 1.6 &  17.9 \\
SLX\,1746$-$331          &   4.6 $\pm$ 0.3 &   4.1 & 13.5 $\pm$ 0.9 &  10.9 \\
XTE\,J1710$-$281         &   6.0 $\pm$ 0.4 &   8.5 & 14.9 $\pm$ 1.2 &  17.1 \\
4U\,1722$-$30            &  16.3 $\pm$ 0.2 &   6.8 & 15.9 $\pm$ 0.8 &  10.5 \\
IGR\,J17200$-$3116       &   4.4 $\pm$ 0.4 &   4.7 & 12.0 $\pm$ 1.3 &  14.0 \\
MXB\,1730$-$335          &   5.7 $\pm$ 0.3 &   5.5 & 14.2 $\pm$ 1.0 &  15.0 \\
\hline
\end{tabular}
\label{average}
\end{table}
\begin{table}[ht]
\begin{tabular}{l|cccc}
\multicolumn{5}{l}{{\bf Table 6} (continued).} \\
\multicolumn{5}{l}{~} \\
\hline
\multicolumn{1}{l|}{} &
\multicolumn{1}{c}{$\overline{F_{\rm A}}$} &
\multicolumn{1}{c}{$\sigma$} &
\multicolumn{1}{c}{$\overline{F_{\rm B}}$} &
\multicolumn{1}{c}{$\sigma$} \\
\multicolumn{1}{l|}{Source} &
\multicolumn{2}{c}{(mCrab)} &
\multicolumn{2}{c}{(mCrab)} \\
\hline
XTE\,J1720$-$318         &   4.4 $\pm$ 0.4 &   4.2 & 12.6 $\pm$ 1.0 &  10.2 \\
GX\,354$-$0              &  27.0 $\pm$ 0.2 &  18.4 & 14.5 $\pm$ 1.0 &  12.3 \\
IGR\,J17254$-$3257       &   4.5 $\pm$ 0.4 &   3.7 & 11.7 $\pm$ 1.2 &  12.7 \\
1A\,1744$-$361           &   5.0 $\pm$ 0.4 &   3.9 & 12.9 $\pm$ 1.2 &  11.5 \\
1H\,1746$-$370           &   6.3 $\pm$ 0.4 &   4.6 & 14.3 $\pm$ 1.3 &  26.4 \\
XTE\,J1743$-$363         &   6.9 $\pm$ 0.3 &   6.3 & 14.1 $\pm$ 1.2 &  21.9 \\
4U\,1705$-$32            &   6.7 $\pm$ 0.4 &   5.6 & 16.4 $\pm$ 1.2 &  12.8 \\
IGR\,J17252$-$3616       &  13.1 $\pm$ 0.3 &  12.6 & 16.6 $\pm$ 1.2 &  12.5 \\
IGR\,J17098$-$3628       &   7.7 $\pm$ 0.5 &  11.3 & 16.7 $\pm$ 1.6 &  19.9 \\
IGR\,J17091$-$3624       &   4.8 $\pm$ 0.5 &   5.4 & 15.1 $\pm$ 1.5 &  19.9 \\
GX\,349+2                &  34.8 $\pm$ 0.4 &  17.4 & 17.8 $\pm$ 1.7 &  17.1 \\
SAX\,J1712.6$-$3739      &   8.5 $\pm$ 0.5 &   7.3 & 21.0 $\pm$ 1.5 &  21.1 \\
4U\,1700$-$377           & 191.9 $\pm$ 0.5 & 197.8 & 91.0 $\pm$ 1.4 &  84.1 \\
GRO\,J1655$-$40          &  50.2 $\pm$ 1.0 &  72.8 & 70.9 $\pm$ 2.6 &  79.4 \\
OAO\,1657$-$415          &  75.5 $\pm$ 1.1 &  49.4 & 47.0 $\pm$ 3.5 &  45.1 \\
\hline
\end{tabular}
\end{table}

In the following subsections we describe our results for individual sources in more detail. We focus mainly on light curves in the 
20--60\,keV band; wherever appropriate we also give information on the 60--150\,keV results.
A study of the long-term soft X-ray ($<$20\,keV) behaviour can be far better done
with, e.g., data from the {\rm RXTE}/ASM or PCA. In this paper we discuss 
the long-term JEM-X results for the few sources that are detected by both
JEM-X and IBIS/ISGRI during either all or at least most single pointings. 

Only sources that exceed a 20--60\,keV 
detection significance of 7 in one or more single hexagonal dither observations are considered here. 
This level was chosen in order to assure that the long-term light curves reveal 
significant variations.
Note that the light curves display all data points, including those where the 
detection significance was lower than 7.
For most of the sources we show the mean
intensities averaged per hexagonal dither observation.
Only when sources are highly variable within an observation are the 
results from the single pointings displayed.

We again grouped the 20--60\,keV light-curve figures and results according to source type, i.e., 
black-hole (candidate) binaries, X-ray bursters, X-ray pulsars, other low-mass X-ray binaries, and miscellaneous sources.
To each group of sources we devote a separate subsection.
Wherever possible, a comparison to previous hard X-ray 
monitoring results is done; these comprise mainly observations
made by MIT/{\em OSO\,7} (1971--1974, Markert et al.\ 1979), 
{\em GRANAT}/SIGMA (1989--1998, see, e.g., Churazov et al.\ 1994; Revnivtsev et al.\ 2004b)
and {\em CGRO}/BATSE (1991--2000, see, e.g., Harmon et al.\ 2004).

\subsubsection{Black-hole (candidate) binaries}
\label{1E1740}

\begin{figure*}
  \includegraphics[height=.3\textheight,angle=-90]{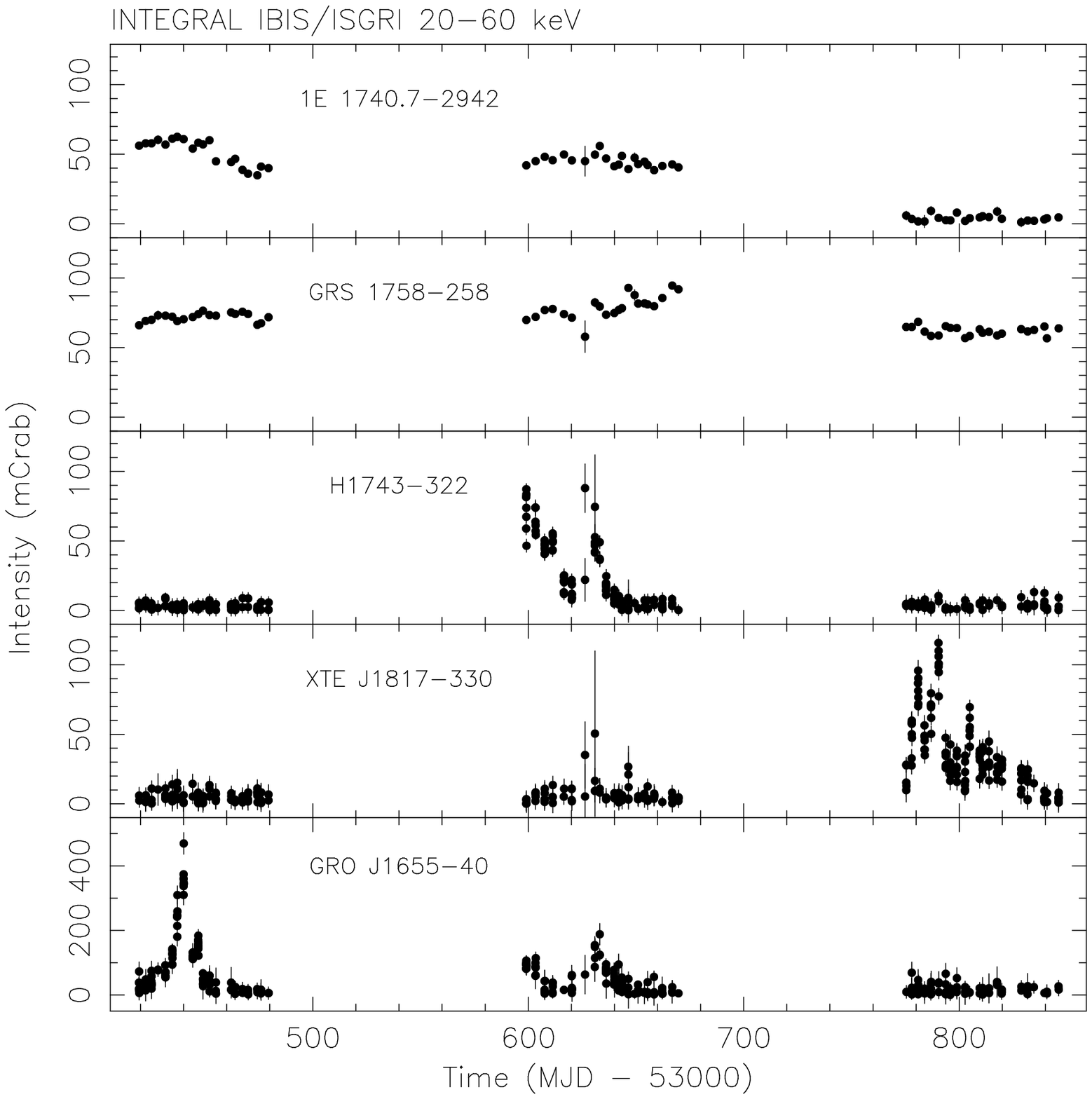}
  \includegraphics[height=.3\textheight,angle=-90]{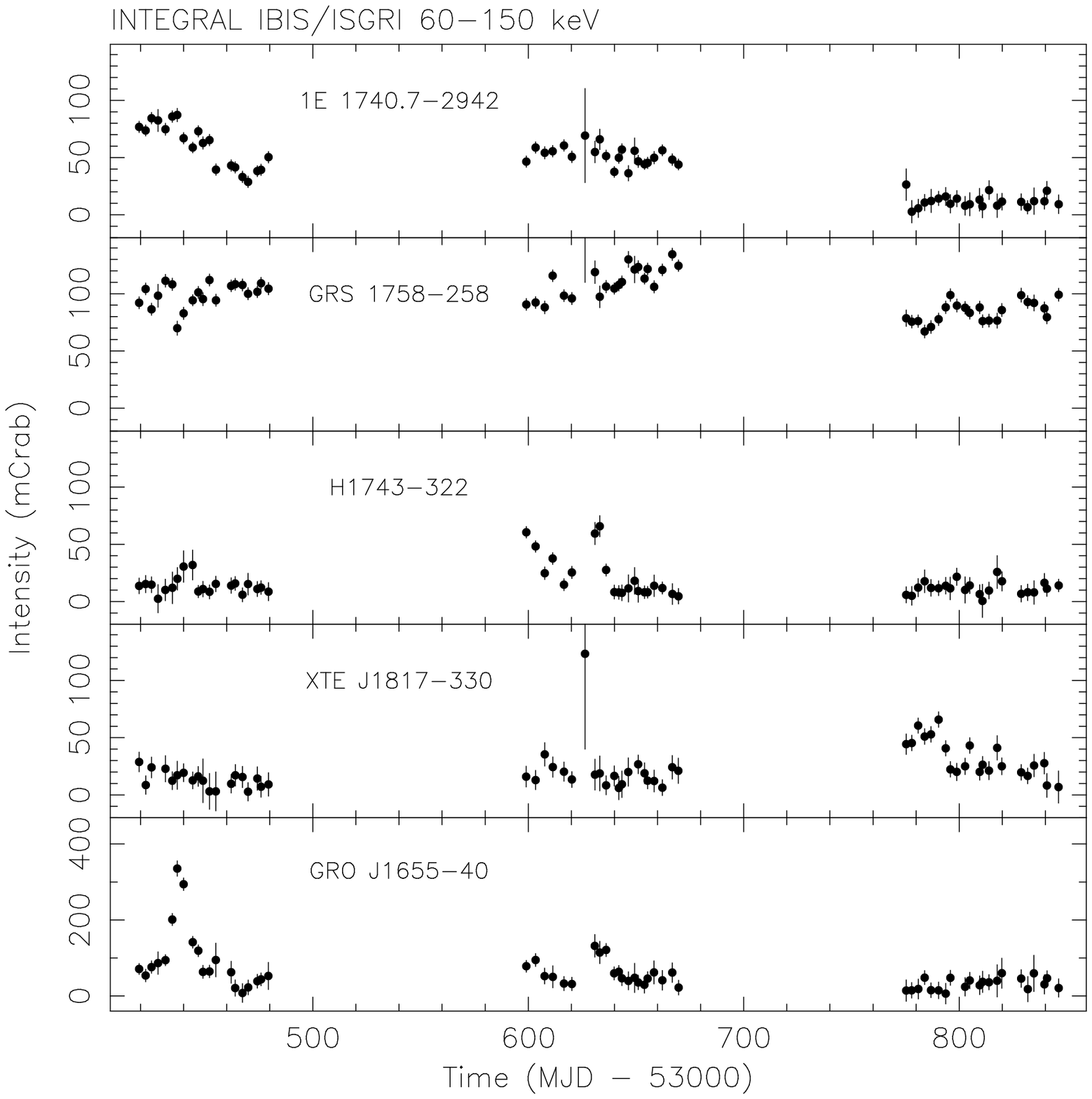}
  \caption{{\em INTEGRAL} IBIS/ISGRI light curves from the three
seasons of the Galactic bulge monitoring program, for the 
energy ranges 20--60\,keV ({\it left}) and 60--150\,keV ({\it right}). Shown are the 
black-hole (candidate) binaries detected during our program:
1E\,1740.7$-$2942, 
GRS\,1758$-$258, 
H1743$-$322, 
XTE\,J1817$-$330, and
GRO\,J1655$-$40. In the 20--60\,keV band, we show for the first two sources the averages per 
hexagonal dither observation (7 pointings), while for the latter we show single pointing
fluxes. The 60--150\,keV data points are averages per hexagonal dither observation.}
\label{bhc}
\end{figure*}

{\it 1E\,1740.7$-$2942.}
Normally, the most dominant source within the few degrees of the Galactic Center is 
the black-hole candidate 1E\,1740.7$-$2942.
It is therefore not surprising that earlier hard X-ray/$\gamma$-ray
($\gtrsim$20\,keV) measurements of the region could only focus on this source, given
the poor spatial resolution of the instruments 
(see, e.g., Cook et al.\ 1991, Bazzano et al.\ 1992, Churazov et al.\ 1994).
Variability in the hard X-ray flux of 1E\,1740.7$-$2942
was already evident from observations with different experiments (see, e.g., Bazzano et al.\ 1992).
In 2005 1E\,1740.7$-$2942 slowly varied on a monthly time scale between 
35--65\,mCrab and 30--90\,mCrab, in the 20--60\,keV and 60-150\,keV bands,
respectively (Fig.~\ref{bhc}). Similar flux levels were observed
previously by {\em GRANAT}/SIGMA (see Mandrou et al.\ 1994; Churazov et al.\ 1994) 
and {\em INTEGRAL}/IBIS (Del Santo et al.\ 2005).
In spring 2006 (MJD 53775--53846) the source went below the detection limits of JEM-X 
($\lesssim$4\,mCrab, 3--25\,keV) and IBIS/ISGRI ($\lesssim$1\,mCrab, 20--60\,keV; 
(see Fig.~\ref{bhc}, Table~\ref{significance}; see also Sect.~\ref{off}). 
Similar `switch-offs' at hard X-rays occurred
in 1991/1992 (e.g., Mandrou et al.\ 1994; Harmon et al.\ 2004), 1994/1995
(Harmon et al.\ 2004) and 2004 (Grebenev et al.\ 2004b; 
Del Santo et al.\ 2005). They seem to occur every $\sim$600~days
(Smith et al.\ 2002).

{\it GRS\,1758$-$258.}
This persistent black-hole candidate, located $\simeq$40' away from GX\,5$-$1
(Sect.~\ref{other}) varies between $\simeq$60--95\,mCrab and $\simeq$60--140\,mCrab
in the 20--60\,keV and 60--150\,keV energy bands, respectively, on weekly to monthly time scales. The source
is more variable at higher energies, especially during the third season
(Fig.~\ref{bhc}). The count rates in the two energy bands show that
the source is harder than the Crab.
Previous {\em INTEGRAL} observations (Pottschmidt et al.\ 2006), as well
as observations with {\em GRANAT}/SIGMA (Gilfanov et al.\ 1993;
Mandrou et al.\ 1994; Kuznetsov et al.\ 1999) and 
{\rm RXTE} and {\em CGRO}/BATSE (Smith et al.\ 2001, 2002) showed the source had similar 
variability, disappearing at various times below the detection limits during a 
whole season, indicating variability by more than a factor of 10. 
Note that this is similar to that seen in 1E\,1740.7$-$2942, as described above.

{\it H1743$-$322.}
In 2003 an outburst of this system was detected by INTEGRAL (Revnivtsev et al.\ 2003)
and it was designated IGR\,J17464$-$3213. The source was soon associated with H1743$-$322
(Markwardt \&\ Swank 2003a,b), from which outbursts had been previously seen (see Kalemci et al.\ 2006, 
and references therein).
INTEGRAL performed various observations throughout that outburst
(Parmar et al.\ 2003; Lutovinov et al.\ 2005; Capitanio et al.\ 2005; Joinet et al.\ 2005).
The source reappeared again in 2004 (Swank 2004) and 2005
(on MJD\,53588; Swank et al.\ 2005). Our program monitored the source from just after the 2005 maximum 
(Kretschmar et al.\ 2005; see Fig.~\ref{bhc}). The source was clearly seen in both the
20--60\,keV and 60--150\,keV bands. After a decay with an e-folding time
scale of $\sim$16~days, the source rebrightened to $\simeq$70\,mCrab,
after which it decayed with a shorter e-folding time, i.e., $\sim$5~days (20--60\,keV).

{\it XTE\,J1817$-$330.}
A new bright X-ray transient and black-hole candidate was reported in January 2006, designated
XTE\,J1817$-$330 (Remillard et al.\ 2006b). {\em INTEGRAL} detected it at the
start of the third season (see Shaw et al.\ 2006; Kuulkers et al.\ 2006a; Goldoni et al.\ 2006), 
and showed the source to be quite variable, up to
$\simeq$120\,mCrab (see Fig.~\ref{bhc}). The hard X-ray spectral shapes were
also seen to vary substantially, especially near the beginning of the season
(Kuulkers et al.\ 2006a). The transient was active during the whole third season.

{\it GRO\,J1655$-$40.}
Precisely at the start of our program the black-hole
X-ray transient GRO\,J1655$-$40 was reported to
become active (on MJD\,53419; Markwardt \&\ Swank 2005). 
Our {\em INTEGRAL} GRO\,J1655$-$40 light curves (Kuulkers et al.\ 2005a; 
Shaw et al.\ 2005c; Fig.~\ref{bhc}) nicely complement observations at soft
X-ray ({\em RXTE}/PCA; see Homan 2005, Shaposhnikov et al.\ 2006), optical/IR
(Buxton et al.\ 2005, Shaposhnikov et al.\ 2006) and radio 
({\em VLA}; see Rupen et al.\ 2005; Shaposhnikov et al.\ 2006) wavelengths. 
The 20--60\,keV and 60--150\,keV fluxes peaked at $\simeq$400 and $\simeq$350\,mCrab,
respectively. Contemporaneous hard X-ray coverage was also provided by
{\em RXTE}/HEXTE (see Homan 2005, Shaposhnikov et al.\ 2006) and {\em Swift/BAT} (Brocksopp et al.\ 2006).
The source was still active during our second season;
it is known for its multiple rebrightenings after the main one
(see, e.g., Harmon et al.\ 2004).

\subsubsection{X-ray bursters}
\label{bursters}

\begin{figure*}
  \includegraphics[height=.3\textheight,angle=-90]{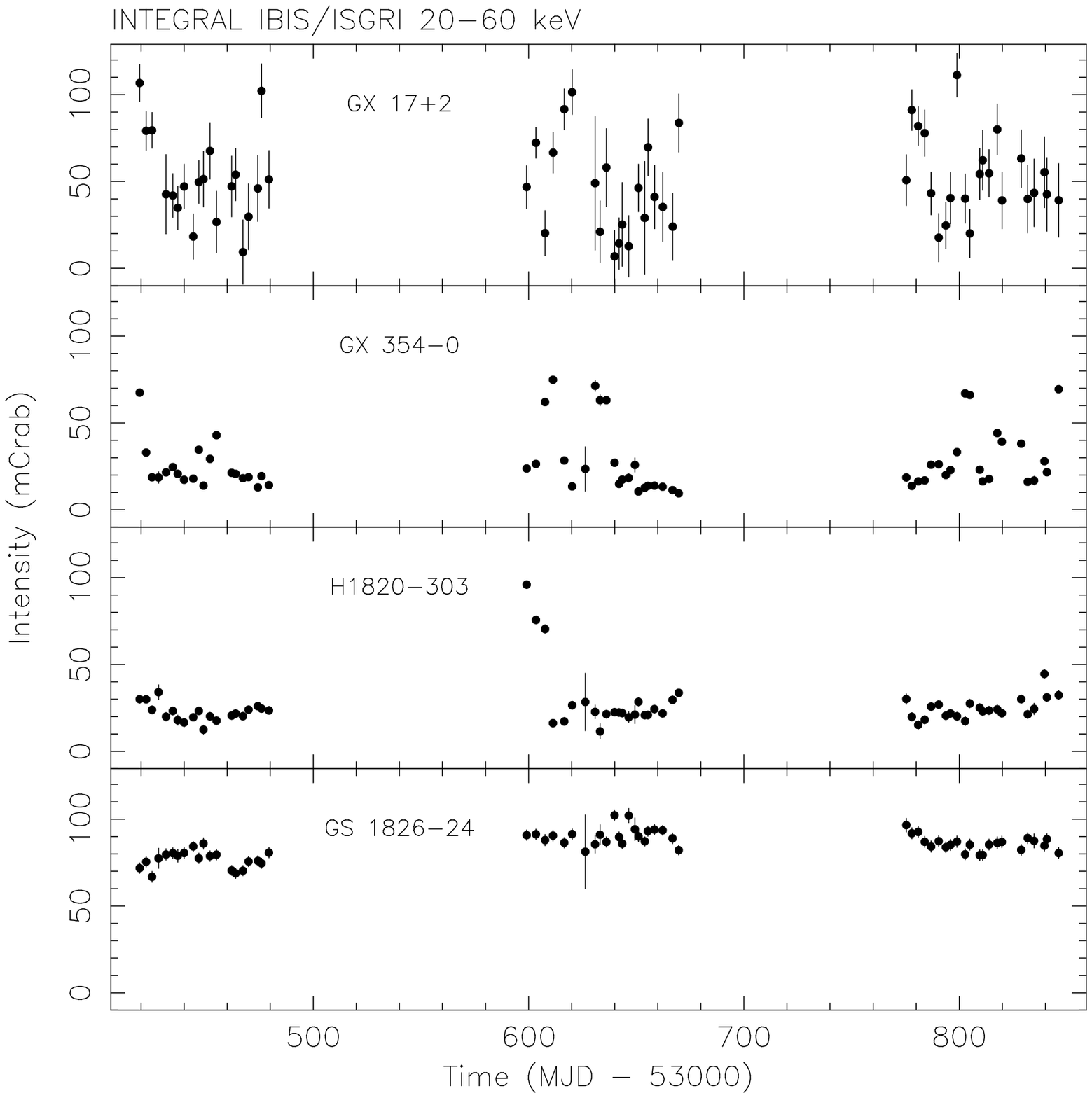}
  \includegraphics[height=.3\textheight,angle=-90]{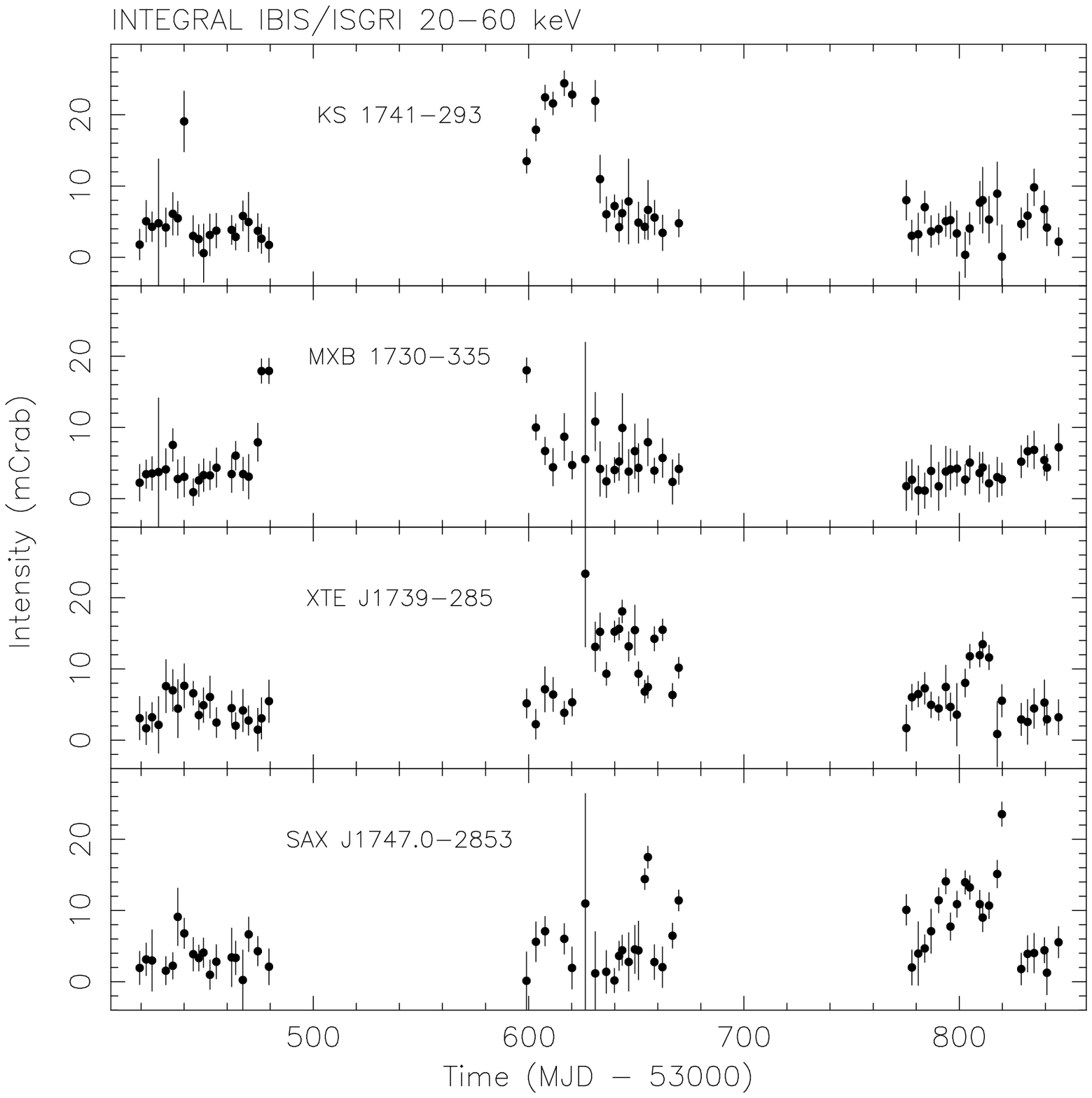}

  \includegraphics[height=.3\textheight,angle=-90]{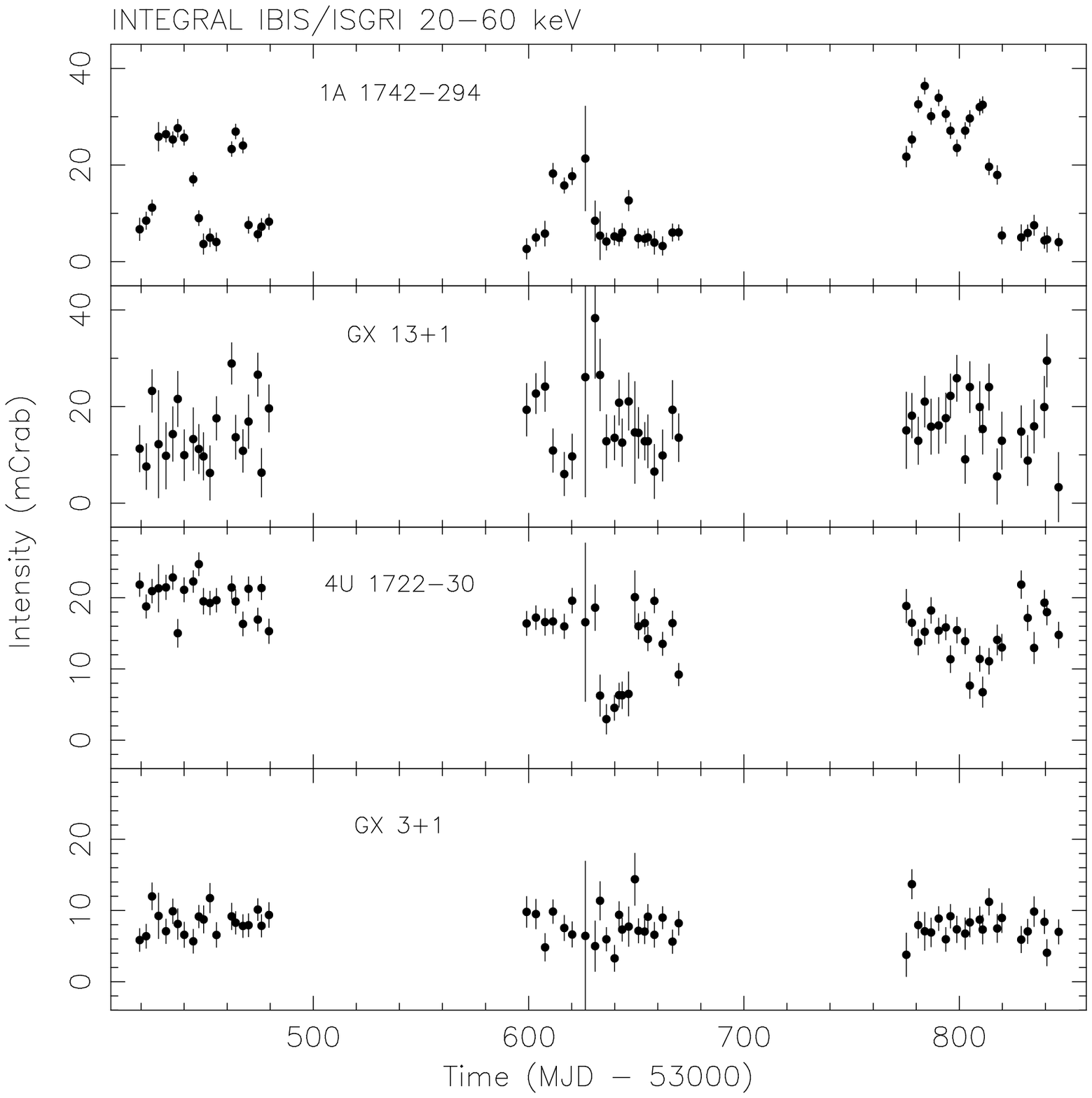}
  \includegraphics[height=.3\textheight,angle=-90]{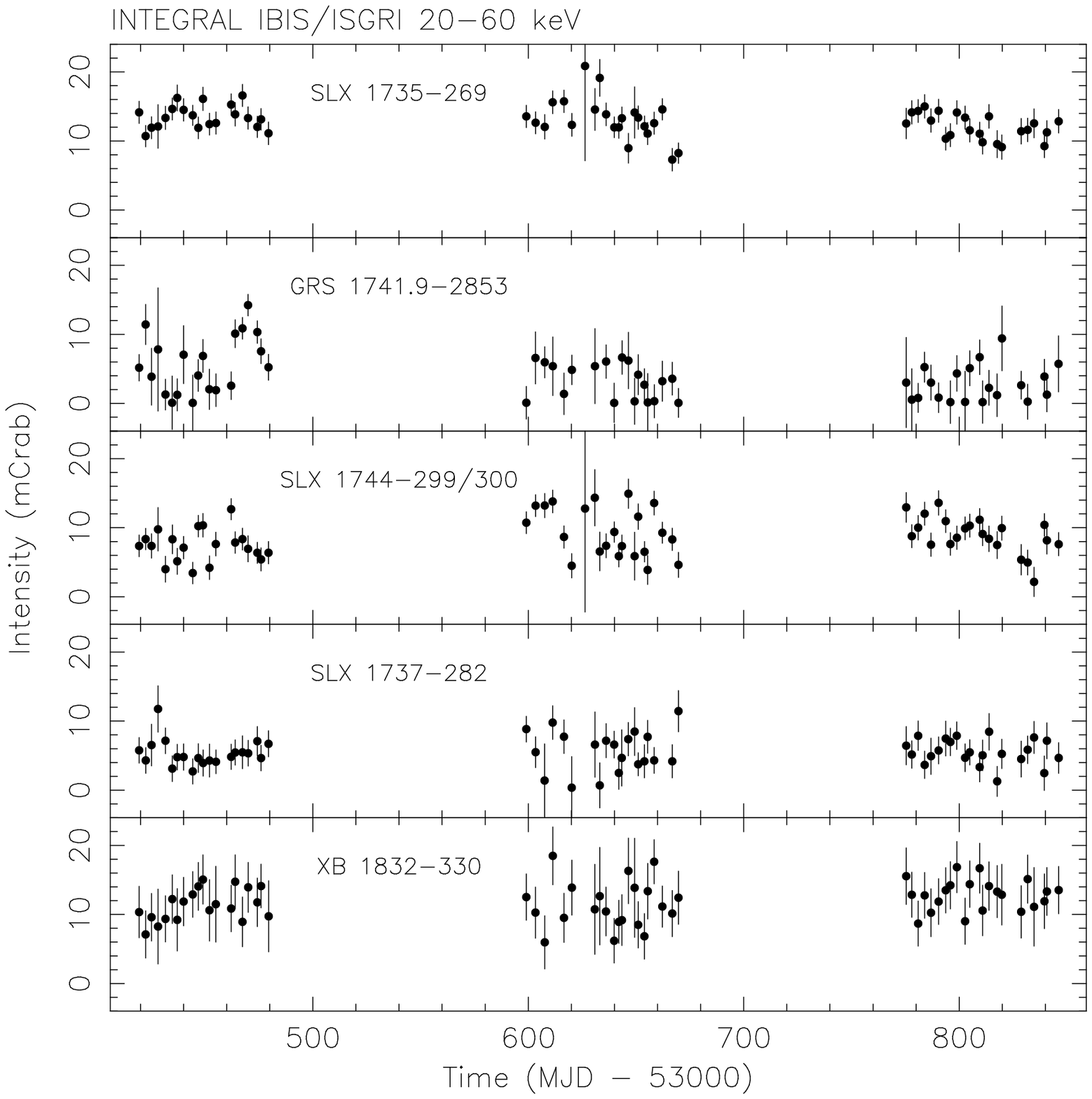}
  \caption{{\em INTEGRAL} IBIS/ISGRI light curves (20--60\,keV) from the three
seasons of the Galactic bulge monitoring program.
Shown are the averages per hexagonal dither observation for the following type I X-ray burst sources:
{\it upper left:}
GX\,17+2,
GX\,354$-$0,
H1820$-$303, and
GS\,1826$-$24; 
{\it upper right:} 
KS\,1741$-$293,
MXB\,1730$-$335,
XTE\,J1739$-$285, and
SAX\,J1747.0$-$2853;
{\it lower left:}
1A\,1742$-$294,
GX\,13+1,
4U\,1722$-$30, and
GX\,3+1;
{\it lower right:}
SLX\,1735$-$269,
GRS\,1741.9$-$2853,
SLX\,1744$-$299/300,
SLX\,1737$-$282, and
XB\,1832$-$330.}
\label{xrb}
\end{figure*}

{\it GX\,17+2.}
GX\,17+2 is a bright persistent (soft) X-ray source and highly variable 
on various time scales, one of the characteristics of Z-sources 
(e.g., Hasinger \&\ van der Klis 1989).
Previous IBIS/ISGRI observations showed the hard X-ray flux (22--40\,keV)
already to be variable on a $\simeq$10-day time scale between $\simeq$25--85\,mCrab 
(Paizis et al.\ 2006). Note that it is on average brighter at hard X-rays than 
the Z-source GX\,5$-$1 (Sect.~\ref{other}), whereas at soft X-rays GX\,5$-$1 is the brightest of the ``GX''-sources
(see, e.g., Hasinger \&\ van der Klis 1989). 
GX\,17+2 can flare up to $\simeq$110\,mCrab (20--60\,keV).
Similar flaring activity can be discerned from previous observations with IBIS/ISGRI
(Piraino et al.\ 2004). Similarly, the {\em MIT}/OSO-7 observations showed
variability on a time scale of months from their detection limits up to $\sim$300\,mCrab
(15--40\,keV; Markert et al.\ 1979).

{\it GX\,354$-$0.}
GX\,354$-$0 (or 4U\,1728$-$34) is generally seen with a 20--60\,keV flux 
between 10--30\,mCrab in our data (Fig.~\ref{xrb}). 
However, every now and then we see it flaring up to 70--80\,mCrab (20--60\,keV)
for a time scale of about a week (Fig.~\ref{xrb}). 
Comparable flux variations on weekly
time scales have been observed earlier by {\em INTEGRAL}/IBIS (Bazzano et al.\ 2004, 20--40\,keV;
Falanga et al.\ 2006, 20--100\,keV), 
as well as by {\em GRANAT}/SIGMA (Claret et al.\ 1994;
see also Mandrou et al.\ 1994). 
Similar variability is also present in the {\em CGRO}/BATSE light curves
(Harmon et al.\ 2004; see also Barret et al.\ 1996).
Flux increases up to 200\,mCrab (20--60\,keV) have been reported during previous 
{\em INTEGRAL}/IBIS observations (Zurita et al.\ 2004).
The {\em MIT}/OSO-7 (15--40\,keV) observations show that similar flux levels were reached 
in the early seventies (Markert et al.\ 1979).

\begin{figure}
\centering
  \includegraphics[height=.3\textheight,angle=-90]{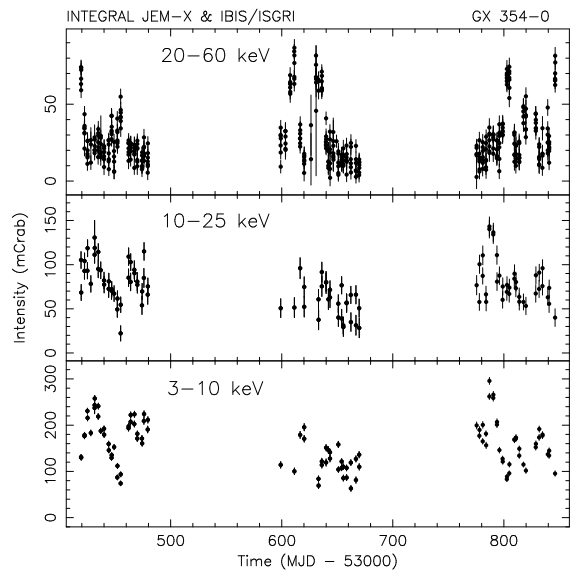}
  \caption{{\em INTEGRAL} IBIS/ISGRI (20--60\,keV, {\it top}) and 
JEM-X (3--10\,keV, {\it bottom}; 10--25\,keV, {\it middle}) light curves
(data from single 1800\,s pointings) of GX\,354$-$0.}
\label{GX354-0_isgri_jmx}
\end{figure}

GX\,354$-$0 is detectable in both JEM-X and IBIS/ISGRI most of the time
(Fig.~\ref{GX354-0_isgri_jmx}). Also in the 3--10 and 10--25\,keV band 
does the flux vary on typically weekly time scales, between $\simeq$60--300\,mCrab
and 20--150\,mCrab, respectively. The fluxes in these two bands are
clearly correlated. There appears, however, an anti-correlation between
the soft (3--10\,keV) and hard X-ray (20--60\,keV) flux: the highest soft X-ray
fluxes are accompanied by low hard X-ray fluxes, while the highest hard X-ray fluxes
are accompanied by low soft X-ray fluxes. However, at times both the soft and hard
X-ray fluxes can be simultaneously low.
Similar behaviour can be discerned from the 3--20\,keV and 20--100\,keV 
{\em INTEGRAL}/JEM-X and IBIS/ISGRI light curves presented by Falanga et al.\ (2006).
We saw 13 type~I X-ray bursts spread over our whole program. They occur
at different IBIS/ISGRI flux levels, similar to that found by Falanga et al.\ (2006).

{\it H1820$-$303} (in NGC\,6624).
The source generally varies slowly between 10--30\,mCrab in the 20--60\,keV band (Fig.~\ref{xrb};
see also {\em INTEGRAL}/IBIS observations presented by Bazzano et al.\ 2004 in the 20--40\,keV band, 
and Tarana et al.\ 2006a in the 20--30\,keV, 30--60\,keV and 60--120\,keV bands). This is consistent
with the 20--100\,keV upper limits derived from {\em CGRO}/BATSE (80\,mCrab and 30\,mCrab for
1-day and 10-day integrations, respectively; Bloser et al.\ 1996).
At the start of our second season (August 2005) the source was bright, i.e., 
$\simeq$100\,mCrab (20--60\,keV) and rather hard ($\simeq$40\,mCrab in the 60--150\,keV
band). It declined within 2 weeks to its normal flux level.
The same hard state was recently reported by Tarana et al.\ (2006a) from 
other {\em INTEGRAL} data taken close in time to our monitoring observations.
This can be connected to the soft ($\lesssim$20\,keV) low-intensity states
which occur roughly every 170~days (e.g., Chou \&\ Grindlay 2001).
Some evidence for variations on time scales of months can be seen
in the {\em MIT}/OSO-7 observations too, with 15--40\,keV 
fluxes generally being below 100\,mCrab (Markert et al.\ 1979).

{\it GS\,1826$-$24.}
During the first years of {\em CGRO}/BATSE (Harmon et al.\ 2004) GS\,1826$-$24 was
below/near its detection limits; later on it gradually became brighter, reaching
up to $\simeq$70\,mCrab near the end of the mission (20--100\,keV; Harmon et al.\ 2004). 
At the moment, the source is one of the brightest persistent type~I X-ray bursters in 
the Galactic bulge region. During our observations, it slowly varies on monthly and 
longer time scales between $\simeq$70 and $\simeq$100\,mCrab at 20--60\,keV 
(Fig.~\ref{xrb}). In the 60--150\,keV band, however, the source varies more on a
weekly time scale between $\simeq$30 and $\simeq$80\,mCrab, around an average
flux of $\simeq$55\,mCrab (Fig.~\ref{misc2}).

{\it KS\,1741$-$293.}
KS\,1741$-$293 was most of the time not significantly detected 
during our monitoring observations. However, it was bright, reaching up to 
$\simeq$25\,mCrab (20--60\,keV), for about a month during the first half of the
second season (August/September 2005, MJD\,53599--53633; Fig.~\ref{xrb}).
KS\,1741$-$293 was earlier seen to be active in March 2003 
and March 2004 (B\'elanger et al.\ 2004; Grebenev et al.\ 2004b; De Cesare et al.\ 2006).
Type~I X-ray bursts were previously found with JEM-X (De Cesare et al.\ 2006); in our program
we did not see any.

{\it MXB\,1730$-$335} (in Liller~1).
We saw MXB\,1730$-$335 
(better known as The Rapid Burster) turning on at the end
of the first season (mid-April 2005; see also Molkov et al.\ 2005b) 
and it was turning off at the beginning of the 
second season (mid-August 2005; Fig.~\ref{xrb}; see also 
Kretschmar et al.\ 2005). Previous outbursts of MXB\,1730$-$335 were
already recorded by IBIS/ISGRI, in February (Falanga et al.\ 2004) and 
August 2003 (Sunyaev et al.\ 2003a). This is consistent
with the outburst recurrence period being roughly 100~days
since 2000 (Masetti 2002). 
Strong burst activity is seen in our data near the end of the hard X-ray 
outburst, between 2005 August 25 and September 3 (MJD\,53607--53616).

{\it XTE\,J1739$-$285.}
In August 2005 (second season), the X-ray transient XTE\,J1739$-$285 
(discovered in 1999, Markwardt et al.\ 1999)
was found by {\em INTEGRAL} to be bright at soft and not detected at
hard X-ray energies (Bodghee et al.\ 2005). About a month
later the situation had reversed; it was bright at hard
and weak at soft X-ray energies (Shaw et al.\ 2005b).
Although at first we attributed the state change to
the compact object being a black hole, we proved it 
to be a neutron star based on the occurrence
of type I X-ray bursts detected with JEM-X (Brandt et al.\ 2005).
During the third season the source was still active (Chenevez et al.\ 2006a).
We saw a total of 13 type~I X-ray bursts when the source was active.

{\it SAX\,J1747.0$-$2853.}
SAX\,J1747.0$-$2853 is active at relatively low flux levels for long periods
(e.g., Wijnands et al.\ 2002, Natalucci et al.\ 2004). True quiescence, however, is reached sometimes 
(Werner et al.\ 2004). The source was active (Fig.~\ref{xrb}) 
during the end of the second visibility season (October 2005;
see Kuulkers et al.\ 2005c), as well during most part of the third visibility season
(e.g., Chenevez et al.\ 2006a). It reached a maximum of about 24\,mCrab
(20--60\,keV) during the third season.
Previous activity, as seen by {\em INTEGRAL}, was reported during March 2004 (Deluit et al.\ 2004).
In our data SAX\,J1747.0$-$2853 showed 6 type~I X-ray bursts, 1 during both the first and second
season, the rest during the third season (see also Kuulkers et al.\ 2005c, Chenevez et al.\ 2006a).

{\it 1A\,1742$-$294.}
This X-ray burster is 32' away from 1E\,1740.7$-$2942 and well
resolved by {\em INTEGRAL}/IBIS (see, e.g., B\'elanger et al.\ 2006). 
We see it clearly varying up to $\simeq$40\,mCrab (20--60\,keV)
on a monthly time scale (Fig.~\ref{xrb}). Similar variability on roughly a half
year time scale was earlier reported by Churazov et al.\ (1995)
using {\em GRANAT}/SIGMA.
This source is the most active type~I X-ray burster in our program. We found 36 type~I X-ray bursts;
most of the bursts were seen when the source was at the highest IBIS/ISGRI flux levels.

\begin{figure}
\centering
  \includegraphics[height=.3\textheight,angle=-90]{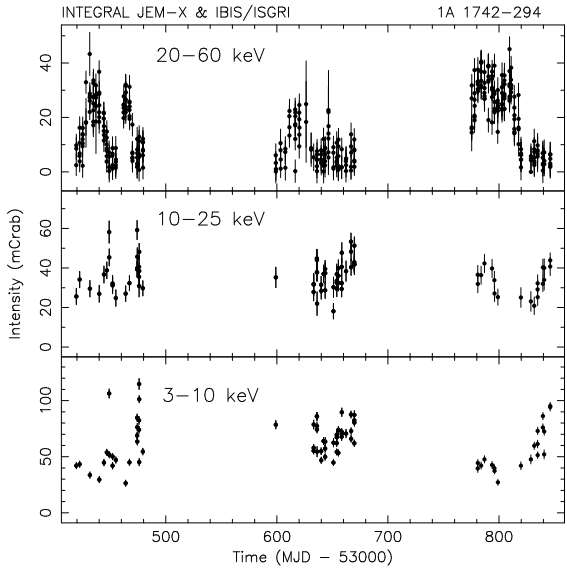}
  \caption{Same as Fig.~\ref{GX354-0_isgri_jmx} but for 
1A\,1742$-$294.}
\label{1A_1742-294_isgri_jmx}
\end{figure}

Due to the fact that at hard X-ray energies
IBIS/ISGRI is able to distinguish the source from neighboring sources, and the simultaneity
of the softer X-ray JEM-X information, we are able to study this behaviour for the first time
clearly in this source. During most of the pointings we detect 1A\,1742$-$294 with JEM-X (Fig.~\ref{1A_1742-294_isgri_jmx}).
The source shows the same behaviour between 3--10\,keV versus 10--25\,keV, and 
3--10\,keV versus 20--60\,keV, as is seen for GX\,354$-$0 (see above). 
Again, most of the time there is an anti-correlation 
between the fluxes in 3--10\,keV and 20--60\,keV energy bands. 

{\it GX\,13+1.}
Previous {\em INTEGRAL} observations showed the 20--40\,keV flux to be quickly
varying from the IBIS/ISGRI detection limits (upper limit typically 5\,mCrab) 
up to $\simeq$35\,mCrab on a $\simeq$10-day time scale (Paizis et al.\ 2006).
In our program the source varies between the IBIS/ISGRI detection limits and 40\,mCrab (20--60\,keV) 
on a revolution to revolution basis (Fig.~\ref{xrb}), consistent with that seen previously.

{\it 4U\,1722$-$30} (in Terzan 2).
The source 4U\,1722$-$30 is persistently
visible between $\simeq$10--25\,mCrab (20--60\,keV). {\em GRANAT}/SIGMA
saw the source varying between $\simeq$10--40\,mCrab
(35--75\,keV) on a $\sim$200-day time scale
(Goldwurm et al.\ 1995; see also Barret et al.\ 1991).
During the middle of our second season it showed a drop to near 
IBIS/ISGRI detection levels for a time period of $\simeq$2~weeks
(September/October 2005; Fig.~\ref{xrb}).
Strong type~I X-ray bursts are seen in all seasons, for a total of 5.

{\it GX\,3+1.}
During the February 2003 to May 2004 period the {\em RXTE}/ASM 2--12\,keV
intensity decreased more or less monotonically from about 400\,mCrab to about
250\,mCrab. The $\simeq$2-month average hard X-ray flux (22--40\,keV) decreased in the same
period from about 15\,mCrab to about 8\,mCrab. Within these 2 months periods the 
flux varied only weakly on a $\simeq$10-day time scale (Paizis et al.\ 2006).
Since May 2004 up to August 2004 the source continued to
decline to about 150\,mCrab in the 2--12\,keV band. Thereafter, it varied
erratically on a roughly 100\,day time scale between 100 and 250\,mCrab up
to the end of 2005. Over the course of 2006 the 2--12\,keV flux has been increasing steadily again;
it was about 250\,mCrab at the end of our third season.
The source is barely detectable in the 20--60\,keV band around $\simeq$10\,mCrab
in our program (Fig.~\ref{xrb}), consistent with the source behaviour in both soft and 
hard X-rays around May 2004 reported by Paizis et al.\ (2006). 
During the whole monitoring program we observed 10 type~I X-ray bursts.

{\it SLX\,1735$-$269.}
We see SLX\,1735$-$269 between 8--20\,mCrab 
(20--60\,keV; Fig.~\ref{xrb}), i.e., 
just above the detection limits. This is typical of the
source (Goldwurm et al.\ 1996), although 
INTEGRAL observations previous to our monitoring program show that
occasionally the hard X-ray flux decreases below 
$\simeq$5\,mCrab (Molkov et al.\ 2005a). We saw no type~I X-ray bursts during our osbervations with JEM-X.

{\it GRS\,1741.9$-$2853.}
GRS\,1741.9$-$2853 is, similar to SAX\,J1747.0$-$2853, a faint X-ray transient source
(e.g., Muno et al.\ 2003b). A hard X-ray outburst, with a peak of $\simeq$15\,mCrab and a duration of
at least a couple of weeks (20--60\,keV) was seen near the end of the first visibility season (April 2005,
MJD\,53464--53479; see Fig.~\ref{xrb}). We note that {\em XMM-Newton} and {\em Chandra} found the source to
be still active in soft X-rays ($<$10\,keV), respectively, two and three months later
(Wijnands et al.\ 2006).
A previous detection at hard energies
(40--100\,keV) was reported by {\em GRANAT}/SIGMA in March/April 1990
(Churazov et al.\ 1993). No type~I X-ray bursts were seen with JEM-X.

{\it SLX\,1744$-$299/300.}
SLX\,1744$-$299/300 weakly varies. Fluxes up to $\simeq$15\,mCrab (20-60\,keV) were reached
during the second season (Fig.~\ref{xrb}). A total of 9 type~I X-ray bursts were seen in our data
to come from them.

{\it SLX\,1737$-$282.}
SLX\,1737$-$282 is a weak persistent X-ray source (Skinner et al.\ 1987, in 't Zand et al.\ 2002),
and detected in the hard X-ray band (3.4$\pm$0.2\,mCrab, 18--60\,keV, Revnivtsev et al.\ 2004a; see also
Bird et al.\ 2004, 2006).
We see it varying between the detection limits up to about 12\,mCrab on a revolution
time scale (20--60\,keV; Fig.~\ref{xrb}). In our JEM-X data we saw no type~I X-ray bursts.

{\it XB\,1832$-$330} (in NGC\,6652).
The globular cluster source XB\,1832$-$330 lies far off-axis from the Galactic Center
($\simeq$11.5$^{\circ}$). We see it is a weak hard X-ray source with 20--60\,keV fluxes between 
$\sim$10--20\,mCrab (Fig.~\ref{xrb}). The source was seen at similar flux levels, without
strong variability, during several {\em INTEGRAL} IBIS/ISGRI observations taken between March 2003 and September 2005
(Tarana et al.\ 2006b). 

\begin{figure}
\centering
  \includegraphics[height=.3\textheight,angle=-90]{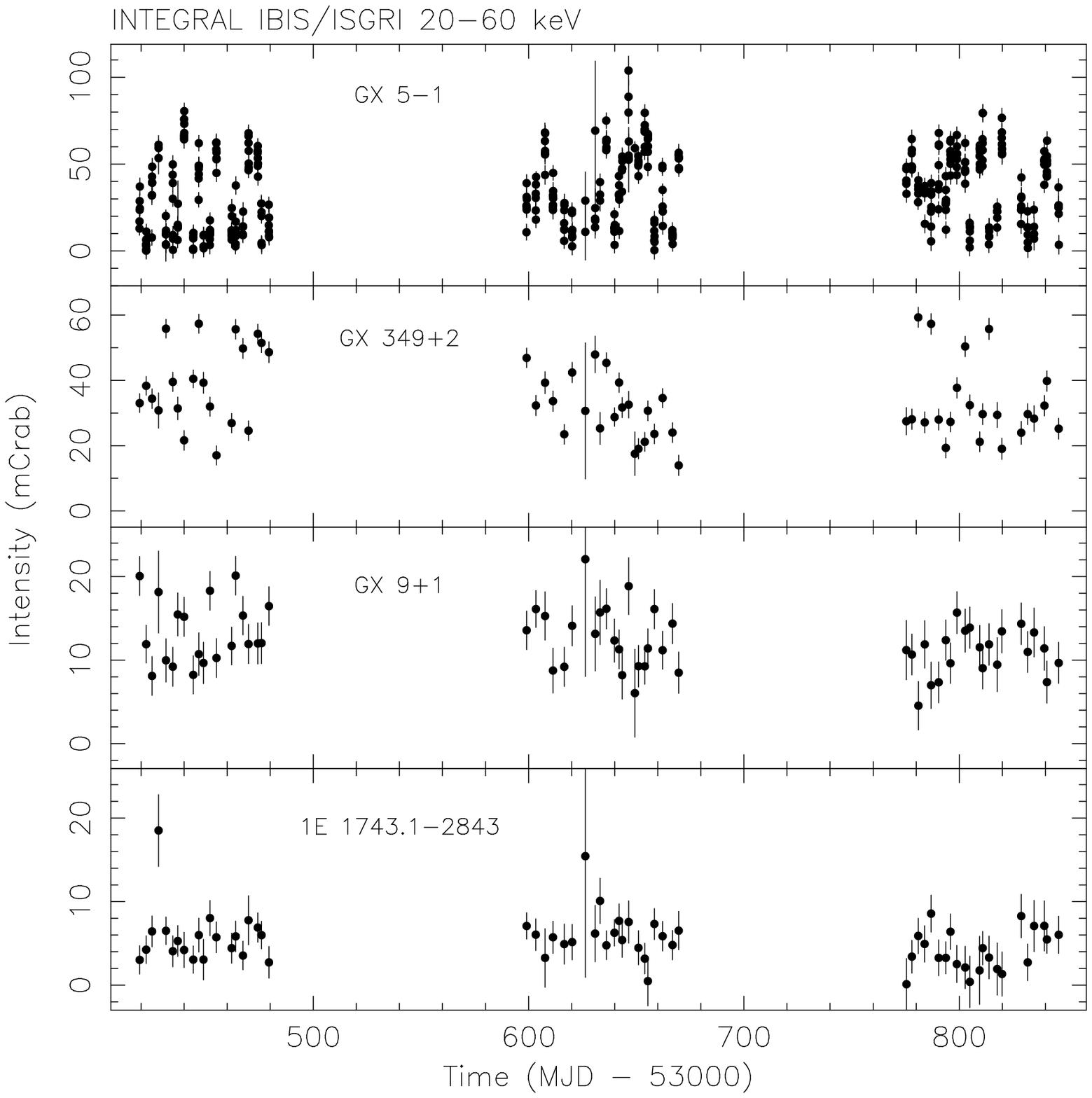}
  \includegraphics[height=.3\textheight,angle=-90]{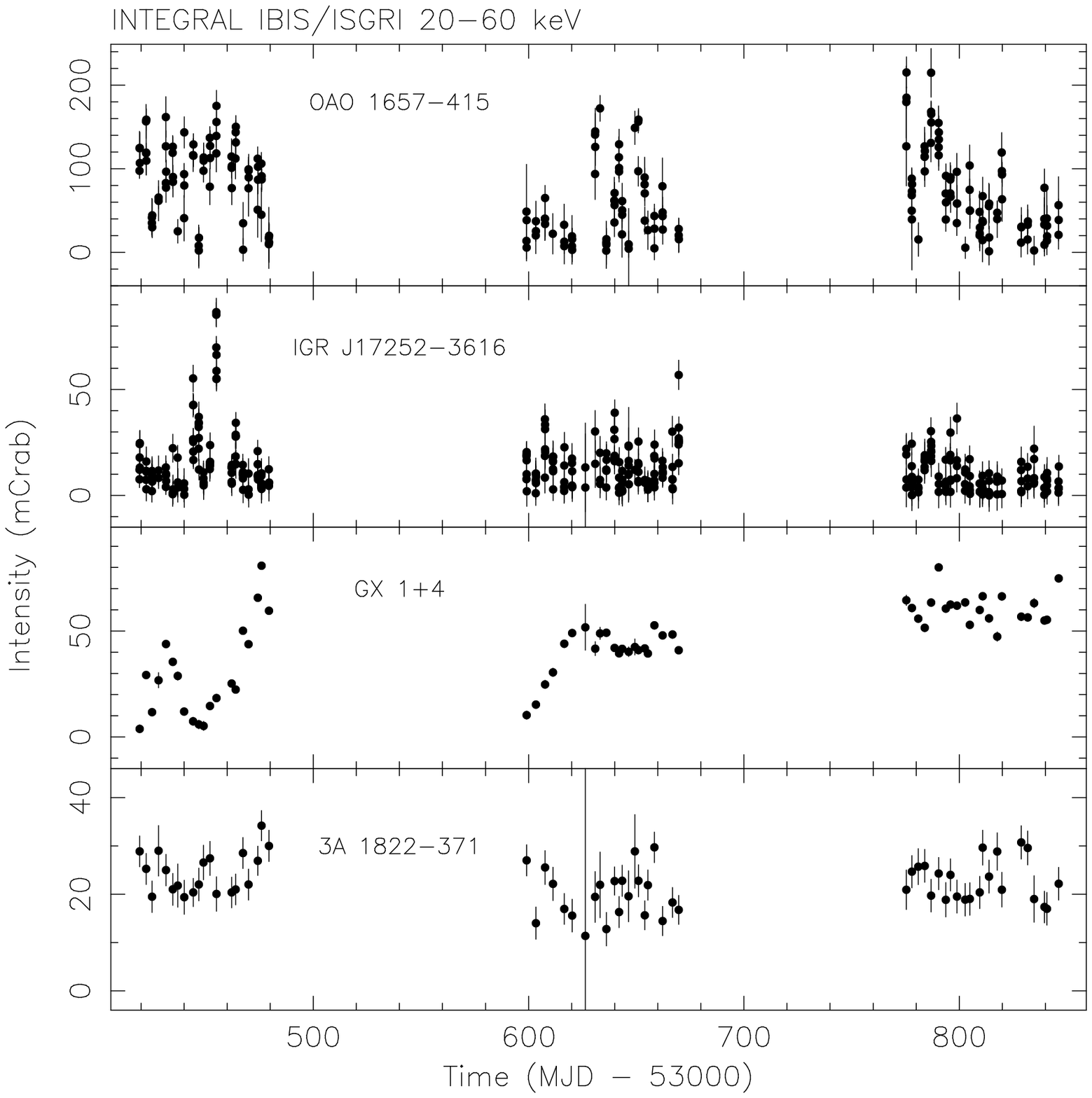}
  \includegraphics[height=.3\textheight,angle=-90]{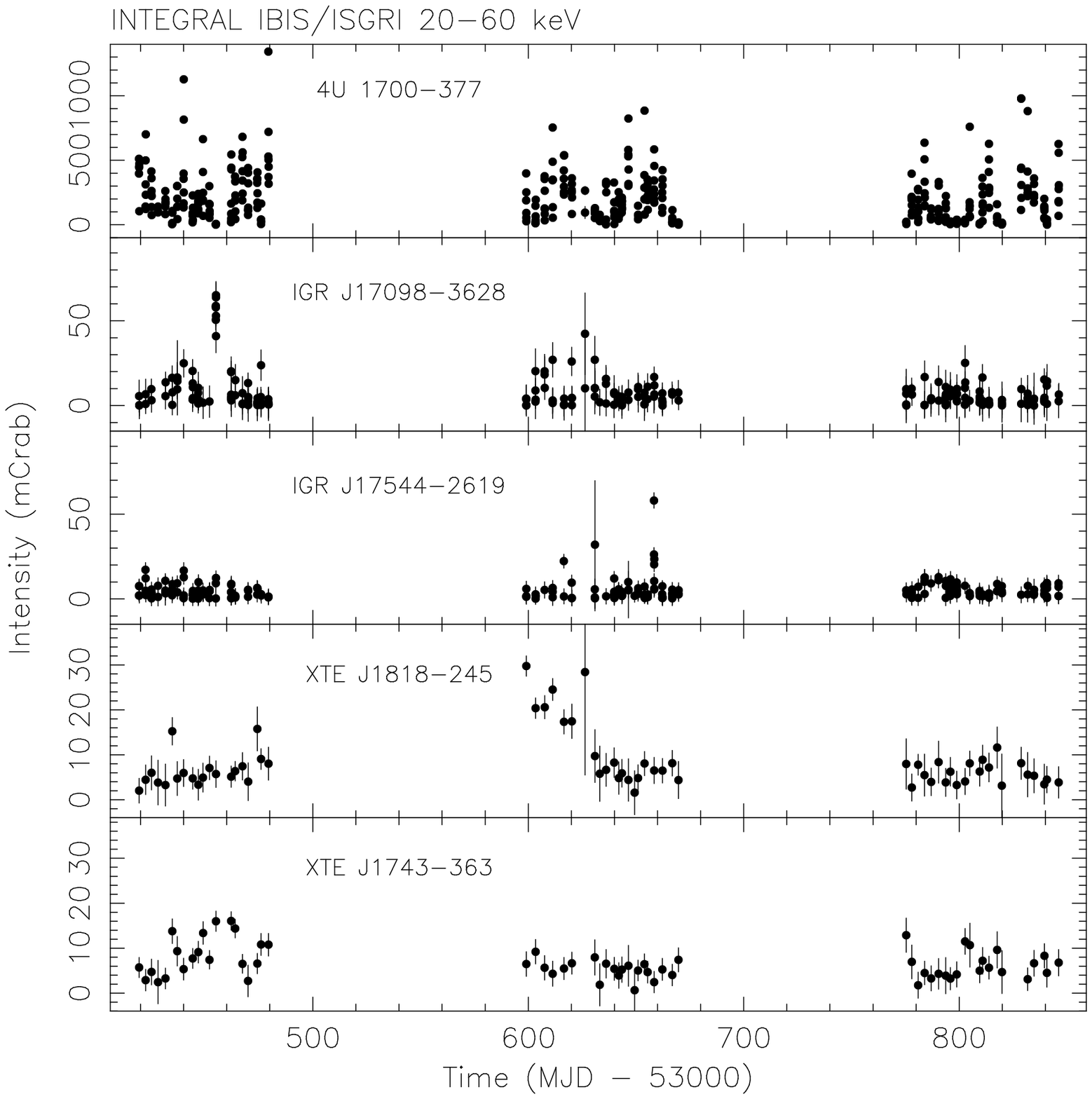}
  \caption{Same as Fig.~\ref{xrb}, but showing the 
following sources:
{\it top:} (other low-mass X-ray binaries)
GX\,5$-$1,
GX\,349+2,
GX\,9+1, and
1E\,1743.1$-$2843;
{\it middle:} (X-ray pulsars)
OAO\,1657$-$415,
IGRJ\,17252$-$3616,
GX\,1+4, and
3A\,1822$-$371;
{\it bottom:} (miscellaneous sources)
4U\,1700$-$377,
IGR\,J17098$-$3628,
IGR\,J17544$-$2619,
XTE\,J1818$-$245, and
XTE\,J1743$-$363.
}
\label{misc}
\end{figure}

\begin{figure}
\centering
  \includegraphics[height=.3\textheight,angle=-90]{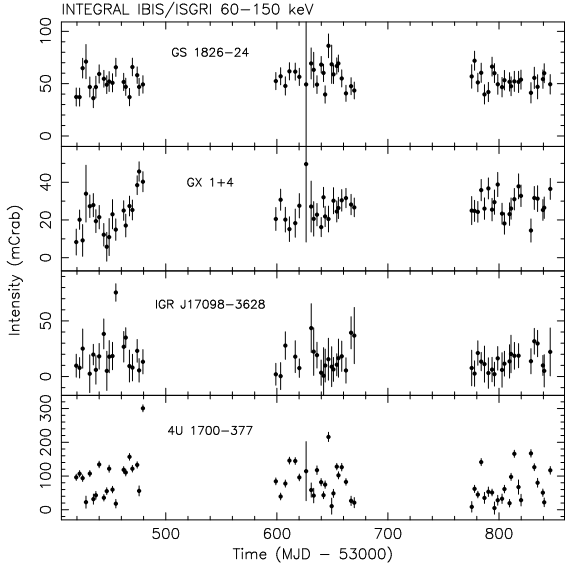}
  \caption{{\em INTEGRAL} IBIS/ISGRI light curves (60--150\,keV) from the three
seasons of the Galactic bulge monitoring program.
Shown are the averages per hexagonal dither observation for the following sources:
GS\,1826$-$24,
GX\,1+4,
IGR\,J17098$-$3628, and
4U\,1700$-$377.}
\label{misc2}
\end{figure}

\subsubsection{Other low-mass X-ray binaries}
\label{other}

\begin{figure}
\centering
  \includegraphics[height=.3\textheight,angle=-90]{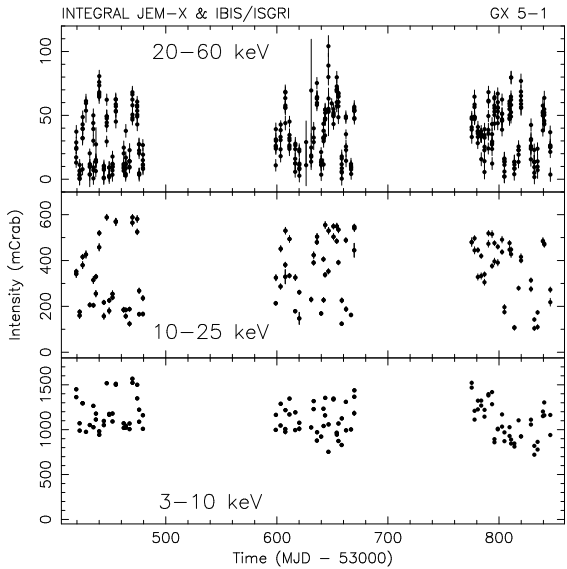}
  \caption{Same as Fig.~\ref{GX354-0_isgri_jmx} but for 
GX\,5$-$1.}
\label{GX_5-1_isgri_jmx}
\end{figure}

{\it GX\,5$-$1.}
Like GX\,17+2 (Sect.~\ref{bursters}), GX\,5$-$1 is highly variable on various time scales
(see Fig.~\ref{misc}).
It is one of the brightest persistent sources in the canonical 2--10\,keV band
(only Sco\,X-1 and Crab are brighter); it is also a Z source (Hasinger \&\ van der Klis 1985). 
Thanks to the high angular resolution of IBIS/ISGRI we are able to 
discriminate clearly its hard-energy radiation from that of the nearby 
($\simeq$40') strong hard X-ray source GRS\,1758$-$258 (Sect.~\ref{1E1740}).
On time scales of half an hour (one pointing) and longer also at harder energies
(20--60\,keV) the flux changes considerably, from the detection limit up to $\simeq$100\,mCrab
(Fig.~\ref{misc}; see also Paizis et al.\ 2005, 2006 for previous {\em INTEGRAL} observations).
Markert et al.\ (1979) do provide 15--40\,keV long-term light curves for GX\,5$-$1,
but they are most likely contaminated by GRS\,1758$-$258. 

\begin{figure*}
  \includegraphics[height=.2\textheight,angle=-90]{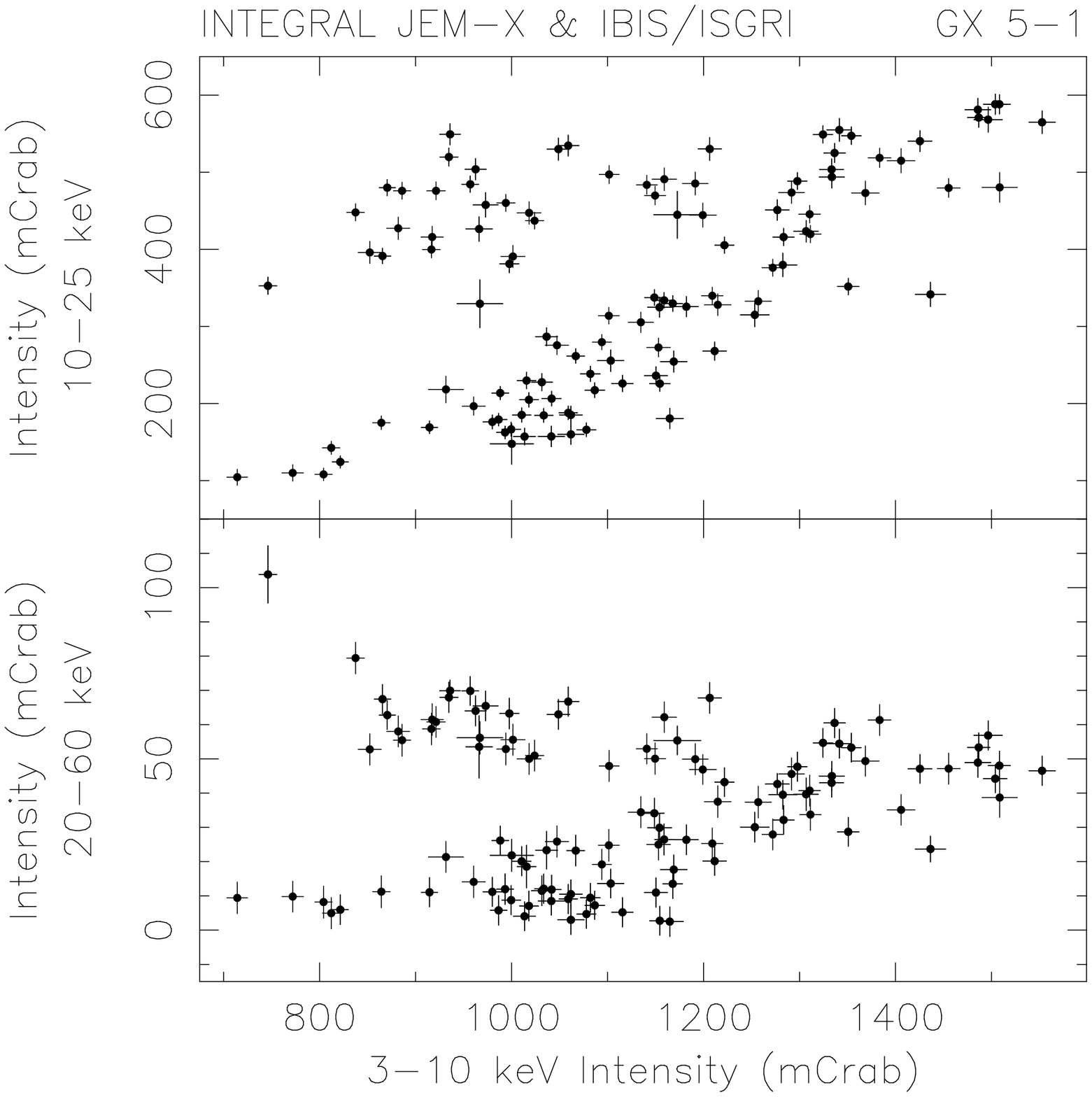}
  \includegraphics[height=.2\textheight,angle=-90]{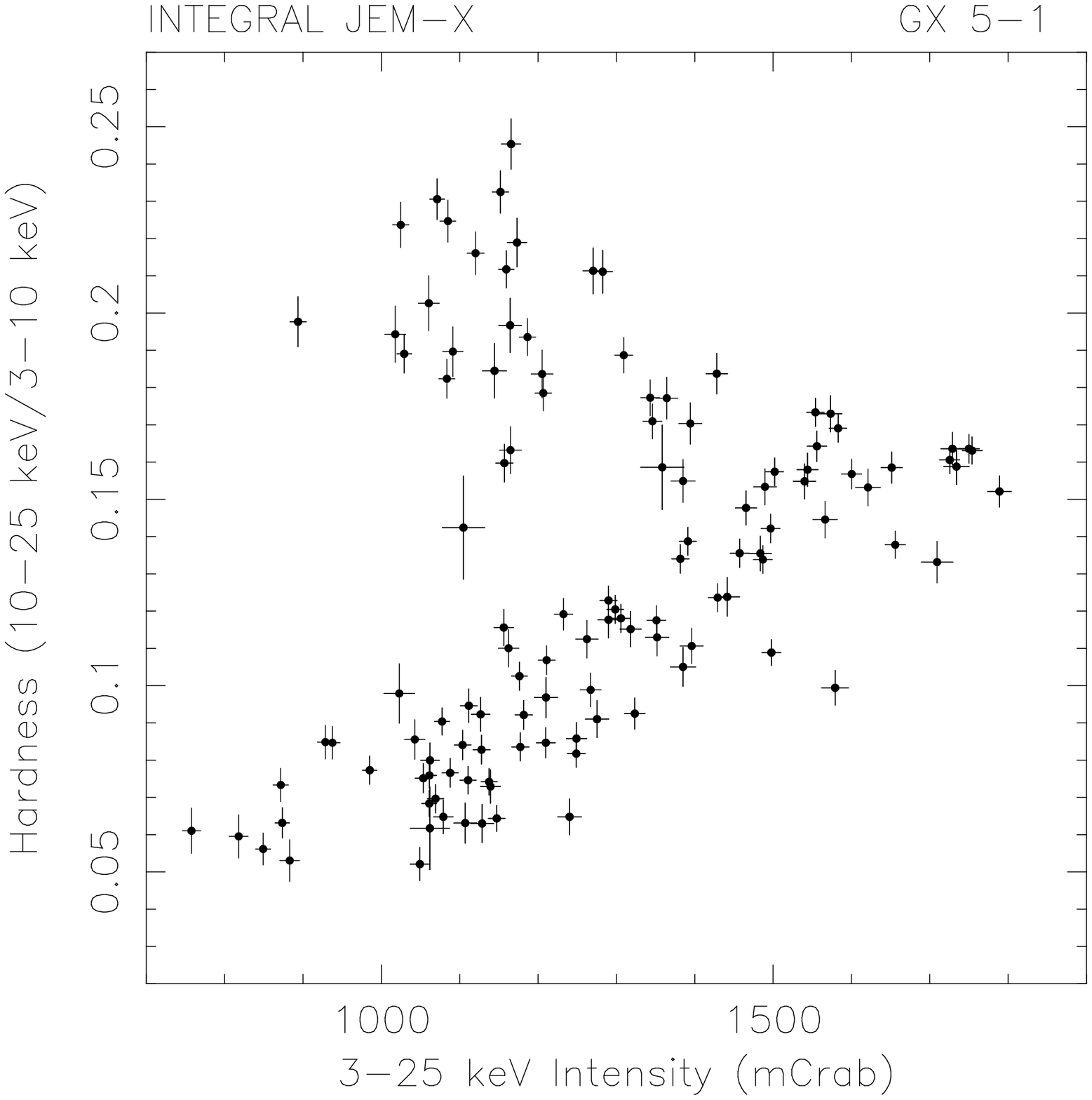}
  \includegraphics[height=.2\textheight,angle=-90]{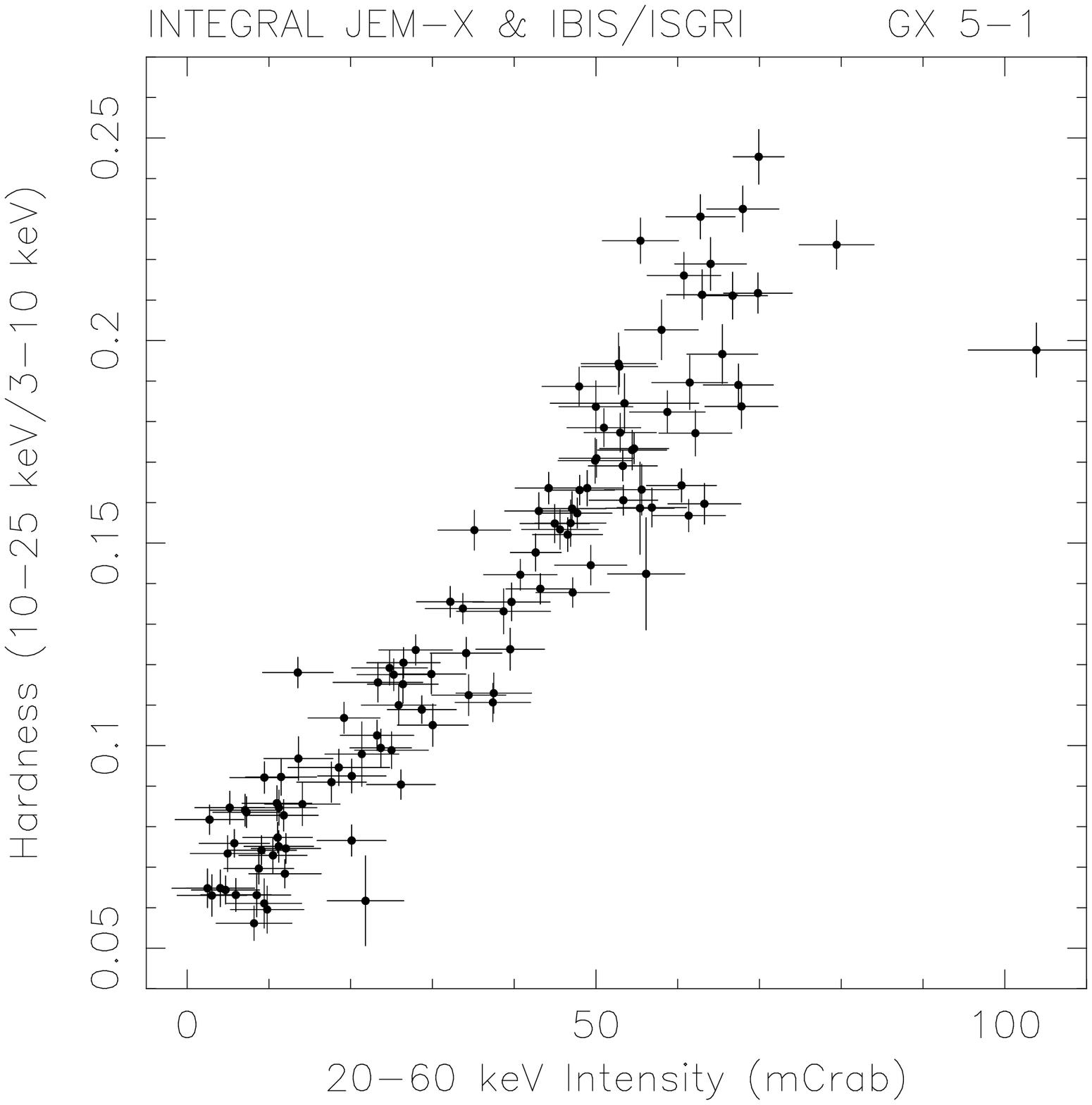}
  \caption{{\it Left:} 10--25\,keV and 20--60\,keV intensities 
versus 3--10\,keV intensity for GX\,5$-$1.
{\it Middle:} Hardness (ratio of the count rates in the 10--25\,keV 
and 3--10\,keV bands) versus intensity (3--25\,keV) for GX\,5$-$1.
{\it Right:} Hardness versus intensity (20--60\,keV) for GX\,5$-$1.}
\label{GX_5-1_jmx_isgri_cd}
\end{figure*}

GX\,5$-$1 is the brightest source seen in the JEM-X field-of-view of the 
Galactic bulge observations (see Fig.~\ref{fig3}). 
It is so bright ($\simeq$700--1600\,mCrab and
$\simeq$100--600\,mCrab, in the 3--10\,keV and 10--25\,keV bands, respectively;
Fig.~\ref{GX_5-1_isgri_jmx})
that it dominates the JEM-X detector count rates, which considerably 
influences the image reconstruction and therefore the quality of the observations.
GX\,5$-$1 shows a two-branch behaviour, both
in the 3--10\,keV versus 10--25\,keV and 3--10\,keV versus 20--60\,keV bands 
(Fig.~\ref{GX_5-1_jmx_isgri_cd}, left panel).
In one branch there is a correlation in the intensities between the 
lowest X-ray band and the higher X-ray bands. In the other branch, which
is connected to the former one at the highest intensities,
there is (almost) no correlation: whereas the 3--10\,keV intensity
varies, the 10--25\,keV and 20--60\,keV stay (almost) constant.
This two-branch behaviour is related to the so-called ``Z'' branches seen 
in this source and other Z-sources. GX\,5$-$1 is mostly seen in the 
so-called horizontal and normal branch (e.g., Kuulkers et al.\ 1994).
This is also reflected in the hardness versus intensity (flux) diagram
(HID; Fig.~\ref{GX_5-1_jmx_isgri_cd}, middle panel). 
The horizontal branch runs from top left to middle right, the
normal branch from middle right to bottom left. The HID is qualitatively similar
to that drawn from other (all-sky) monitoring observations
(Blom et al.\ 1993; van der Klis et al.\ 1991; Paizis et al.\ 2005).
There is no evidence for such branch behaviour in the hardness versus the 20--60\,keV
flux (Fig.~\ref{GX_5-1_jmx_isgri_cd}, right panel); in that case the 
hardness just increases linearly with increasing 20--60\,keV flux.

{\it GX\,349+2.}
Like GX\,5$-$1 and GX\,17+2, also GX\,349+2 is a Z-source (Hasinger \&\ van der Klis 1989).
It is highly variable between 15 and 60\,mCrab (20--60\,keV; Fig.~\ref{misc}).
Previous IBIS/ISGRI observations show similar flux variations on a $\simeq$10-day
time scale (22--40\,keV; Paizis et al.\ 2006).

{\it GX\,9+1.}
GX\,9+1 varies from revolution to revolution, up to $\simeq$20\,mCrab (20--60\,keV; 
Fig.~\ref{misc}). On a $\simeq$10-day time scale the source has been seen to weakly
vary between $\simeq$10 and $\simeq$20\,mCrab in the 22--40\,keV band,
using previous {\em INTEGRAL} observations (Paizis et al.\ 2006).

{\it 1E\,1743.1$-$2843.}
1E\,1743.1$-$2843 is a fairly persistent source in the Galactic Center region
with a 20--40\,keV flux of $\simeq$5\,mCrab; it only
shows marginal variability over a few months time scale 
(Del Santo et al.\ 2006). Our monitoring is consistent with this (Fig.~\ref{misc});
1E\,1743.1$-$2843 is not significantly detected in the average of
the first and third season, but it is during the second season (see Fig.~\ref{fig2}, Table~\ref{significance}).
The average flux over the second season is 4.4$\pm$0.2\,mCrab (20--60\,keV),
similar to that reported by Del Santo et al.\ (2006).
We note that the high data point near the end of February 2005 (MJD\,53428) is instrumental;
the source was not significantly detected during that particular revolution
(290).

\subsubsection{X-ray pulsars}

{\it OAO\,1657$-$415.}
The source is a high-mass X-ray binary with a pulse period of 38\,s and an orbital period of 10.4~days
(Chakrabarty et al.\ 1993). It is highly variable on monthly and longer time scales
as seen by {\em CGRO}/BATSE, reaching up to $\simeq$200\,mCrab 
(20--100\,keV; Harmon et al.\ 2004). Like 4U\,1700$-$377, the 
hard X-ray flux modulates with the orbital period (e.g., Laycock et al.\ 2003).
We see OAO\,1657$-$415 far off-axis from the Galactic Center ($\simeq$15.6$^{\circ}$); 
the 20--60\,keV and 60--150\,keV fluxes vary on a single pointing basis, and range 
between the IBIS/ISGRI detection limits and $\simeq$200\,mCrab (Fig.~\ref{misc}) 
and $\simeq$100\,mCrab, respectively.
During the first few weeks of the second season the source was
not very active, compared to the rest of the observations.

{\it IGR\,J17252$-$3616.}
IGR\,J17252$-$3616 was discovered in February 2004
(Walter et al.\ 2004). It has been found to show a pulse period of 414\,s and an 
orbital period of 9.72~days (Zurita Heras et al.\ 2006).
{\em INTEGRAL}/IBIS monitoring of IGR\,J17252$-$3616
indicates a mean 20--60\,keV flux of $\simeq$6.4\,mCrab; the source
was not detected in the 60--150\,keV band with
3$\sigma$ upper limits of typically $\simeq$7\,mCrab.
Every now and then IGR\,J17252$-$3616 flares on
$<$1~day time scales up to about 70\,mCrab
(Zurita Heras et al.\ 2006).
This is consistent with the fact that most of the time we
do not see the source, and our detection of a couple of flares, one of 
which occurred on March 26, 2005 (MJD\,53820), with fluxes up to 90\,mCrab
(20--60\,keV; Fig.~\ref{misc}).

{\it GX\,1+4.}
GX\,1+4 is a symbiotic binary composed of
a giant star and a neutron star (Chakrabarty \&\ Roche 1997, Belczy\'nski et al.\ 2000), 
with an orbital period of $\simeq$304~days (Pereira et al.\ 1999)
and a spin period of about 2\,min (e.g., Lewin et al.\ 1971).
{\em INTEGRAL} observations between March 2003 and October 2004
showed the source evolving from a weak intensity state in the beginning 
at about 20\,mCrab to a brighter intensity state at the end at about 120\,mCrab (20--40\,keV). 
The source showed strong variability by a factor of $\simeq$10 
on a few 1000\,s time scale on some occassions (Ferrigno et al.\ 2006). 
We detect the source at least up to 150\,keV (see also Ferrigno et al.\ 2006);
it clearly varies on weekly and longer time scales from 
$\simeq$5--85\,mCrab and $\simeq$5--50\,mCrab
(20--60\,keV and 60--150\,keV, respectively; see Figs.~\ref{misc} and \ref{misc2}). 
Note that the correlated variability between the 20--60\,keV and 60--150\,keV bands, present
during the first season, is absent during the first part of the second season.
{\em GRANAT}/SIGMA observations already showed similar variability 
at energies $\gtrsim$40\,keV at monthly time scales 
(Cordier et al.\ 1993; Mandrou et al.\ 1994; David et al.\ 1998) and
half-a-year time scales (Goldwurm et al.\ 1995).
{\em CGRO}/BATSE shows clearly the 20--100\,keV variability on 
time scales longer than a month with fluxes between below the 
detection limit to up to $\simeq$200\,mCrab (Harmon et al.\ 2004).
During the {\em MIT}/OSO-7 observations the 15--40\,keV flux varied
mostly within about 200 and 300\,mCrab on months time scales,
with one possible flaring period up to about 600\,mCrab
(Markert et al.\ 1979).

{\it 3A\,1822$-$371.}
3A\,1822$-$371 is a 5.57\,hr dipping and eclipsing accretion-disk corona source
(e.g., White et al.\ 1981), with a pulse period of 0.59\,s
(Jonker \&\ van der Klis 2001). It is a persistent source in the 20--60\,keV band,
and we see it varying on time scales of typically
a revolution to a couple of revolutions between $\simeq$12 and 35\,mCrab
(20--60\,keV; Fig.~\ref{misc}).
Previous {\em INTEGRAL}/IBIS and {\em BeppoSAX}/PDS observations 
showed that the 15--40\,keV flux is clearly modulated with the orbital period
(Williams et al.\ 2004).

\subsubsection{Miscellaneous sources}

{\it 4U\,1700$-$37.}
The high-mass X-ray binary 4U\,1700$-$377 has an orbital period of 3.41~days (Jones et al.\ 1973).
The nature of the compact object is still unknown
(Gottwald et al.\ 1986, Clark et al.\ 2002).
It shows the strongest flaring activity in our sample, with fluxes from near the 
IBIS/ISGRI detection limit to generally $\simeq$500\,mCrab (20--60\,keV) within
one observation. Flares with fluxes up to about 1500\,mCrab are seen in our program
(Fig.~\ref{misc}). In the 60--150\,keV band the source varies 
generally between the IBIS/ISGRI detection limits and $\simeq$200\,mCrab
(Fig.~\ref{misc2}), occasionally flaring reaching up to $\simeq$500\,mCrab
in single pointings.
This is similar to the hard X-ray behaviour seen in previous observations 
by {\em INTEGRAL}, {\em CGRO}/BATSE, {\em GRANAT}/SIGMA, as well
as older experiments; the lowest fluxes are reached during eclipse
(Orr et al.\ 2004, Laycock et al.\ 2003, Kudryavtsev et al.\ 2001, Rubin et al.\ 1996, 
Laurent et al.\ 1992, Pietsch et al.\ 1980, Markert et al.\ 1979).

{\it IGR\,J17098$-$3628.}
IGR\,J17098$-$3628 was discovered end of March 2005 (MJD\,53453) by {\em INTEGRAL}
with 18--45\,keV and 45--80\,keV fluxes of $\simeq$28 and $\simeq$39\,mCrab,
respectively (Grebenev et al.\ 2005a). Near the peak
the source spectrum changed significantly (Grebenev et al.\ 2005b).
Our observations caught the source near the peak and we saw it fading away 
(Mowlavi et al.\ 2005). 
On March 26 (MJD\,53455) the source showed considerable variability, up to 65\,mCrab
(20--60\,keV; Fig.~\ref{misc}); on that date a significant detection was also made in the 
60--150\,keV band, with an average flux of $\simeq$75\,mCrab 
(Fig.~\ref{misc2}; see also Mowlavi et al.\ 2005).

{\it IGR\,J17544$-$2619.}
On 17 September, 2003 {\em INTEGRAL} discovered IGR\,J17544$-$2619 
(Sunyaev et al.\ 2003b).
It flared on time scales of hours with fluxes up to $\simeq$160\,mCrab, 60\,mCrab and 
$\lesssim$15\,mCrab in the 18--25, 25--50 and 50--100\,keV IBIS/ISGRI bands (Sunyaev et al.\ 2003b;
Grebenev et al.\ 2003; see also Walter et al.\ 2006). 
Earlier activity was seen by IBIS/ISGRI in April 2003;
it was found flaring again near the end of February 2004 (Walter et al.\ 2006), 
as well as on March 8, 2004 (Grebenev et al.\ 2004a).
On the latter date the source reached 17--45\,keV peak fluxes of $\simeq$160\,mCrab
(Grebenev et al.\ 2004a). Walter et al.\ (2006) reported flux increases 
from the detection limits up to $\simeq$1000\,mCrab (15--30\,keV) within 5\,min
during the periods of activity in 2003 and 2004.
Our monitoring shows that it flared again 
on October 15, 2005 (MJD\,53658), up to $\simeq$60\,mCrab (20--60\,keV; Fig.~\ref{misc}).
Walter et al.\ (2006) suggested a period of 165$\pm$5~days between activity
(based on three flaring periods);
however, our observations do not support this.

{\it XTE\,J1818$-$245.}
On August 12, 2005 (MJD\,53594) a new source, XTE\,J1818$-$245, was reported (Levine et al.\ 2005).
This was just before the start of the second season, and 
the first observations showed the source to be bright, $\simeq$30\,mCrab 
(20--60\,keV; Shaw et al.\ 2005).
The transient faded more or less monitonically and went below
the detection limits within a month (Fig.~\ref{misc}).
During the first few monitoring observations of the same period the source was seen
at 60--150\,keV as well, with fluxes between $\simeq$30--40\,mCrab.

{\it XTE\,J1743$-$363.}
This source was first detected in outburst by the {\it RXTE}/PCA in 1999
(Markwardt et al.\ 1999). 
It was marginally (1.7$\pm$0.2\,mCrab) detected by IBIS/ISGRI in September 2003
(Revnivtsev et al.\ 2004a), but it became active again in September
2004, with a mean flux of $\simeq$10\,mCrab (18--45\,keV; Grebenev \&\ Sunyaev 2004a).
We see clear activity up to $\simeq$18\,mCrab (20--60\,keV) in the period 
March--April 2005 (MJD\,53435--53479), during the first season (Fig.~\ref{misc}).

\subsection{April 2006: a quiet Galactic Center region}
\label{off}

\begin{figure*}
  \includegraphics[height=.207\textheight]{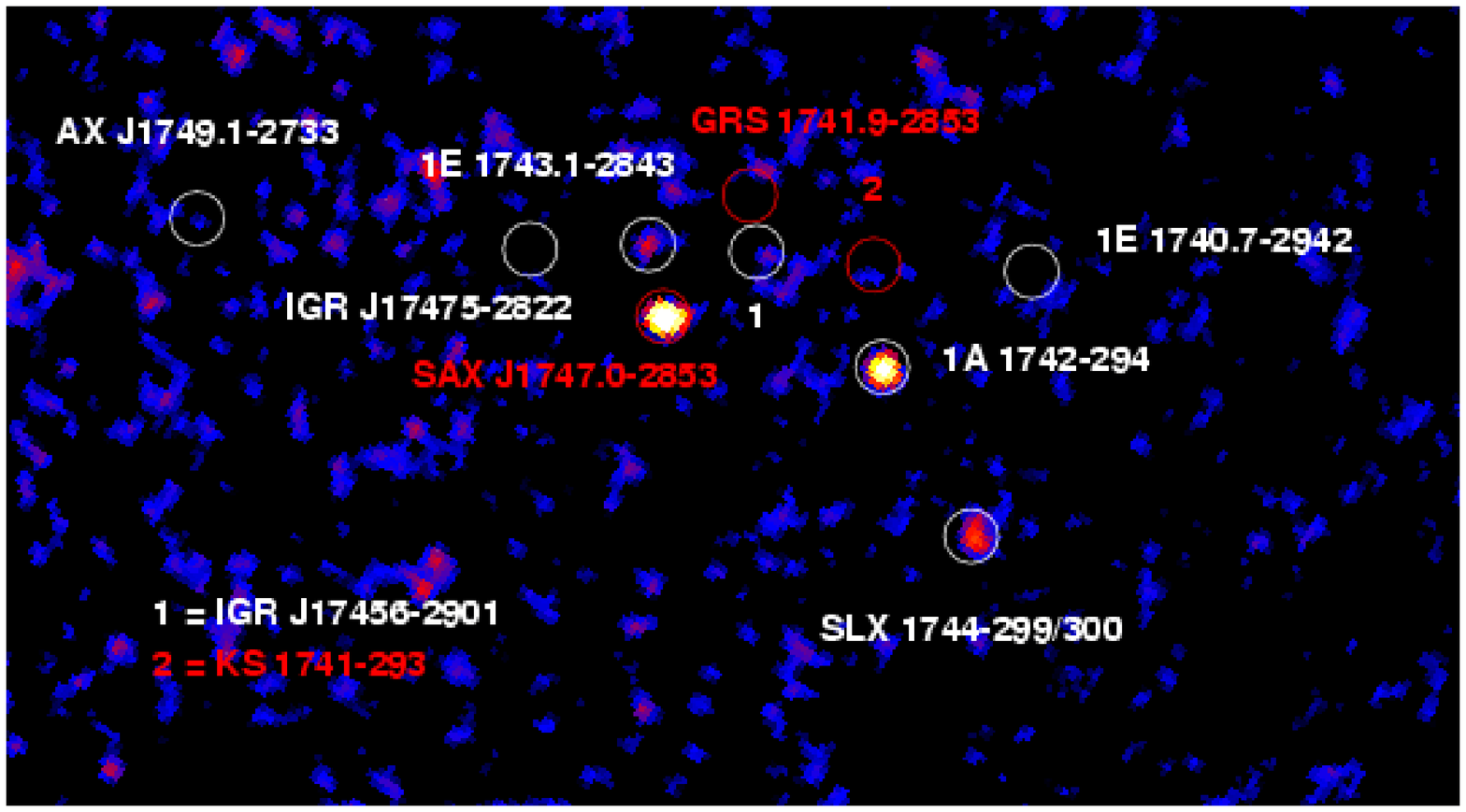}
  \includegraphics[height=.207\textheight]{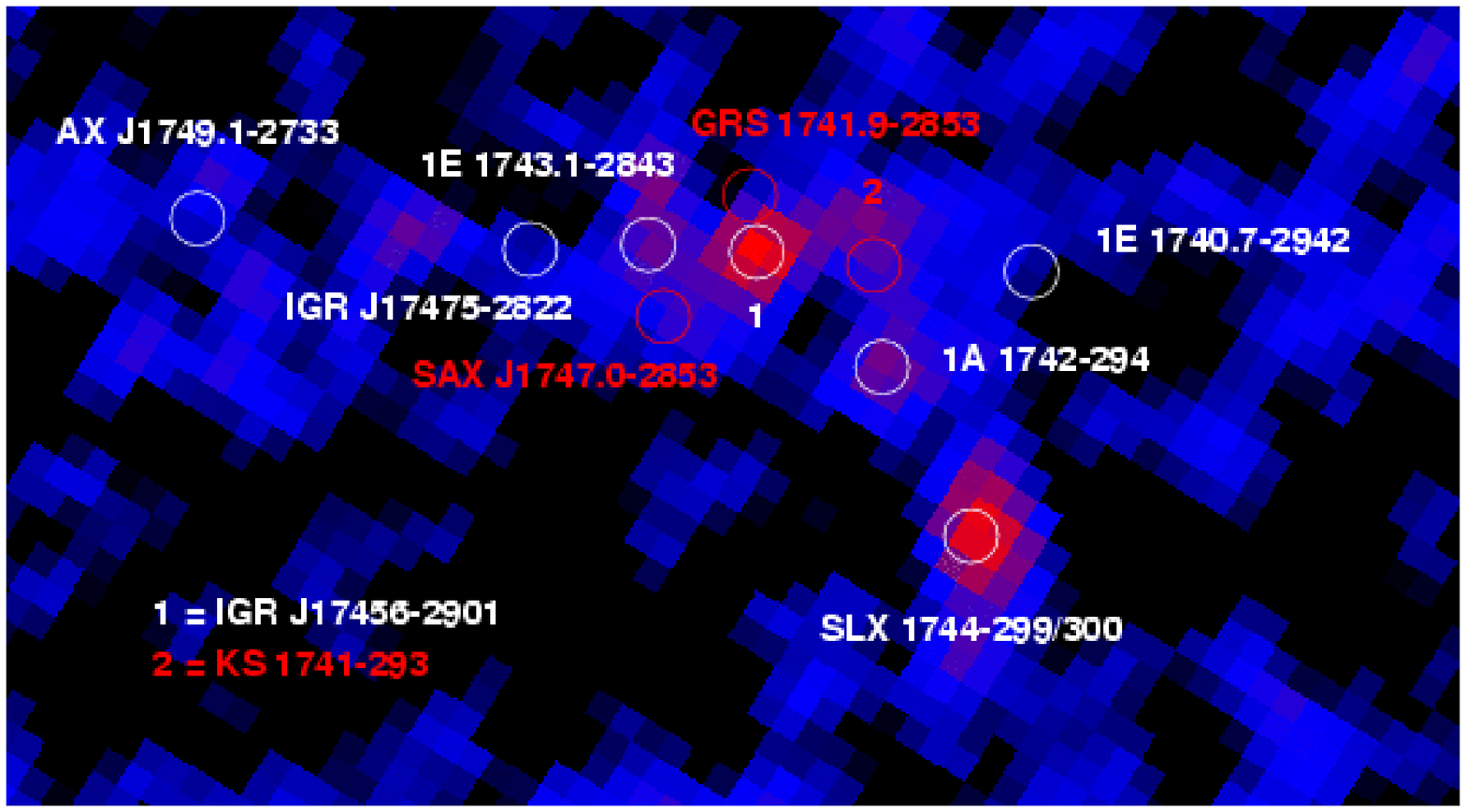}
  \caption{{\em INTEGRAL} JEM-X (3--20\,keV) ({\it left}) and 
IBIS/ISGRI (20--60\,keV) ({\it right}) mosaic significance images of
the Galactic Center region during revolutions 424--429 (April 3--21, 2006; MJD\,53828--53846),
for a total exposure of 69\,ks. Note that the same scales are used
as in Figs.~1--4. The annoted sources are those similar to Fig.~\ref{fig2}.
Clearly, most of the sources are ``off'' during that time.
With JEM-X only SAX\,J1747.0$-$2853, 1A\,1742$-$294 and SLX\,1744$-$299/300 are significantly
detected, whereas with IBIS/ISGRI only SLX\,1744$-$299/300 and a source positionally 
coincident with IGR\,J17456$-$2901 are significantly detected.
We refer to Sect.~\ref{off} for a more detailed discussion.
}
\label{424-429}
\end{figure*}

It was already noted (Sect.~\ref{1E1740}) that 1E\,1740.7$-$2942 was not detected during
the third monitoring season. A close inspection of the 20--60\,keV light curve 
reveals that most of the other bright and/or transient X-ray sources in the Galactic Center region
(1A\,1742$-$294, SLX\,1744$-$299/300, KS\,1741$-$293, GRS\,1741.9$-$2853, SAX\,J1747.0$-$2853)
had faded during our monitoring observations performed on April 3--21, 2006. 
These observations clearly show the dynamic range in source activity present in the
Galactic Center region.

In Fig.~\ref{424-429} we show the average mosaic image of the region in the 
3--20\,keV and 20--60\,keV bands. Clearly, 1A\,1742$-$294 and SAX\,J1747.0$-$2853
were in a soft X-ray state during that time, whereas the transients 
KS\,1741$-$293 and GRS\,1741.9$-$2853 were off. 
A source in a hard X-ray state is seen 
($\lesssim$6\,mCrab, 3--20\,keV;
9$\sigma$, 4.6$\pm$0.5\,mCrab, 20--60\,keV), 
positionally coincident in our source sample with IGR\,J17456$-$2901. 
We derive a position of this source of (J2000.0) RA, Dec = 266.410$^{\circ}$, 
$-$29.029$^{\circ}$, with a 90\%\ confidence error of 2.9'.
Within the quoted error of this Galactic Center source various (transient) X-ray binaries are known: A1742$-$289
(Eyles et al.\ 1975), AX\,J1745.6$-$2901 (Sakano et al.\ 2002), 
CXOGC\,J174538.0$-$290022, CXOGC\,J174541.0$-$290014, CXOGC\,J174535.5$-$290124 (Muno et al.\ 2003a),
CXOGC\,J174540.0$-$290005 (Muno et al.\ 2005a),
CXOGC\,J174540.0$-$290031 (Muno et al.\ 2005b),
SWIFT\,J174535.5$-$290135.6 (Kennea et al.\ 2006a)
and SWIFT\,J17454.0$-$290005.3 (Kennea et al.\ 2006b, 2006c).
We have no (soft X-ray) information available of which one of these sources was active
during our April 2006 observations, and thus can not securely identify the 
origin of the hard X-ray emission with one of them.
On the other hand, the 20--60\,keV flux of our source is similar
to that seen from IGR\,J17456$-$2901, which has been constant over 
the two years before our monitoring observations 
(B\'elanger et al.\ 2006). Note that none of the above mentioned 
X-ray binaries and X-ray transients fall within the error circle of 
IGR\,J17456$-$2901 (see B\'elanger et al.\ 2006). 
We, therefore, tentatively identify our source with the steady source
IGR\,J17456$-$2901, which we could detect significantly using a shorter
exposure time, since no other closeby (bright) sources could contaminate
the observations.

\subsection{New {\em INTEGRAL} sources}
\label{IGR}

Apart from the sources described in the previous Sections, during the course of the third season
we found several new X-ray sources (see Table~\ref{new}). Also, a re-analysis of the whole Galactic bulge monitoring data 
set is in progress, and in the course of this exercise we also find new X-ray sources.
Since these sources were not part (yet) of our catalog of sources, they are not included in the 
discussion on the images (Sect.~\ref{images}) and long-term light curves
(Sect.~\ref{light_curves}). Here we only shortly summarize their properties
as described in various ATels, or report on first results from the re-analysis on
the new {\em INTEGRAL} sources.

\begin{table}
\caption[]{New {\em INTEGRAL} sources detected through our monitoring program.
We give the source position (J2000.0) with its 90\%\ confidence error,
whether the source was detected by JEM-X (J) and/or IBIS/ISGRI (I) or not, and the
reference in which the detection was first noted. References: [1] Kuulkers et al.\ (2006c),
[2] this paper, [3] Turler et al.\ (2006), [4] Chenevez et al.\ (2006b).}
\begin{tabular}{l|ccccc}
\hline
 & RA & Dec & err & J/I? & Reference \\
\multicolumn{1}{l|}{Source} & \multicolumn{2}{c}{(degrees)} &  \multicolumn{3}{c}{} \\
\hline
IGR\,J17354$-$3255 & 263.854 & $-$32.922 & 4' & no$^{a}$/yes & [1] \\
IGR\,J17454$-$2703 & 266.329 & $-$27.039 & 1' & yes/no & [2] \\
IGR\,J17453$-$2853 & 266.328 & $-$28.891 & 2' & no/yes & [2] \\
IGR\,J17536$-$2339 & 268.409 & $-$23.654 & 4' & no$^{b}$/yes & [3] \\
IGR\,J17541$-$2252 & 268.517 & $-$22.871 & 4' & no$^{b}$/yes & [3] \\
IGR\,J17456$-$2901b & 266.412 & $-$29.029 & 2' & yes/yes & [4] \\
\hline
\multicolumn{6}{l}{\footnotesize $^{a}$\,JEM-X instrument was off.} \\
\multicolumn{6}{l}{\footnotesize $^{b}$\,Not in the JEM-X field of view.} \\
\end{tabular}
\label{new}
\end{table}

\begin{figure}
\centering
  \includegraphics[height=.3\textheight,angle=-90]{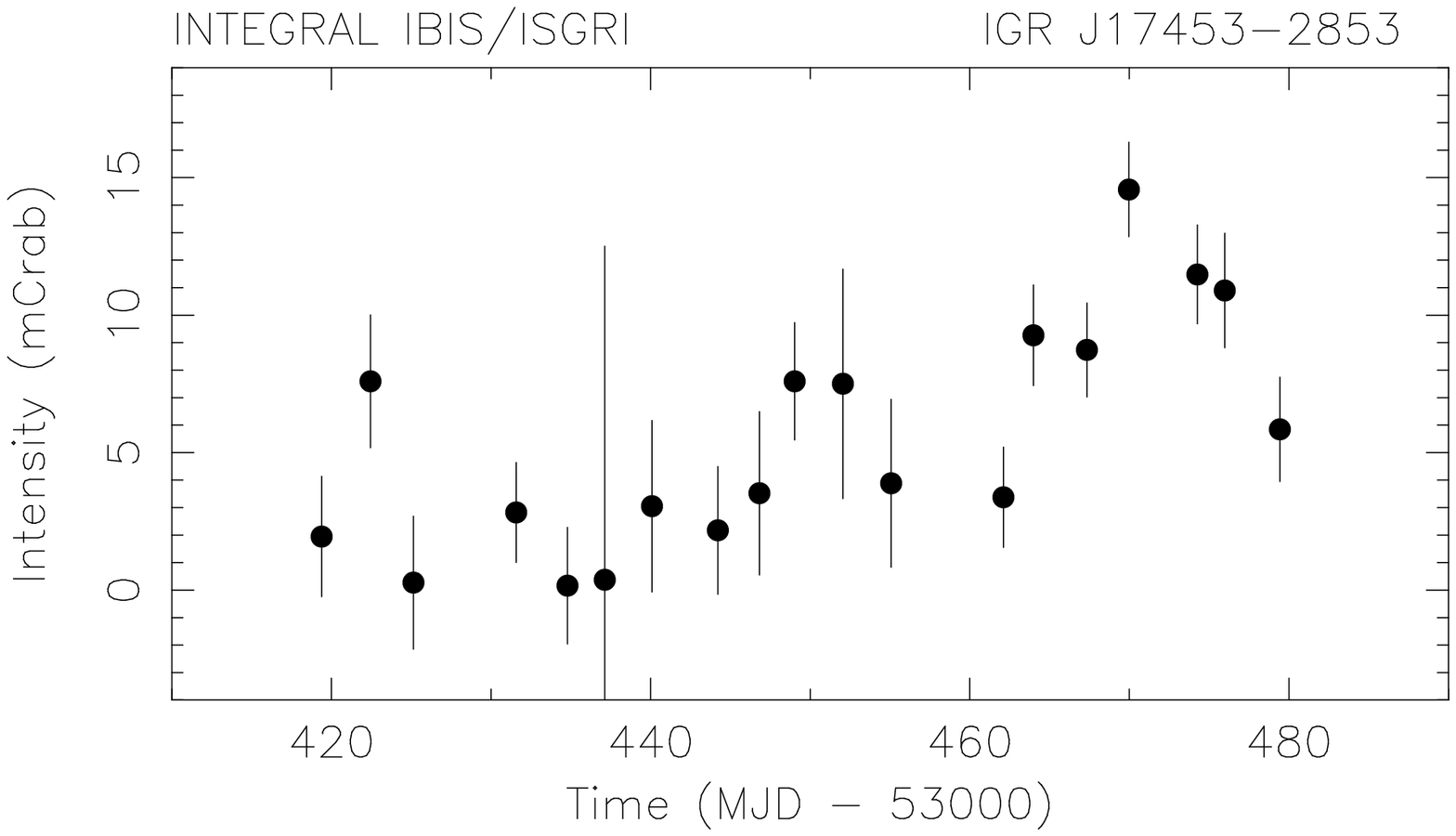}
  \includegraphics[height=.3\textheight,angle=-90]{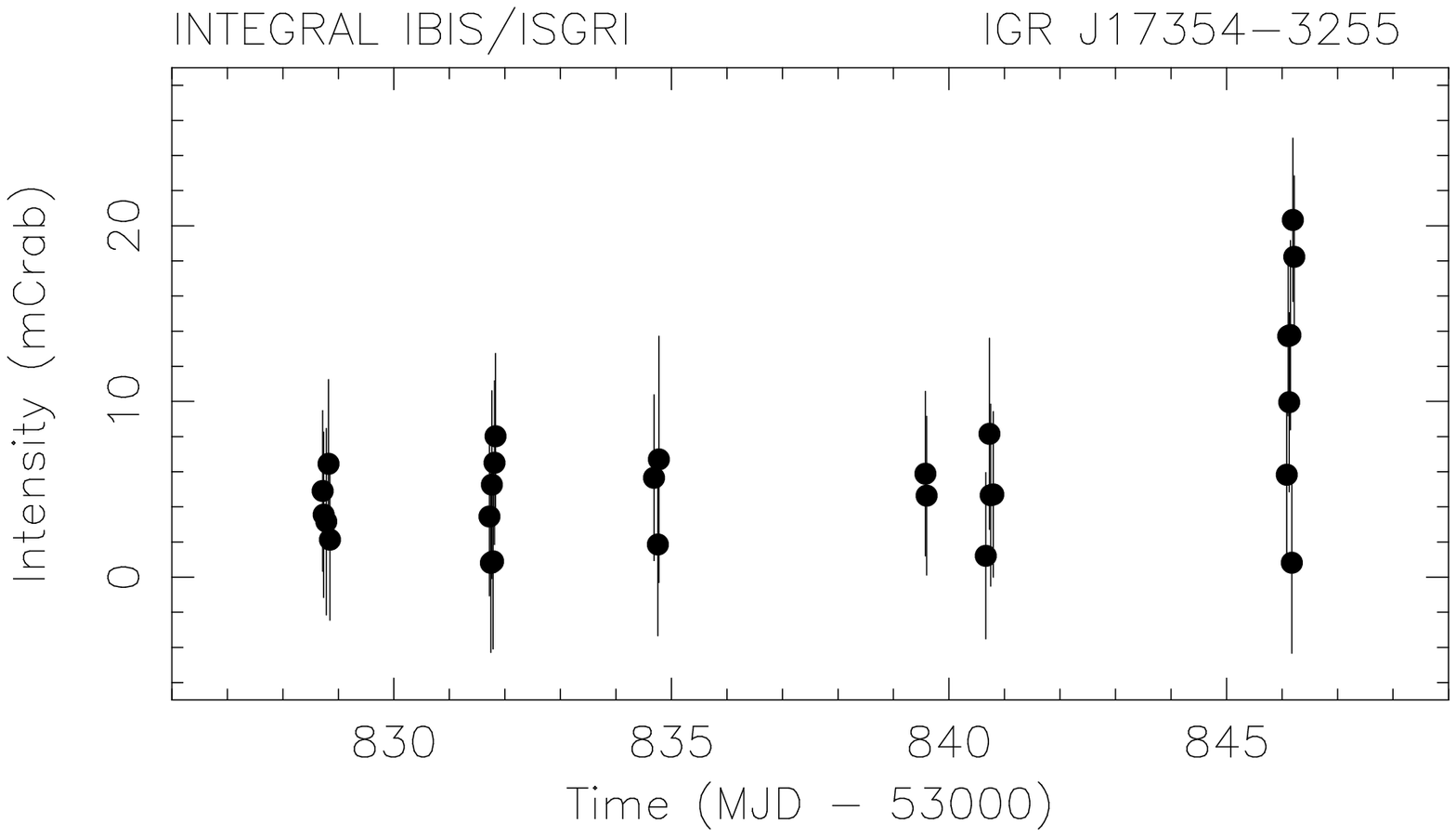}
  \caption{{\em INTEGRAL} IBIS/ISGRI 20--60\,keV light curves of IGR\,J17453$-$2853 ({\it Top}) and 
IGR\,J17354$-$3255 ({\it Bottom}) around and during the time of their detection.}
\label{IGR_J17453-2853}
\end{figure}

{\it IGR\,J17453$-$2853.} 
In the observations performed on April 3, 2005 (MJD\,53463) a new source is detected by IBIS/ISGRI
at the $\simeq$6$\sigma$ level with a 20--60\,keV flux of $\simeq$7\,mCrab. 
It was seen on average at the same level in the following observations, up to April 14, 2005
(MJD\,53474; see Fig.~\ref{IGR_J17453-2853}). 
The best-fit position is given in Table~\ref{new}.
We designate this source IGR\,J17453$-$2853. 
Despite being at a favourable off-axis angle from the Galactic Center ($\simeq$2$^{\circ}$), IGR\,J17453$-$2853 is not seen 
with JEM-X ($\lesssim$6\,mCrab and $\lesssim$7.5\,mCrab, for 3--10\,keV and 10--25\,keV, respectively), 
potentially indicating high intrinsic absorption or 
that it underwent a hard X-ray outburst. The source is rapidly variable, 
with the 20--60\,keV flux changing by a factor of $\simeq$2 on a time scale of about half an hour.
It reached a peak flux 27$\pm$4\,mCrab on April 10, 2005 (MJD\,53470; Fig.~\ref{IGR_J17453-2853}).
A preliminary spectral analysis shows that the source spectrum can be described by
a power-law with index of $\simeq$2.2, indeed indicating a hard spectrum.
We note that if the non-detection with JEM-X is due to absorption in the line of sight,
N$_{\rm H}$ should be larger than about 10$^{23}$\,atoms\,cm$^{-2}$.

\begin{figure}
\centering
  \includegraphics[height=.3\textheight,angle=-90]{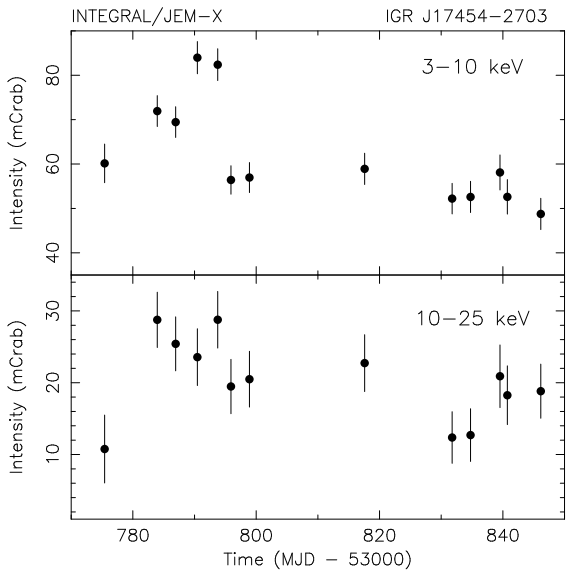}
  \caption{{\em INTEGRAL} JEM-X 3--10\,keV ({\it Top}) and 10--25\,keV ({\it Bottom}) light curve of IGR\,J17454$-$2703 during the third visibility season.
Data points are from single pointings whenever the source was within the JEM-X field of view.}
\label{lc_IGR_J17454-2703}
\end{figure}

{\it IGR\,J17454$-$2703.}
Combining our monitoring observations performed between February 17 and March 4, 2006 (MJD\,53783--53798),
a new source is found with JEM-X. The position is given in Table~\ref{new}.
We designate the source IGR\,J17454$-$2703. 
No previously known X-ray source is found within 5' of this position in the Simbad database.
The light curve of this new X-ray transient shows that it was at its maximum between 
February 24 and 27 (MJD\,53790--53793; Fig.~\ref{lc_IGR_J17454-2703}). In one of the single pointings on February 27
the source reached a peak flux of $\simeq$82\,mCrab (3--10\,keV) and $\simeq$29\,mCrab (10--25\,keV).
We note that the 3--20\,keV JEM-X spectrum can be modeled by an absorbed power law with an 
absorption column, $N_{\rm H}$ of $\simeq$1.2$\times$10$^{23}$\,atoms\,cm$^{-2}$ and 
photon index of $\simeq$3.5. 
We do not detect the source with IBIS/ISGRI, either by combining the observations between February 17 and March 4
($\lesssim$2\,mCrab, 20--60\,keV), or
in single hexagonal dither observations ($\lesssim$4\,mCrab, 20--60\,keV), consistent with the steep soft X-ray spectrum.

{\it IGR\,J17456$-$2901b.}
A new X-ray transient, SWIFT\,J174535.5$-$290135.6, 
was detected on February 25, 2006 (MJD\,53791). It was not seen the previous day, and showed 
a power-law-like decay in flux the days after (Kennea et al.\ 2006a).
We (Chenevez et al.\ 2006b) reported on a source coincident with this new {\em Swift} transient, which was
seen about a week earlier with JEM-X and IBIS/ISGRI with fluxes of about 9\,mCrab 
(6--10\,keV) and about 6\,mCrab (20--60\,keV), and, therefore, 
associated with SWIFT\,J174535.5$-$290135.6. 
However, the {\em Swift} transient was not seen on February 24, and
more X-ray (transient) sources are known (see Sect.~\ref{off}) within the JEM-X error circle
(Table~\ref{new}). Most of our detected hard X-ray emission may come from IGR\,J17456$-$2901, but the transient soft X-ray 
emission does not (see Sect.~\ref{off}; B\'elanger et al.\ 2006). We 
conclude that at least our detected soft X-ray emission is due to a transient source, which we designate IGR\,J17456$-$2901b,
but we can not attribute with confidence this transient emission to either SWIFT\,J174535.5$-$290135.6
or any other known source within the error circle.

{\it IGR\,J17536$-$2339, IGR\,J17541$-$2252.}
In the beginning of April 2006 (MJD\,53828) two new faint closeby ($\simeq$0.8$^{\circ}$) 
sources were detected by IBIS/ISGRI (see Table~\ref{new} for the position): IGR\,J17536$-$2339 and
IGR\,J17541$-$2252. They had fluxes of $\simeq$11 and $\simeq$10\,mCrab, respectively, in the 
20--60\,keV band. Within the positional errors IGR\,J17536$-$2339 is
coincident with the nearby X-ray burster SAX\,J1753.5$-$2349 (Turler et al.\ 2006).

{\it IGR\,J17354$-$3255.}
Another new source (see Table~\ref{new} for the position), IGR\,J17354$-$3255, showed up in the last single pointings of our AO-3 
program, i.e., those performed on April 21, 2006 (MJD\,53846; Kuulkers et al.\ 2006c). It peaked to a 20--60\,keV flux of $\simeq$20\,mCrab
(Fig.~\ref{IGR_J17453-2853}). 

\subsection{OMC results}
\label{omc}

Many of the sources we study do not have optical/IR counterparts 
or their optical counterparts are fainter than the OMC detection limit
(mainly due to interstellar absorption).
Since the Galactic bulge region is crowded, when observing a 
high-energy source which has a relatively large positional error circle,
one may observe various possible counterparts, or the right counterpart may not
even be visible. This is the case for most of our targets.
However, whenever a source flares up in brightness, with concurrent brightening 
in hard X-rays/$\gamma$-rays, we may be able to detect them with our
analysis method.

\begin{figure}
\centering
  \includegraphics[height=.3\textheight,angle=-90]{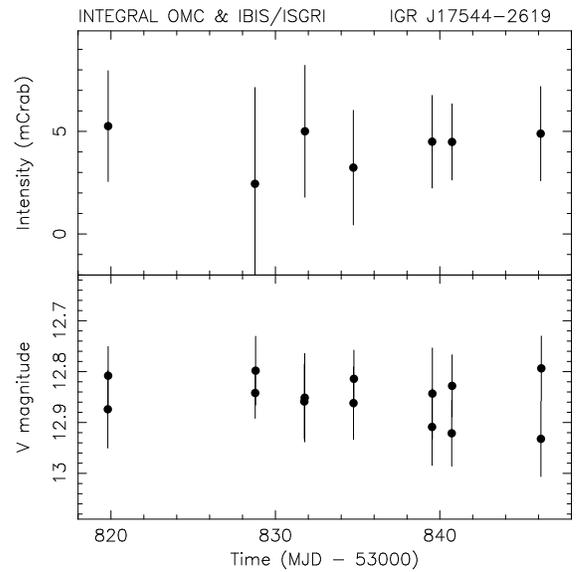}
  \caption{{\it Top}: IBIS/ISGRI 20--60\,keV light curve; data points are averages per hexagonal dither observation.
{\it Bottom}: OMC light curve of the optical counterpart of IGR\,J17544$-$2619, i.e.,
2MASS\,J17542527$-$2619526 (= IOMC\,6849000050).}
\label{OMC_lightcurve}
\end{figure}

One of the few examples, however, of a relatively clear field with an optical/IR counterpart is
that of IGR\,J17544$-$2619. The
OMC light curve of the counterpart (2MASS\,J17542527$-$2619526;
Rodriguez 2003, Pellizza et al.\ 2006) is given in Fig.~\ref{OMC_lightcurve}. 
The source does not vary much on a daily time scale,
and is consistent with being constant near V$\simeq$12.9~mag, close to that seen
previously (Pellizza et al.\ 2006). As shown by the simultaneous 20--60\,keV IBIS/ISGRI
light curve in the same plot, the hard X-ray source was not active at that time.

\section{Summary and conclusions}
\label{discussions}

In this paper we have shown the results of the first one and a half year
of monitoring of sources in and around the Galactic bulge region 
at soft (3--25\,keV) and hard (20--150\,keV) X-ray energies, with a focus 
on the short, medium and long-term source variability within this period.
For the first time we have performed a high-sensitivity $>$20\,keV study of variability at
hourly to weekly to monthly time scales, especially for the sources in the 
Galactic Center region, which has not been possible before due to
source confusion of previous hard X-ray instruments. 
Our program has been succesfull in monitoring the hard X-ray variability of 
bright persistent sources, as well as characterising the (long-term) properties of
old and new X-ray transients. Moreover, due to the frequent monitoring 
we are able to detect fast X-ray transients, i.e., those which are
typically active for less than a day. We also found six new {\em INTEGRAL} sources:
IGR\,J17354$-$3255, IGR\,J17453$-$2853, IGR\,J17454$-$2703, IGR\,J17456$-$2901b, IGR\,J17536$-$2339, and
IGR\,J17541$-$2252. 

The fact that {\em INTEGRAL} has various instruments aboard which are
sensitive in the soft and hard X-ray region, allows us to {\it simultaneously} monitor 
these energy ranges for (anti-) correlations, as well as fast phenomena such as X-ray bursts.
Moreover, due to the wide field of view of the instruments we are able to
monitor a large sample of X-ray sources all at the same time. It thus allows
a homogeneous analysis of all the sources, similar to what was possible 
with the {\it BeppoSAX}/WFC at soft X-rays (see, e.g., in 't Zand 2001),
and, to a lesser extent, with {\em GRANAT}/SIGMA at hard X-rays (see, e.g., Mandrou et al.\ 1994).

In the Galactic bulge and its surroundings we see about 15 persistent neutron-star low-mass X-ray binaries, 
10 of which are X-ray bursters, active over the whole monitoring period
at energies 20--60\,keV. Per season we detect 22/23 sources above a detection significance of 7.
Of the black-hole (candidate) binaries in our sample we see the two persistent sources for 
most of the time, 1E\,1740.7$-$2942 and 
GRS\,1758$-$258. The former one was seen to turn off for a period of at least about three months. 
So far, we have seen one bright (exceeding 100\,mCrab, 20--60\,keV) black-hole (candidate) 
X-ray transient (such as GRO\,J1655$-$40) in outburst per season, for a period of months. 
We see on average about three transient X-ray bursters active per season
for periods ranging between weeks (e.g., GRS\,1741.9$-$2853) to months 
(e.g., XTE\,J1739$-$285) with peak fluxes of about 25\,mCrab (20--60\,keV). The average
number of fast X-ray transients (i.e., visible for only a couple of hours
up to a day) per season is about one; they peak
to about 100\,mCrab (20--60\,keV; e.g., IGR\,J17544$-$2619). 
Most of the sources we detect on a hours to daily
time scales are (transient) low-mass or high-mass X-ray binaries.
Those detected only using averages per season, i.e., the weaker sources with fluxes
typically around 5\,mCrab (20--60\,keV), also include cataclysmic variables and AGN.

Whereas, for example, GX\,3+1 (X-ray burster) does not vary significantly, 
sources like GX\,1+4 (symbiotic neutron-star binary)
and 1A\,1742$-$294 (X-ray burster) vary
on monthly times scales, while GRS\,1758$-$258 (black-hole candidate) 
and GS\,1826$-$24 (X-ray burster) generally vary smoothly
on even longer time scales. As noted above, some sources clearly show transient
behaviour, i.e., they show outbursts with durations exceeding
months (e.g., H1743$-$322, a black-hole candidate), weeks
(e.g., MXB\,1730$-$335, an X-ray burster) or
flaring on time scales of hours to days (e.g., IGR\,J17252$-$3616,
an X-ray pulsar). For the X-ray transient XTE\,J1739$-$285
we secured its identification to host a neutron star
through our monitoring program (Brandt et al.\ 2005).
Some sources vary on all time scales accessible through our program,
as displayed by, e.g., the high-mass X-ray binary 4U\,1700$-$377.
Hard X-ray flares/outbursts in X-ray burst sources (e.g., GX\,354$-$0, H1820$-$30) 
are accompanied by soft X-ray drops, i.e., the spectra pivot somewhere between 10--20\,keV
when the sources become harder (see also Tarana et al.\ 2006a). 
The reverse is also seen (e.g., 4U\,1722$-$30). 
GX\,5$-$1, a bright non-bursting low-mass X-ray binary, on the other hand, shows no such 
anti-correlation. The soft X-ray emission is either correlated 
with the hard X-ray emission, or not, depending on the position
in the `Z'-diagram.

The temporal behaviour of hard X-ray emission from X-ray bursters
has led to a division into three classes (see, e.g., Tavani \&\ Barret 1997):
1) X-ray bursters with persistent (though variable on a monthly time scale) hard X-ray emission
(such as GS\,1826$-$24),
2) X-ray bursters showing episodic (for $\sim$10--20~days) hard X-ray emission
(such as GX\,354$-$0), and
3) X-ray bursters which are transient (such as Aql\,X-1). The behaviour we describe above
confirms this division, except that some of the persistent X-ray bursters
may show temporary (for $\sim$10--20~days) drops in their hard X-ray emission
(e.g., 4U\,1722$-$30).
The sparse examples of long-term simultaneous soft and hard X-ray observations of
X-ray bursters (in the pre-{\em INTEGRAL} literature) indicate that
hard X-rays are generally anti-correlated with soft X-rays
(e.g., 4U\,0614+091, Ford et al.\ 1996).
However, we show that sometimes no (anti-)correlation is seen between
soft and hard X-ray emission in X-ray bursters, i.e., the soft and hard X-ray can 
{\it both} be low.

At higher energies, i.e., 60--150\,keV, we detect about 8/9 sources per season
above a detection significance of 7,
and about a dozen in the data covering the whole monitoring period.
These are either transient or persistent black-hole candidates, X-ray bursters or high-mass X-ray binaries.
This is in accordance with the hard X-ray ($>$100\,keV) {\em INTEGRAL}/IBIS survey 
(which includes the Galactic bulge region) by Bazzano et al.\ (2006), who concluded that 
the 100--150\,keV band is dominated by low and high-mass X-ray binaries.

The next step will be, apart from the continuation of the Galactic bulge monitoring 
in the future, the investigation of (hard X-ray) variability at smaller
time scales (milliseconds to minutes) and the relation to long-term
variability, such as time-resolved pulse timing and detailed 
X-ray burst analysis. We also plan to study the long-term energy spectral behaviour 
of our sources, as well as a comparison with other ongoing monitoring programs (e.g., with
the {\em RXTE}/ASM and PCA). 

In this introductory paper we have shown that most of the hard X-ray sources in the field of view of the 
{\em INTEGRAL} instruments included in our program clearly vary
on time scales of a few hours to days to months.
It is therefore no surprise that the Galactic bulge is a region to stay tuned to. 


\begin{acknowledgements}
Based on observations with {\em INTEGRAL}, an ESA project with instruments and science data centre funded
by ESA member states (especially the PI countries: Denmark, France, Germany, Italy, Switzerland, Spain),
Czech Republic and Poland, and with the participation of Russia and the USA.
This research has made use of the SIMBAD database, operated at CDS, Strasbourg, France.
EK thanks Angela Bazzano and Guillaume B\'elanger for comments on an earlier version of the paper.
\end{acknowledgements}

\appendix

\section{The Crab as the reference source}
\label{crab}

\begin{figure*}
  \includegraphics[height=.3\textheight,angle=-90]{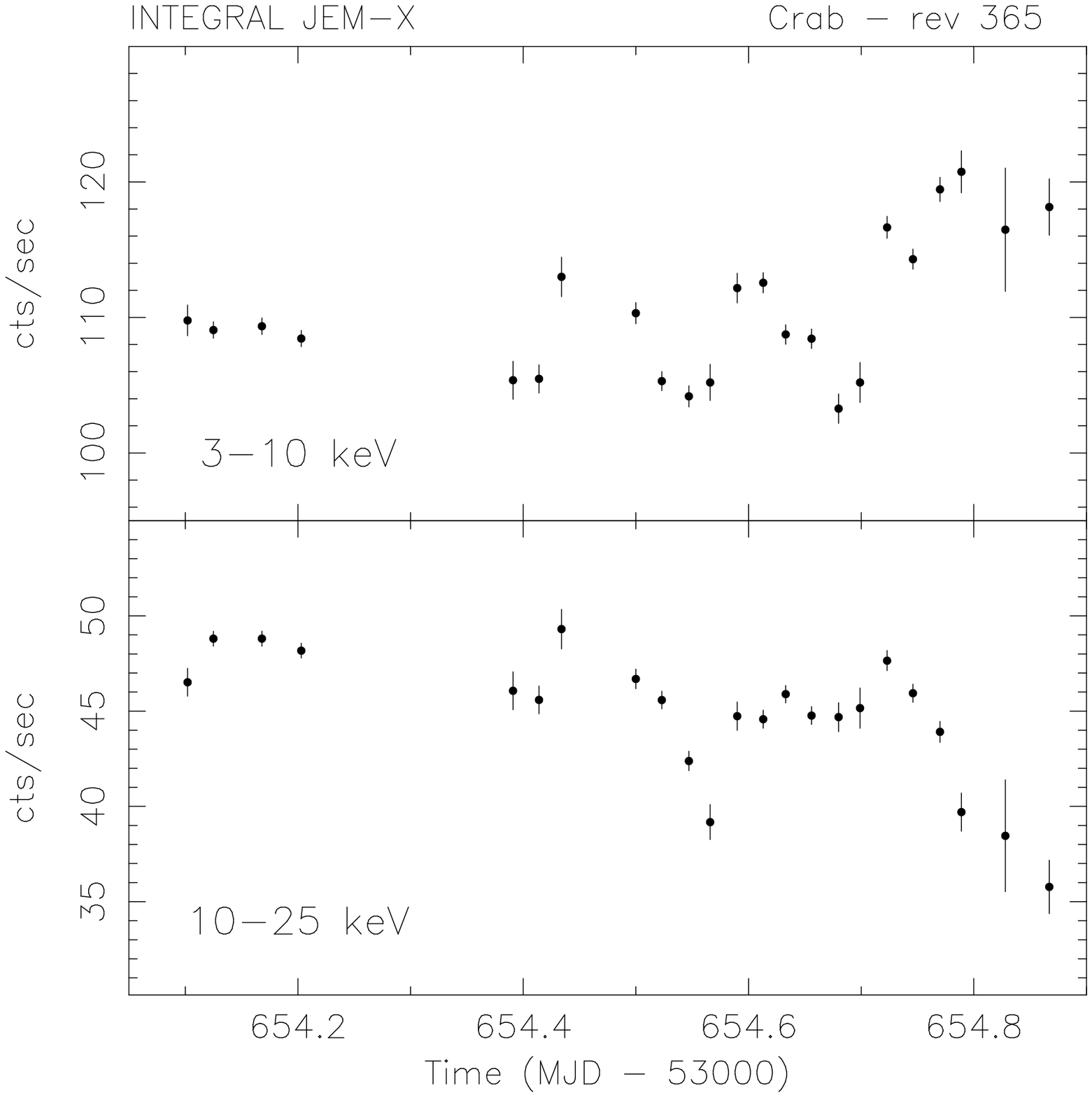}
  \includegraphics[height=.3\textheight,angle=-90]{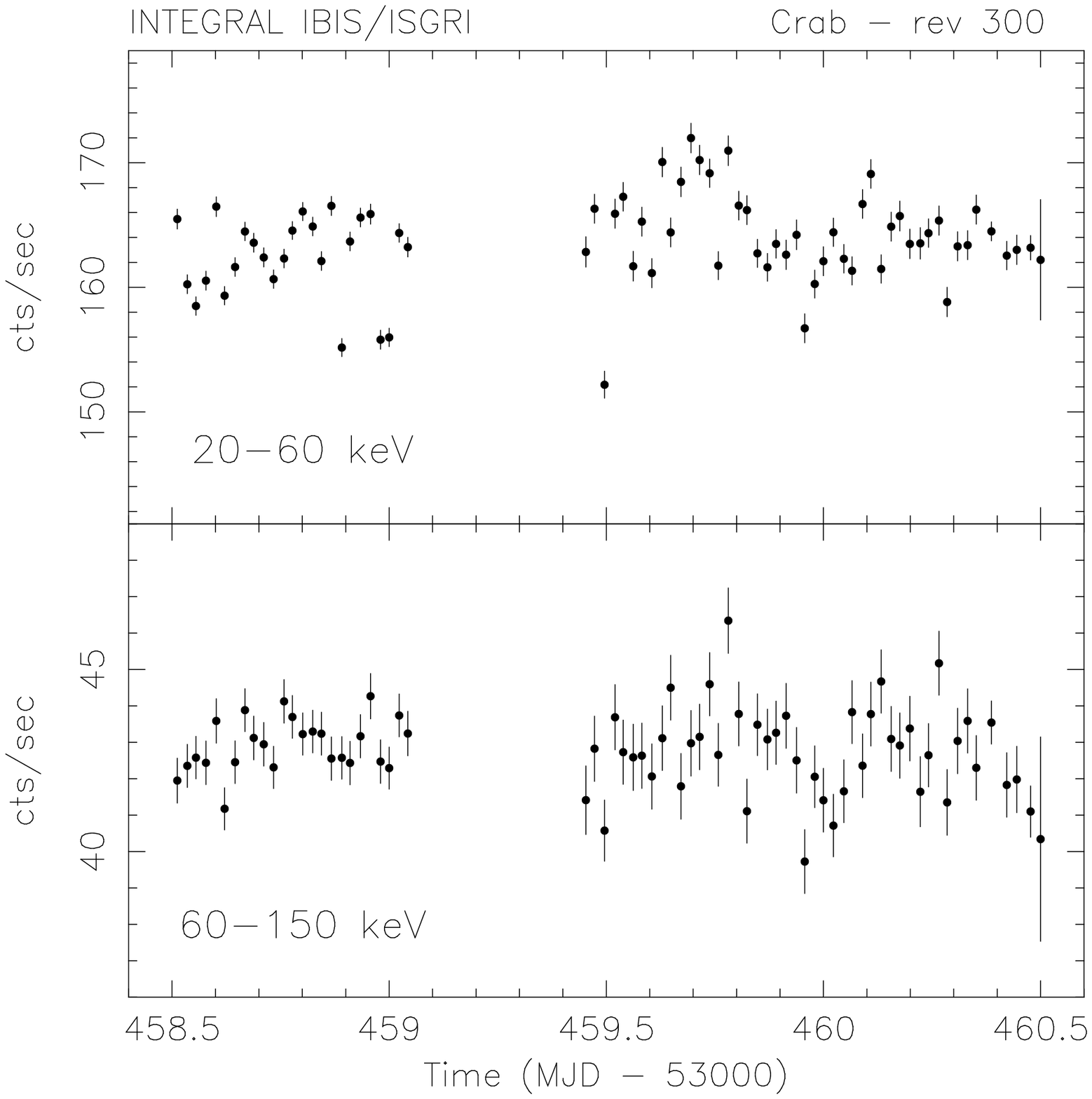}

  \includegraphics[height=.3\textheight,angle=-90]{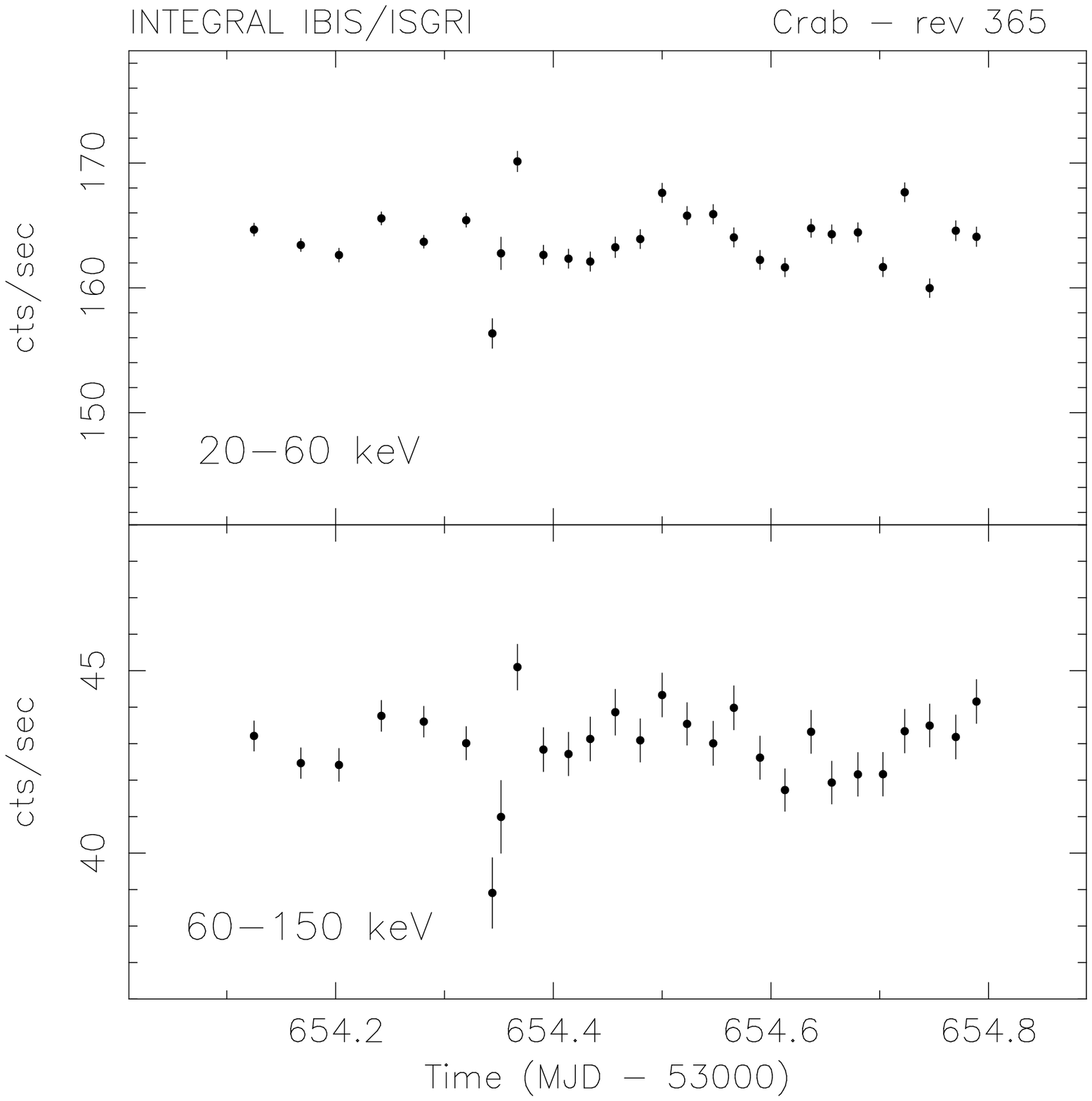}
  \includegraphics[height=.3\textheight,angle=-90]{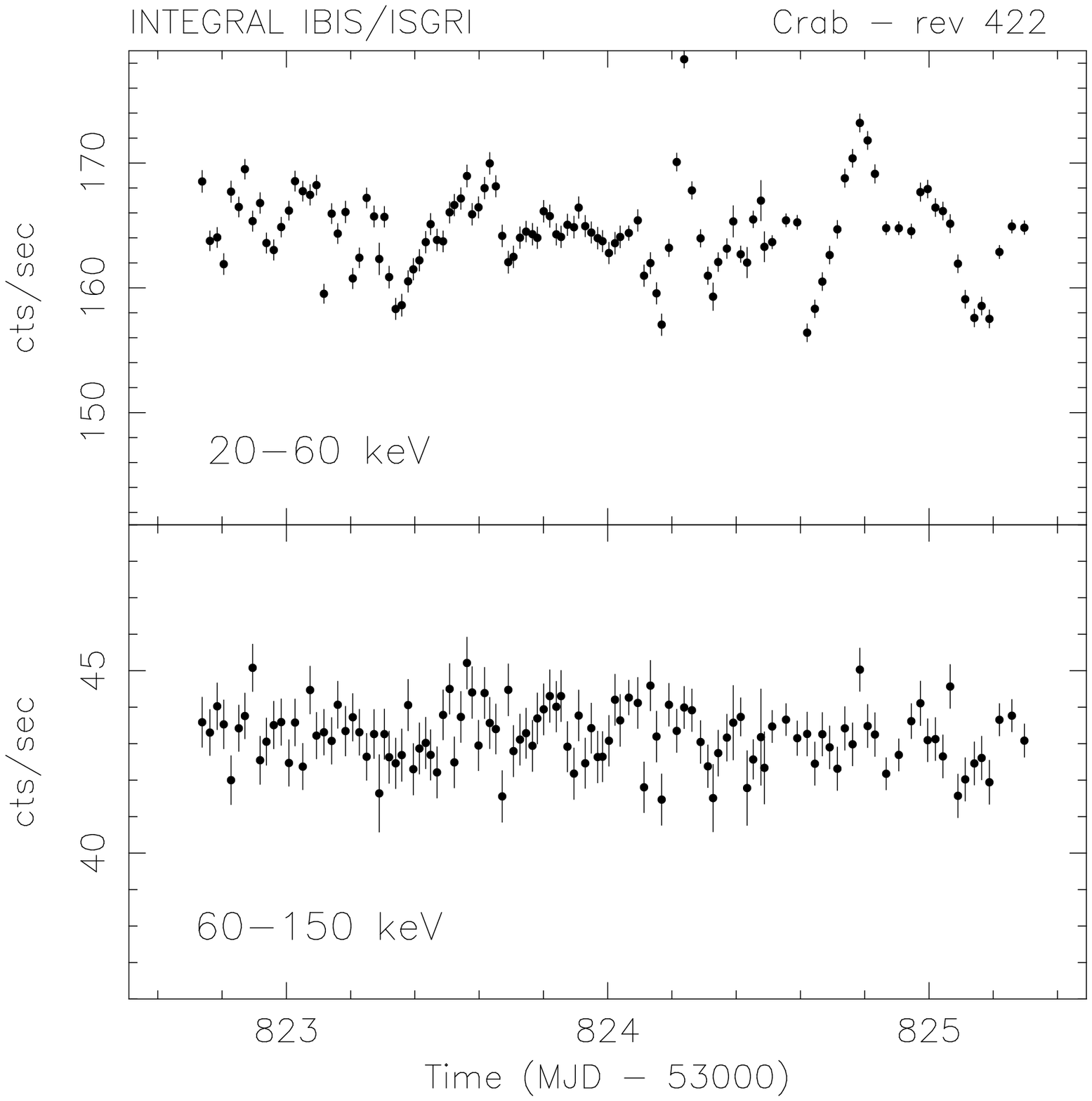}
  \caption{{\em INTEGRAL} JEM-X light curves of the Crab during the 
calibration in revolution 365 ({\it top left}) and IBIS/ISGRI light curves of the Crab during the 
calibration in revolutions 300 ({\it top right}), 365 ({\it bottom left}) 
and 422 ({\it bottom right}). Plotted per panel are the two energy bands
used: JEM-X 3--10\,keV ({\it panel top}), 10--25\,keV ({\it panel bottom}), and
IBIS/ISGRI 20--60\,keV ({\it panel top}), 60-150\,keV ({\it panel bottom}).}
\label{Crab_jmx_isgri}
\end{figure*}

\begin{figure}
\centering
  \includegraphics[height=.3\textheight,angle=-90]{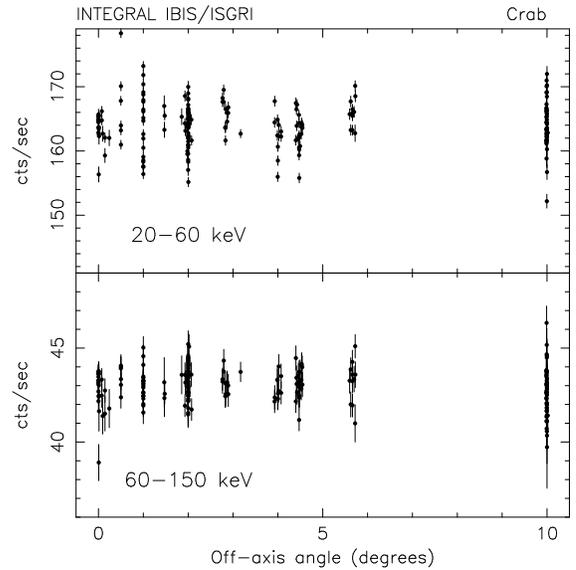}
  \caption{IBIS/ISGRI 20--60\,keV ({\it top}) and 60--150\,keV ({\it bottom}) 
Crab count rates as a function of the off-axis angle.}
\label{Crab_isgri_ang}
\end{figure}

In order to derive the average Crab rates we used the Crab calibration observations,
performed during revolution 300 (March 2005; MJD\,53457--53459), 365 (October 2005; MJD\,53654) 
and 422 (March 2006; MJD\,53822--53824).
They were performed during the same seasons as our monitoring observations.
We only used those observations with similar instrument settings as our monitoring observations;
they were analysed in the same way as described for our program (Sect.~3).
The JEM-X and IBIS/ISGRI light curves are shown in Fig.~\ref{Crab_jmx_isgri}.
The IBIS/ISGRI 20--60\,keV and 60--150\,keV count rates are largely constant 
(i.e., within about 2\%\/) across the central 
part of the field of view (see Fig.~\ref{Crab_isgri_ang}). 
The JEM-X count rates are constant within about 5\%\ (3--10\,keV) and (10--25\,keV).
The average count rates in the JEM-X 3--10\,keV and 10--25\,keV, and IBIS/ISGRI 
20--60\,keV and 60--150\,keV, are 109.8$\pm$0.2\,cts\,s$^{-1}$, 46.0$\pm$0.1\,cts\,s$^{-1}$,
164.13$\pm$0.05\,cts\,s$^{-1}$ and 43.11$\pm$0.04\,cts\,s$^{-1}$, respectively.

\section{IBIS/ISGRI 20--60\,keV sensitivity}
\label{sens}

\begin{figure}
\centering
  \includegraphics[height=.35\textheight,angle=-90]{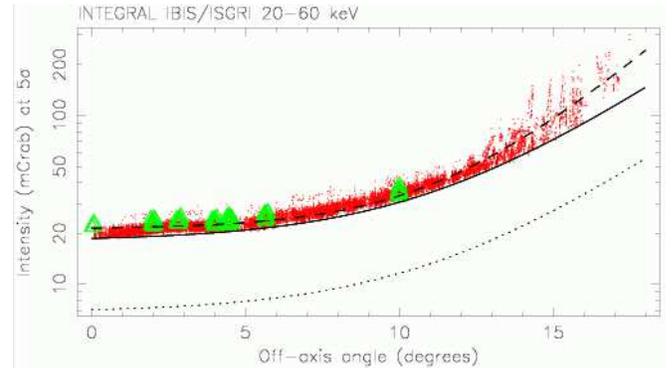}
  \caption{IBIS/ISGRI 20--60\,keV 5$\sigma$ sensitivity information. We refer to the text (Sect.~\ref{sens}) 
for a detailed explanation of the various data points and curves. 
}
\label{crab-bulge-sens}
\end{figure}

The 5$\sigma$ sensitivity of our Galactic bulge monitoring observations for the IBIS/ISGRI instrument
in the 20--60\,keV band was determined using Crab observations and as well as the monitoring 
observations themselves.

The Crab calibration observations, with exposure lengths in the range 1700--1800\,s
(i.e., similar to our monitoring observations), were selected from revolutions 300, 365 and 422
(see also Appendix~\ref{crab}). 
For each Crab detection, the value of the 1$\sigma$ uncertainty on the Crab flux 
detection was multiplied by 5 and scaled by the average Crab flux to give an estimate of the 5$\sigma$ sensitivity 
of the observations as a function of the off-axis angle (the angular distance of the Crab from the 
center of the IBIS/ISGRI field of view). These points are shown as the green triangles Fig.~\ref{crab-bulge-sens}.

In order to increase the sampling of the sensitivity curve over a wider range of off-axis angles, the same 
exercise was performed by taking the 1$\sigma$ uncertainty on every detection made in the Galactic bulge 
observations, with similar exposure times of 1800\,s. 
These are shown as red dots in Fig.~\ref{crab-bulge-sens}. It can be clearly seen that estimating 
the sensitivity from all detections agrees well with the data from the Crab, although there is some evidence 
that the sensitivity for the Crab observations could be systematically lower (i.e., the limiting fluxes are 
higher) than the Galactic bulge observations by a small amount. This may be because the brightness of the Crab 
leads to an increase in the background of the IBIS/ISGRI images.

The top dashed line in Fig.~\ref{crab-bulge-sens} is a fit of an exponential curve to the data ($y = 0.329 e^{0.362x}+21.2$, where
$y$ is the 5$\sigma$ sensitivity in mCrab and $x$ the off-axis angle in degrees).  
Below that the lower envelope of the points has been estimated to provide a lower limit to the sensitivity for a 
single 1800\,s pointing; again an exponential has been used ($y = 0.7 e^{0.29x}+18$) and is shown by the continuous line.
 
An estimate of the sensitivity for a full hexagonal dither observation (i.e., 7$\times$1800\,s observations) has been performed 
by scaling the latter curve by $\surd$7 (the signal-to-noise ratio scales with the square-root of the exposure time;
see, e.g., Goldwurm et al.\ 2003); this is shown as a dotted line in Fig.~\ref{crab-bulge-sens}. 
Note that this should only be used as a guide to the observation sensitivity, 
since the hexagonal dither pattern means that each pointing within a pattern will have off-axis angles varying 
by $\pm$2$^{\circ}$. This indicates a 5$\sigma$ sensitivity, in the 20--60\,keV band of about 7\,mCrab for 
on-axis observations, rising to 8\,mCrab at 5$^{\circ}$, 12\,mCrab at 10$^{\circ}$ and above 27\,mCrab 
outside of 15$^{\circ}$.
We note that these sensitivity fluxes are consistent with the observed fluxes for sources detected 
around 5$\sigma$ in the mosaics per revolution.
For more information on the IBIS/ISGRI sensitivities we refer to, e.g., Lebrun et al.\ (2003) and Bird et al.\ (2006).

\section{Mean fluxes from individual pointings versus average fluxes from overall mosaic image}
\label{ave_mosa}

\begin{figure}
\centering
  \includegraphics[height=.3\textheight,angle=-90]{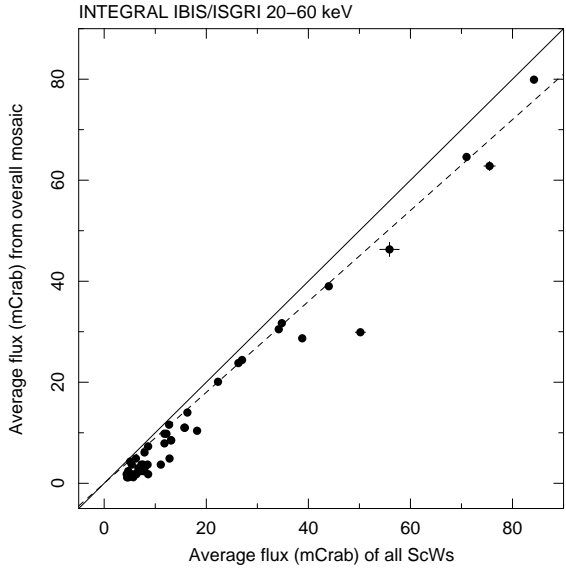}
  \caption{Average flux from the mosaic image of all pointings (ScWs) versus the mean flux over all single pointings (20--60\,keV).
Data are taken from Tables~\ref{significance} and \ref{average}; plotted are those sources which reached a significance higher than 7. 
The values for which the mosaic average flux equals the mean flux are indicated by a continuous line; the dashed line indicates
where the mosaic flux equals 90\%\ of the mean flux.}
\label{average_mosaic}
\end{figure}

As noted in Sect.~\ref{light_curves} the mean fluxes computed by averaging the flux values (weighted by their errors) 
from all the single pointings are different with respect to flux values obtained from the mosaic image
of all pointings together. There are mainly two reasons for this.
While the mosaic image merges \emph{all} the single pointings of a source (even those pointings where the source is too weak to be 
detected and contributes with a negative fluctuation), the single-pointing 
averaged fluxes are obtained by averaging only those single pointings where the source had a non-negative 
significance (and hence flux) value. This results in an overestimate of the mean flux, especially for weak sources or (bright) 
transient sources which are off for some time.
Additionally, in the mosaic images the counts associated to a source 
are spread around a single central peak, resulting in a better source location. However, the source flux is then 
also somewhat reduced (by $\simeq$10\%; see, e.g., Chernyakova 2005). 
In Fig.~\ref{average_mosaic} we show these effects by plotting the mosaic flux (from Table~\ref{significance}) 
versus the mean flux (from Table~\ref{average}) values for those source where the significance value
of detection is higher than 7. The continuous line represents where the mean flux equals the mosaic flux, whereas the 
dotted line represents where the mosaic flux is 10\%\ lower than the mean flux. Permanently bright (typically $\gtrsim$20\,mCrab)
sources indeed show a mosaic flux which is about 10\%\ lower than the mean flux. The weaker sources and bright transient sources 
(notably GRO\,J1655$-$40 and 1E\,1740.7$-$2942) clearly show the additional offset, due to our method of averaging,
in the mosaic flux with respect to the mean flux. 

\end{document}